\documentclass[12pt]{article}
\pdfoutput=1 

\usepackage{epsf}
\usepackage{epsfig,graphics}
\usepackage{graphicx}
\usepackage{epsfig}
\usepackage{subfigure}
\usepackage{tabularx}
\usepackage{rotate}	
\usepackage{slashed}

\usepackage{amsmath}
\usepackage{amsfonts}
\usepackage{amssymb}
\usepackage{graphicx}
\usepackage{cite}

\usepackage{fancyhdr}
\usepackage{hyperref}

\def\be{\begin{equation}}
\def\ee{\end{equation}}
\def\bea{\begin{eqnarray}}
\def\eea{\end{eqnarray}}
\def\bes{\begin{subequations}}
\def\ees{\end{subequations}}
\def\raw{\rightarrow}

\def\CR {{\cal R}}

\def\CF {{\cal F}}

\def\CO {{\cal O}}

\def\id{{\bf 1}}
\def\ido{{\bf 0}}
\def\IC{\mathbb{C}}
\def\sig{{\sigma}}

\newcommand{\bmat}{\left(\begin{array}}
\newcommand{\emat}{\end{array}\right)}

\def\yzero{\smash{\hbox{$y\kern-4pt\raise1pt\hbox{${}^\circ$}$}}}
\def\p{\partial}
\def\a{\alpha}

\def\g{\gamma}
\def\d{\delta}
\def\beq{\begin{equation}}
\def\eeq{\end{equation}}
\def\beqa{\begin{eqnarray}}
\def\eeqa{\end{eqnarray}}

\def\th{\theta}

\def\-{\hphantom{-}}

\def\s2{\frac{1}{\sqrt2}}

\def\oh{\frac{1}{2}}
\def\beq{\begin{equation}}
\def\eeq{\end{equation}}
\def\beqa{\begin{eqnarray}}
\def\eeqa{\end{eqnarray}}
\def\tr{{\rm tr \,}}

\def\IF{\relax{\rm I\kern-.18em F}}
\def\II{\relax{\rm I\kern-.18em I}}
\def\IP{\relax{\rm I\kern-.18em P}}
\def\IC{\relax\hbox{\kern.25em$\inbar\kern-.3em{\rm C}$}}
\def\IR{\relax{\rm I\kern-.18em R}}

\def\Dsl{\,\raise.15ex\hbox{/}\mkern-13.5mu D} 
\def\IZ{Z\kern-.4em  Z}
\def\id{{\rm I}}

\def\raw{\rightarrow}

\def\ep{\epsilon}

\catcode`\@=11   
\newdimen\@rotdimen
\newbox\@rotbox  

\def\@vspec#1{\special{ps:#1}}
\def\@rotstart#1{\@vspec{gsave currentpoint currentpoint translate
   #1 neg exch neg exch translate}}
\def\@rotfinish{\@vspec{currentpoint grestore moveto}}
\def\@rotr#1{\@rotdimen=\ht#1\advance\@rotdimen by\dp#1%
   \hbox to\@rotdimen{\hskip\ht#1\vbox to\wd#1{\@rotstart{90 rotate}%
   \box#1\vss}\hss}\@rotfinish}
\def\@rotl#1{\@rotdimen=\ht#1\advance\@rotdimen by\dp#1%
   \hbox to\@rotdimen{\vbox to\wd#1{\vskip\wd#1\@rotstart{270 rotate}%
   \box#1\vss}\hss}\@rotfinish}%
\def\@rotu#1{\@rotdimen=\ht#1\advance\@rotdimen by\dp#1%
   \hbox to\wd#1{\hskip\wd#1\vbox to\@rotdimen{\vskip\@rotdimen
   \@rotstart{-1 dup scale}\box#1\vss}\hss}\@rotfinish}%
\def\@rotf#1{\hbox to\wd#1{\hskip\wd#1\@rotstart{-1 1 scale}%
   \box#1\hss}\@rotfinish}%
\def\rotate{\@ifnextchar[{\@rotate}{\@rotate[l]}}
\def\@rotate[#1]#2{\setbox\@rotbox=\hbox{#2}\@nameuse{@rot#1}\@rotbox}

\catcode`\@=12

\topmargin
-1.5cm
\textwidth
15.5cm
\textheight
23.5cm
\oddsidemargin
0.7cm
\evensidemargin
0.7cm

\begin{document}

\makeatletter
\@addtoreset{equation}{section}
\makeatother
\renewcommand{\theequation}{\thesection.\arabic{equation}}
\pagestyle{empty}
\vspace{-0.2cm}
\rightline{IFT-UAM/CSIC-14-122}
\vspace{1.2cm}
\begin{center}


\LARGE{Higgs-otic Inflation and String Theory} 
\\[13mm]
  \large{Luis E. Ib\'a\~nez,$^{a,b}$ Fernando Marchesano$^{b}$ and Irene Valenzuela$^{a,b}$ }
  \\[13mm]
  \small{
 $^a$\,Departamento de F\'{\i}sica Te\'orica UAM\quad \quad 
$^b$\,Instituto de F\'{\i}sica Te\'orica UAM/CSIC \\ 
Universidad Aut\'onoma de Madrid \\
Cantoblanco, 28049 Madrid, Spain 
\\[8mm]}
\small{\bf Abstract} \\[7mm]
\end{center}
\begin{center}
\begin{minipage}[h]{15.22cm}
We propose that inflation is driven by a (complex) neutral Higgs of the MSSM extension 
of the SM, in a chaotic-like inflation setting. 
The SUSY breaking soft term masses  
are   of order $10^{12}-10^{13}$ GeV, which is identified with the inflaton mass scale and
is just enough to stabilise  the SM Higgs potential. 
The fine-tuned SM Higgs  has then a mass around 126 GeV, in
agreement with LHC results.  We point out that the required large field excursions  of chaotic inflation may be
realised in string theory with the   (complex) inflaton/Higgs  identified with  a continuous Wilson line or 
D-brane position. We show specific examples and study in detail a IIB orientifold with 
D7-branes at singularities, with SM gauge group and MSSM Higgs sector.
In this case the inflaton/Higgs fields correspond to D7-brane positions along a two-torus transverse to them.
Masses and monodromy are induced by closed string $G_3$ fluxes, and the inflaton potential 
can be computed directly from the DBI+CS action. We show how this action sums over Planck suppressed 
corrections,  which amount to a field dependent rescaling of the inflaton fields, leading to a linear potential in the 
large field regime. We study the evolution of the two components of the Higgs/inflaton and compute the
slow-roll parameters for purely adiabatic perturbations. For large regions of initial conditions slow roll
inflation occurs and 50-60 efolds are obtained with   $r>0.07$, testable in forthcoming experiments.
Our scheme is economical in the sense that both EWSB and inflation originate in the same sector
of the theory, all  inflaton couplings are known and reheating occurs efficiently.

\end{minipage}
\end{center}
\newpage
\setcounter{page}{1}
\pagestyle{plain}
\renewcommand{\thefootnote}{\arabic{footnote}}
\setcounter{footnote}{0}

\tableofcontents

\section{Introduction}

The discovery \cite{Higgs} at LHC of a scalar particle with the properties of the Standard Model (SM) Higgs boson 
has completed the minimum set of particles required for a consistent understanding of the 
properties of the SM.  Still, it has also triggered new questions and made more evident the 
existence of  a hierarchy problem of the fundamental  scales of physics. One of the issues raised is the stability of the 
Higgs potential \cite{elias}.
The Higgs mass, around $m_h\simeq 126$ GeV, corresponds to 
a value of the Higgs self coupling $\lambda$ such that, when extrapolated to higher energies,  
implies a metastable second minimum at a scale $10^{11}-10^{13}$ GeV. Although such
a metastable vacuum may not be necessarily problematic, it may lead to some difficulties
in the cosmological evolution of the universe. 

One elegant way to avoid any vacuum instability is to consider a SUSY extension of the 
SM like the MSSM. The scalar potential is then always positive definite in the ultraviolet and
no instabilities appear. In fact the usual MSSM with low scale SUSY breaking soft terms predict a
Higgs mass $m_h\leq 130$ GeV, in agreement with observations. So in principle one could say that a 
Higgs mass around 126 GeV could be good news for SUSY. However this value is a  bit high, and 
implies squarks and gluino masses into the multi-TeV region, very likely out of reach of the
LHC. Furthermore, a fine-tuning in the range $1\%-0.1\%$ in the SUSY parameters is required.
Although this is consistent with the fact that no trace of SUSY particles has been observed as yet 
at LHC, this high level of fine-tuning casts some doubts on the presence of SUSY at low scales $\simeq 1$ TeV.

The theoretical motivations for supersymmetry go beyond the solution of the hierarchy problem
in terms of low-energy SUSY. 
Admitting the possible presence of Higgs mass fine-tuning, one can consider leaving the scale
of soft masses $M_{SS}$ as a free parameter 
and ask for consistency with the measured Higgs mass  \cite{hebecker1,imrv,Ibanez:2013gf,hebecker2}
(see also \cite{otrosinter}).
It was remarked in ref.\cite{Ibanez:2013gf} that if the MSSM SUSY-breaking scale $M_{SS}\simeq 10^{9}-10^{13}$ GeV, 
and a fine-tuned SM Higgs survives below that scale, then necessarily one obtains  $m_h\simeq 126$ GeV, 
consistent  with LHC data. This is true if one assumes a unification boundary condition for
the two MSSM doublets $m_{H_u}=m_{H_d}$.
 One could then perhaps interpret the observed Higgs mass  as a hint  for large scale SUSY breaking in
 a unification scheme.

It is natural to discuss a possible fine-tuning of a light SM Higgs in the context of the string landscape.
In the latter  an enormous set of string solutions allow for some of them which are selected on anthropic grounds,
allowing for a sufficiently light SM Higgs. 
On the other hand SUSY is a fundament symmetry of string theory and guarantees the absence of
tachyons in string compactifications. Since string theory is at present our only complete candidate 
as a unified theory, one could consider a scenario in which SUSY could  be still present  at a higher scale 
but  not be relevant for the understanding of the hierarchy problem.

In a different direction,  evidence is mounting in favour  of the existence of a second fundamental scalar in the
theory, the inflaton.  Simple models of inflation are able to reproduce more and more qualitative and quantitative 
cosmological data (for reviews in the context of string theory see e.g.\cite{Baumann:2014nda,Westphal:2014ana,Silverstein:2013wua,Burgess:2011fa,Burgess:2007pz}). 
The  description of the CMB anisotropies in terms of primordial perturbations induced
by an inflaton is outstanding.  One of the simplest inflation models is chaotic inflation \cite{chaotic}, which features a simple
polynomial potential in which the slow roll regime is achieved due to trans-Planckian excursions of the
inflaton.  An interesting property of these models is that they generically predict large tensor perturbations at
a level detectable in future measurements.  If the BICEP2 hints \cite{bicep2}  for large tensor perturbations were confirmed,
chaotic inflation would be a favoured class of models. On the theoretical side, the requirement of trans-Planckian inflaton
excursions requires good control of Planck scale physics, i.e., a theory of quantum gravity like string theory.
In fact in the last decade a framework to embed large  trans-Planckian excursions into string theory has been
worked out in terms of the so-called {\it monodromy inflation} \cite{Silverstein:2008sg,McAllister:2008hb}, 
 see \cite{Baumann:2014nda,Westphal:2014ana} for reviews and further references.

 Given these two inputs, an obvious question has been around for some  time 
: {\it Can the Higgs boson be identified with the inflaton?}.  Before we knew the value of the Higgs 
boson mass this possibility looked unlikely, since the
Higgs potential  is quartic with no obvious region which could lead to slow roll inflation
(see e.g. \cite{higgsflation} for a review).  However, as we said,  for a Higgs mass value around 
$126$ GeV the Higgs self coupling $\lambda$ evolves down to zero at a scale $10^{11}-10^{13}$ GeV.
In fact, if one takes a $2\sigma$ uncertainty for the measured value of the top-quark mass and $\alpha_{strong}$, 
it could still be possible that we have $\lambda\simeq 0$ close the the Planck scale $M_p$. It has been proposed that
this could  be the signal of some new conformally invariant physics \cite{Shaposhnikov:2009pv}, \cite{Bezrukov:2012sa,Bezrukov:2012hx,Hamada:2014iga,Bezrukov:2014bra}. In this case  inflation could also appear with the
 inflaton identified with the SM Higgs if non-minimal gravitational couplings of type  $\int \alpha  |h|^2R$ are assumed.  While it has
 been debated whether this scheme has problems with unitarity (see e.g. \cite{Barbon:2009ya} and
 references therein),  for appropriate values
 of the parameters one may still obtain a Starobinsky-like inflation with negligible tensor perturbations. 
 See also \cite{Chatterjee:2011qr} for a SUSY Higgs inflation with small field leading also to small tensor
 perturbations.

In ref.\cite{Ibanez:2014kia}  two of the present authors suggested that the Higgs sector of the MSSM can in fact lead to a variety of chaotic inflation.
The idea is to consider a MSSM structure with a large SUSY breaking scale, with soft terms in the 
region $M_{SS}\simeq 10^{11}-10^{13}$ GeV.  As we said, such large values are consistent with a measured Higgs mass 
around $126$ GeV and, on the other hand, guarantee the stability of the SM Higgs potential. The idea was to
identify also such large scale with that of the inflaton mass $m_I\simeq M_{SS}$ \cite{Ibanez:2014zsa}
(see also \cite{Alvarez-Gaume:2013oha,Chatterjee:2011qr}).
The  SUSY breaking soft terms, induced by  string fluxes, 
give a quadratic potential to the inflaton/Higgs boson, leading to a variation of chaotic quadratic inflation.
Since chaotic inflation requires large trans-Planckian inflaton excursions, the proposal was to embed the MSSM
Higgs system into string theory with the Higgs/inflaton identified with the position of a D-brane in a  IIB orientifold model. 
Such a model could in principle give rise to large tensor perturbations, as indicated in the early reports by the 
BICEP2 collaboration.

In the present paper we complete and extend this proposal in several  ways. We show specific string heterotic and type IIB constructions in which the Higgs
bosons of the MSSM are identified with either Wilson lines or D-brane positions. We study a particular local toy model constructed from 
D7-branes at singularities in which the Higgs/inflaton corresponds to the motion of a D7  brane in a torus. 
Closed string $G_3$ fluxes induce SUSY breaking and, at the same time,  a potential for the inflaton which we obtain
from the Dirac-Born-Infeld and Chern-Simmons (DBI+CS) action  of the D7.  Both the DBI and CS contribute equal pieces to the
scalar potential.
The potential is initially quadratic along the 
D-flat direction with a structure akin to that of double chaotic inflation.
 However once the kinetic terms are normalised, the potential at large fields tends to a  linear behaviour.
One can also describe the system in terms of an $N=1$ supergravity potential, under the assumption of
SUSY breaking induced by the auxiliary field of the overall K\"ahler modulus. The potential obtained is analogous to
the one obtained from the DBI+CS actions, but it fails to capture the higher order terms in $\alpha '$ given
by the latter, terms which are responsible for the linear flattening.

One of the issues of large field inflation models is the stability of the inflaton potential against 
Planck suppressed corrections of the form $\simeq (\phi^{4+2n}/M_p^{2n})$, $n>0$.  In the 
present case the DBI+CS action sums all such corrections in a controlled manner and shows how they give rise to a
flattening of the potential. We also show how the stability of the potential may be understood in terms of
a Kaloper-Sorbo description of the effective action. Alternatively, in terms of the $N=1$ supergravity action the 
absence of additional corrections can be understood in terms of the modular symmetry of the torus.
We also show how the setting does not induce RR D3-brane tadpoles, which are often a nuisance in other
monodromy inflation models. 

The model is a 2-field inflaton model since  the D7-brane position lives on ${\bf T}^2$ and is therefore parametrised by a complex field. 
This matches with the fact that along  the $|H_u|=|H_d|$  D-flat direction of the MSSM only a complex neutral scalar remains massless before 
SUSY breaking. We study the cosmological evolution of this complex inflaton and in a first simplified analysis 
concentrate on the induced adiabatic perturbations. We compute the spectral index and tensor to scalar ratio
for a set of initial boundary conditions and its dependence also on the closed string fluxes.
Interestingly, known Higgs physics have an influence on the shape of the potential.
Indeed, the fluxes inducing soft terms must be restricted  so that  a fine-tuned massless SM Higgs survives 
below the SUSY-breaking/inflaton scale $M_{SS}\simeq 10^{11}-10^{13}$ GeV.  We then find that slow roll inflation is
obtained for wide ranges of inflaton/Higgs initial values. A number of e-folds $N_{e}=50-60$ is obtained with 
sizeable tensor perturbations $r>0.07$  depending on the 
initial values. Such large tensor perturbations should be soon tested by forthcoming data.

This paper has the following structure. 
In the next section we discuss the MSSM Higgs system and how the fine-tuning of a light SM Higgs 
can be described. There we identify the neutral complex field whose dynamics will induce inflation
in subsequent chapters. After giving a brief review of some aspects of chaotic inflation
and the structure of scales  in section \ref{inflation},
we show in section \ref{embedding} how a  minimal MSSM Higgs sector may be obtained in string compactifications.
We first describe a specific heterotic orbifold model and then a type IIB orientifold local model with
D7-branes at singularities. In the latter we show how the Higgs vevs are described in terms of the
motion of a D7-brane in a two-torus. In section \ref{fluxpotential} we describe how ISD  $G_3$ fluxes induce soft terms 
on the Higgs/inflaton fields. We obtain the induced  inflaton scalar potential  starting from the 
DBI+CS action and also show its corresponding $N=1$ supergravity description.  
We compute the slow roll parameters for the cases in which the  inflaton field redefinition is neglected.
The  final results  including the flattening effect are presented in section \ref{slow}.
Some issues regarding the stability of the Higgs potential, back-reaction and D3-brane RR tadpole cancellation 
are described in section \ref{others}. 
In section 8 further comments on issues like reheating and isocurvature perturbations are given,
while final comments and  conclusions are presented in section \ref{final}.  Details of the 
computation of the scalar potential from the DBI+CS action are given in Appendix \ref{DBI}.

\section{The Higgs mass and high scale SUSY-breaking}\label{sechiggs}

\begin{figure}[h!]
\begin{center}
{\includegraphics[width=0.49\textwidth]{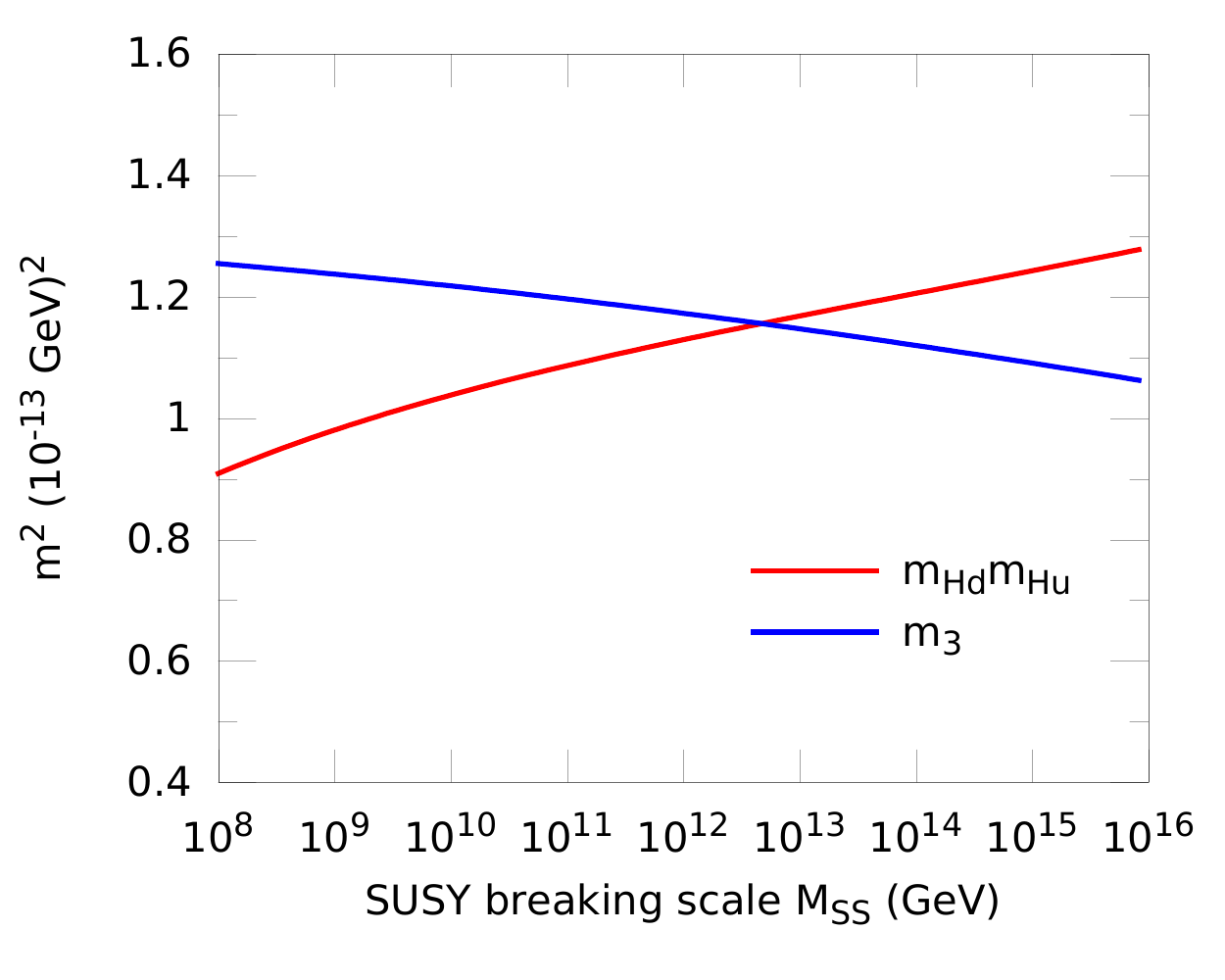}}
{\includegraphics[width=0.49\textwidth]{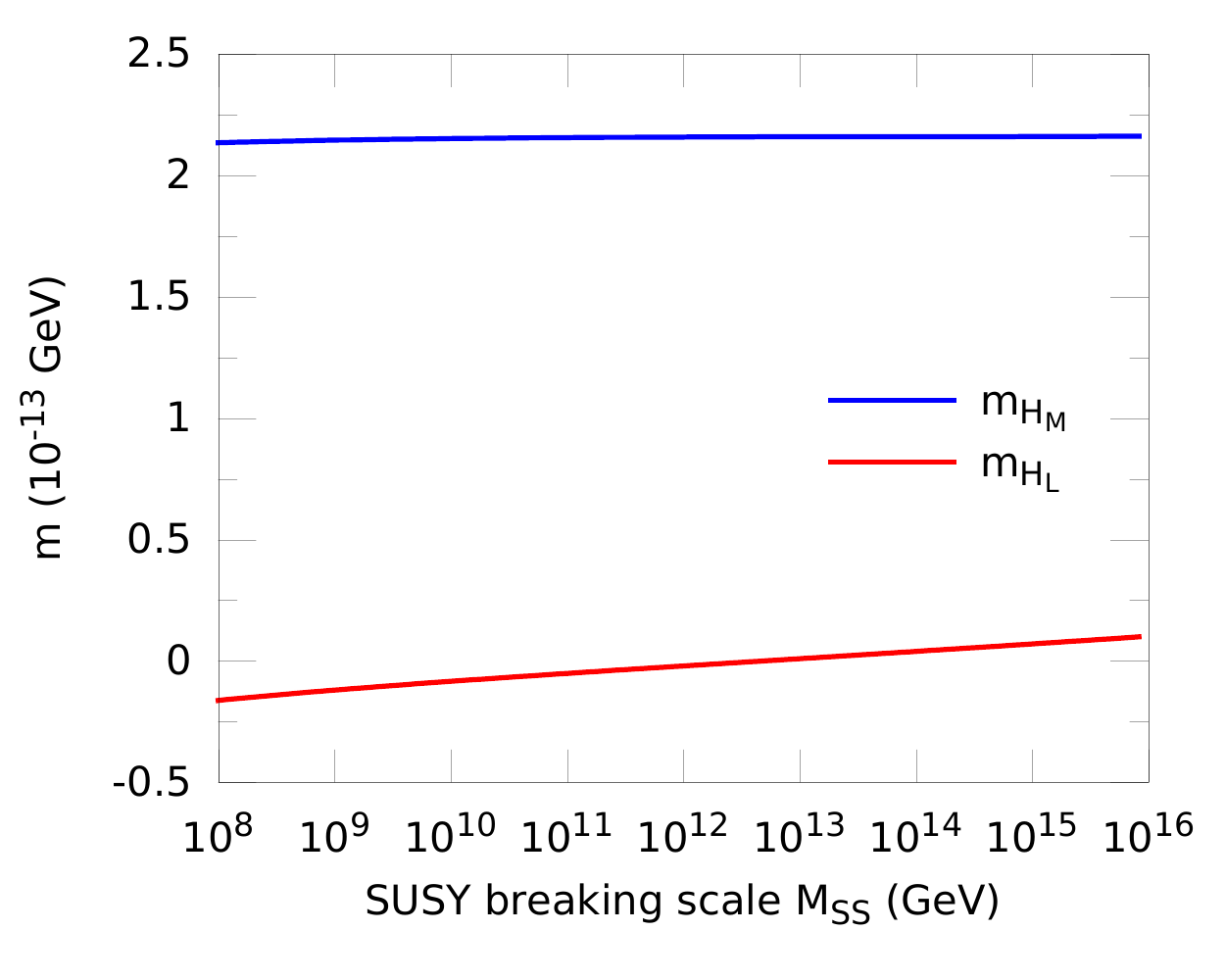}}
\end{center}
\vspace{-15pt}
\caption{Running from $M_c$ down to $M_{SS}$ of the parameters of the Higgs mass matrix (left) and of the mass eigenvalues $m_{H_M}$ and $m_{H_L}$ (right).}
\label{running}
\end{figure}

As we mentioned above,
admitting the possible presence of Higgs mass fine-tuning, one can consider leaving the scale
of soft masses $M_{SS}$ as a free parameter and ask for consistency with the measured Higgs mass.
It was remarked in ref.\cite{Ibanez:2013gf}  that if the MSSM SUSY-breaking scale is $M_{SS}\simeq 10^{9}-10^{13}$ GeV, 
and a fine-tuned SM Higgs survives below that scale, then one necessarily gets  $m_h=126\pm 3$ GeV, in 
 agreement with LHC data. Imposing gauge coupling unification and flux-induced isotropic SUSY breaking 
further points to a Higgs with a mass around 126 GeV\cite{Ibanez:2013gf}. 
This is true if one assumes the unification boundary condition for
the two MSSM doublets $m_{H_u}=m_{H_d}$, but no other further input.
 One could then  interpret the observed Higgs mass as indirect evidence for large scale SUSY breaking in
 a unification scheme.  The fine-tuned light SM Higgs  is obtained from the
 general MSSM Higgs mass matrix
 \beq
\left(
\begin{array}{cc}
 {{H_u}} \!\!\! & ,\  {{ H}_d^*}\\
\end{array}
\right)
\left(
\begin{array}{cc}
 { m_{H_u}^2} &   m_{3}\\   
  { m_3^*} & {m_{H_d}^2 }\\
\end{array}
\right)
\left(
\begin{array}{c}
  {{ H_u^*}} \\
  {{ H}_d}\\
\end{array}
\right) \,.
\label{matrizmasas}
\eeq
If one fine-tunes $|m_3|^2=m_{H_u}^2m_{H_d}^2$,
there are massless ($H_L$) and massive ($H_M$)  eigenstates
\beq
H_L \ =\ \text{sin} \beta\, e^{i\g/2} H_u - \text{cos}\beta\, e^{-i\g/2} H_d^* \ \ ,\ \ H_M\ =\  \text{cos}\beta\, e^{i\g/2}  H_u+\text{sin}\beta\, e^{-i\g/2} H_d^* \ ,
\eeq
with
\beq
\text{tan}\beta \ =\  \frac {|m_{H_d}|}{|m_{H_u}|} 
\label{tanbetacal}
\eeq
and $\g = {\rm Arg}\, m_3$. All these quantities must be evaluated at the  soft mass scale $M_{SS}\simeq 10^{10}-10^{13}$, below which all the SUSY spectrum decouples
and just the SM survives.  Note in particular that at some unification scale  $M_c > M_{SS}$ one might expect $m_{H_u}(M_c)=m_{H_d}(M_c)$ (i.e. $\text{tan}\beta= 1$), and that then the running from $M_c$ down to $M_{SS}$ will make $|\text{tan} \beta(M_{SS})|$ slightly larger than one. 
Moreover at such scale $M_c$ both scalars $H_L,H_M$ will be massive, although one will still have $m_{H_M}\gg m_{H_L}$ due to the short running in between $M_c$ and $M_{SS}$.
In fig.\ref{running} we plot the running of the Higgs mass parameters from $M_c$ down to $M_{SS}$. In the left plot we see the running of $|m_3|$ and $m_{H_u}m_{H_d}$. When both curves intersect the fine-tuning condition is satisfied and we have a massless eigenvalue at the SUSY breaking scale $M_{SS}$. This is also depicted in the right plot, in which although both mass eigenstates are massive at $M_c$, one of them ($H_L$) becomes massless after the running from $M_c$ down to $M_{SS}$.
To correctly interpret these figures recall that the running stops at the point $M_{SS}$ in which all SUSY-particles become massive and one is left just with the SM at
energies below that  given value of $M_{SS}$.
%

 In addition to the mass terms there is the  $SU(2)\times U(1)$ D-term contribution to the
scalar potential given by
\beqa
V_{SU(2)} \  &=&  \frac {g_2^2}{8} \left( |H_u|^4\ +\ |H_d|^4\ +2|H_u|^2|H_d|^2\ -\ 4|H_uH_d|^2\right) \\
V_{U(1)}\ &=&  \frac {g_1^2}{8} \left(|H_u|^4 \ +\ |H_d|^4\ -\ 2|H_u|^2|H_d|^2 \right)
\eeqa
where we have 
\beq
 H_u = \begin{pmatrix}
 H_u^+\\  H_u^0\end{pmatrix}
 \ , \qquad   H_d = \begin{pmatrix} H_d^0\\  H_d^-\end{pmatrix}
 \eeq
 all four fields being complex. Note that here $H_uH_d=(H_u^+H_d^- -H_u^0H_d^0)$,
 so in general $|H_uH_d|^2\not=|H_u|^2|H_d|^2$. The $SU(2)$ piece of the potential is however minimised if
 the charged fields have no vev, in which case $|H_uH_d|^2=|H_u|^2|H_d|^2$, so that the complete potential is then given
 by (with now only neutral components included)
\beqa
V \ &=&\ m_{H_M}^2|H_M|^2 \ +\ m_{H_L}^2|H_L|^2\ +\ \frac {g_1^2+g_2^2}{8}\left(|H_u|^2-|H_d|^2\right)^2 \\ \nonumber
 &=& m_{H_M}^2|H_M|^2 \  +\ m_{H_L}^2|H_L|^2\\ \nonumber
 &  + & \frac {g_1^2+g_2^2}{8} \left(\text{cos}2\beta (|H_M|^2-|H_L|^2) + 2{\rm sin} 2\beta {\rm Re} (H_L H_M^*) \right)^2 
\eeqa
with $m_{H_L}(M_{SS})\simeq 0$.
At this level the  $H_L$ eigenvalue is (approximately) massless and $H_M$ decouples below $M_{SS}$,
leading to the following SM quartic potential at $M_{SS}$ 
\beq
V\ =\  \frac {g_1^2+g_2^2}{8}\,\text{cos}^22\beta |H_L|^4  \  .
\label{Dterm}
\eeq
For $\text{tan}\beta (M_{SS}) \simeq 1$, as implied by the $m_{H_u}(M_c)=m_{H_d}(M_c)$ boundary condition,  one has $\text{cos}2\beta \simeq 0$, explaining
why the SM Higgs self-coupling seems to vanish at the $M_{SS}$ scale. This in turn explains, after running the Higgs self coupling down to the
EW scale, why $m_{H_L}\simeq 126$ GeV.

Note that the D-term potential has a general neutral  flat direction given by
\beq
\sigma=|H_u|=|H_d| \quad \quad \quad H_u =   e^{i\theta} H_d^*
\label{sigthetadef}
\eeq
with $\sigma \in {\bf R^+}$ and $\theta$ the relative phase of $H_u$ and $H_d^*$. Denoting $H_u=|H_u|e^{i\theta_u}$ and $H_d=|H_d|e^{i\theta_d}$ then  $\theta=\theta_u+\theta_d$. Since at $M_{SS}$ one has $\text{tan}\, \beta\simeq  1$, 
it is useful to define the  doublet linear combinations
\beq
h\ =\ \frac {e^{i\g/2}H_u - e^{-i\g/2}H_d^*}{\sqrt{2}}  \ \ ,\ \ H\ =\ \frac {e^{i\g/2}H_u+e^{-i\g/2}H_d^*}{\sqrt{2}} \ .
\eeq
Then at $M_{SS}$ the SM doublet is approximately given by $h\simeq H_L$ whereas $H\simeq H_M$  is massive.
Note that for the neutral components of $h$ and $H$ one has
\beq
H\ =\ \sqrt{2}\sigma  \text{cos}\left(\frac{\theta+\g}{2}\right)e^{i(\theta_u-\theta_d)/2} \ \ ,\ \ 
 h\ =\ i \sqrt{2}  \sigma  \text{sin}\left(\frac{\theta+\g}{2}\right) e^{i(\theta_u-\theta_d)/2}\ ,
\eeq
where $\theta=\theta_u+\theta_d$,  and the universal phase on both fields may be rotated away through a hypercharge rotation.
Then
\beq 
|H|\,+\, i |h| \ =\ \sqrt{2}\sigma e^{i\frac{\theta+\g}{2}} \ .
\eeq
Along the above mentioned  flat direction 
the potential is reduced to quadratic terms. This  suggests to consider these 
neutral  Higgs fields  $|h|$, $|H|$ (or $\sigma,\theta$) as candidates to give rise to inflation  in the manner prescribed by chaotic inflation, as we will describe below. 

\section{Large field inflation, string  theory and the Higgs}\label{inflation}

The fact that  large quadratic terms appear for the Higgs fields  above $M_{SS}$ suggests to study 
whether such fields can indeed  lead to some form of chaotic inflation.  If that were the case, the inflaton would  have a
large mass of order $M_{SS}\simeq 10^{10}-10^{13}$ GeV.
This question is interesting in itself,  but would become
particularly relevant if the indications of BICEP2 of large tensor perturbations \cite{bicep2}  were confirmed.
A straightforward interpretation of this experiment is consistent with 
chaotic large field inflation. The  inflation scale would be $V^{1/4}\simeq 10^{16}$ GeV and the inflaton mass
$m_I\simeq 10^{13}$ GeV. It was proposed in ref.\cite{Ibanez:2014zsa} to identify the large SUSY breaking scale 
suggested by the measured Higgs mass with the inflaton mass suggested by the BICEP2 data.
This indeed would be very attractive and economical, connecting two apparently totally independent 
physical phenomena, the Higgs mass with possible cosmological tensor perturbations.

Before trying to answer the above questions let us for completeness briefly review the main
ingredients in large single field chaotic inflation. 
One has a polynomial potential of the form
\beq
V(\phi) \ =\ \mu^{4-p}\phi^p \ .
\eeq
The standard slow-roll parameters for one-field
inflation are 
\beq
\epsilon\  =\ \frac {M_p^2}{2}\left(\frac
{V'}{V}\right)^2 \ll 1
\ \ \ ,\ \ \ 
\eta \ = \  M_p^2  \frac
{|V''|}{V}  \ll 1 \ ,
\label{en}
\eeq
with $M_p$ the reduced Planck scale, 
and the spectral index and tensor to scalar ratio are
\beq
n_s-1 \ =\ 2\eta -6\epsilon \ \ ,\ \  r\ =\ 16\epsilon .
\label{nsr}
\eeq
The number of e-folds is given by
\beq
N_{\rm efolds}=\frac {1}{M_p} \int^{\phi_*}_{\phi_{end}}
\frac {d\phi}{\sqrt{2\epsilon}} \ ,
\eeq
with $\phi_*$ the {\it pivot}  inflaton value and $\phi_{end}$ the inflaton
value at the end of inflation. With these definitions one obtains  the standard
chaotic inflation results
\beq
n_s-1\ =\ -\ \frac {(2+p)}{2N_{\rm efolds}}
\ \ ,\ \ r=\frac {4p}{N_{\rm efolds}} \ ,
\eeq
with a number of e-folds given by
\beq
N_{\rm efolds}\simeq \frac {1}{2p}\left(\frac {\phi_*}{M_p}\right)^2 \ .
\eeq
Obtaining of order 50-60 e-folds requires large inflaton values of order $\simeq 10-15\ M_p$,
implying, as is well known, large inflaton excursions.  For $p= 2(1)$ one obtains
large tensor perturbations with $r=0.15(0.1)$. If as hinted by BICEP2 such large tensor perturbations were indeed 
produced, they would suggest a large scale of inflation
\beq
V^{1/4} \simeq  \left(\frac {r}{0.01}\right)^{1/4}\times 10^{16}  GeV
 \simeq \ 10^{16} \text{GeV} \ \ ,\  \ 
H_I   \simeq   \left(\frac {r}{0.20}\right)^{1/2}\times 10^{14} \ \text{GeV} \ ,
\eeq
with an inflaton mass $m_I\simeq 10^{13}$ GeV.

The scheme of chaotic inflation is simple and attractive, but requires an implementation in which trans-Planckian inflation excursions make sense,\footnote{For a review with suggestions to avoid trans-Planckian excursions see \cite{peloso}.} which in turn requires a consistent theory of quantum gravity. Our most firm candidate for such a theory is string theory, and indeed string models with large field inflation have been constructed in the last decade, see \cite{Baumann:2014nda,Westphal:2014ana} for reviews. 
Natural candidates for large field inflatons in string theory are axion-like fields, which are abundant in string compactifications. Typical examples of such axions are the imaginary part of K\"ahler $T^i$ or complex structure $U^a$ moduli in type II orientifold vacua. Such axions live in a periodic moduli space, which can be nevertheless be unfolded due to extra ingredients like space-time filling branes, allowing for the required large field excursions. They moreover feature shift symmetries which also keep under control the appearance of Planck suppressed terms in the potential, i.e.
$V_p\simeq M_p^{(4-p)} \phi^p$, $p>4$ that may otherwise easily spoil inflation. 
This class of models go under the name of {\it axion monodromy inflation} because the corresponding potential grows as the axionic inflaton
completes a cycle \cite{Silverstein:2008sg,Gur-Ari:2013sba}. Some of the first such axion models \cite{McAllister:2008hb,Berg:2009tg} made use of non-SUSY configurations of NS-brane-antibrane pairs in type IIB theory (see \cite{Palti:2014kza} for a related F-theory construction). This structure was required in order to cancel unwanted D3 tadpoles and makes the stability of these models difficult to handle. 

More recently it has been realised that the same idea can be implemented in SUSY configurations if the monodromy is induced by an F-term potential for the axion \cite{Marchesano:2014mla}. Typical examples of this framework, dubbed {\em F-term axion monodromy inflation}, involve closed string axions whose potential is created by the presence of closed string background fluxes, see \cite{Marchesano:2014mla,Blumenhagen:2014gta,Franco:2014hsa,McAllister:2014mpa,Blumenhagen:2014nba,Hayashi:2014aua} for concrete realisations. A further novelty of this framework is that one can also implement the monodromy idea in models identifying the inflaton with either continuous Wilson lines or their T-dual, D-brane position moduli, see \cite{Marchesano:2014mla,Hebecker:2014eua,Ibanez:2014kia,Arends:2014qca,Hebecker:2014kva}. In the latter case large inflaton excursions correspond to a D-brane position going around some cycle in the internal compact space. Since in type II models Higgs fields arise from open string degrees of freedom, this is the path that we will follow in order to identify the inflaton with Higgs field within string theory. Namely, Higgs vevs will appear either as continuous Wilson lines or D-brane position moduli. 

Another advantage of F-term axion monodromy is that it allows to connect with the 4d axion monodromy framework developed in \cite{Kaloper:2008fb,Kaloper:2011jz,Kaloper:2014zba}. Indeed, it was found in \cite{Marchesano:2014mla} that upon dimensional reduction one obtains an effective Lagrangian of the form
\beq
-\oh \int d^4x \left[ (\p \phi)^2 + |F_4|^2 - \mu \phi *_4F_4 \right]
\label{KSLag}
\eeq
where $\phi$ is the inflationary axion and $F_4$ is a non-dynamical four-form whose presence creates a quadratic potential for it. As discussed in \cite{Kaloper:2008fb,Kaloper:2011jz,Kaloper:2014zba} (see also \cite{Kallosh:1995hi,Dvali:2005ws,Dvali:2005an,Dvali:2013cpa}) this Lagrangian is protected against dangerous corrections to the slow-roll potential that arise upon UV completion of the theory. Up to now, the Lagrangian (\ref{KSLag}) has been obtained from F-term axion monodromy constructions involving either closed string axions or open string axions arising from massive Wilson lines \cite{Marchesano:2014mla} (see also \cite{Dudas:2014pva}). As part of our analysis we will see that (\ref{KSLag}) can also be reproduced from models where the inflaton is a D-brane position modulus, which is one specific realisation of our scenario.

\begin{figure}[ht]
\centering
{\includegraphics[width=1\textwidth]{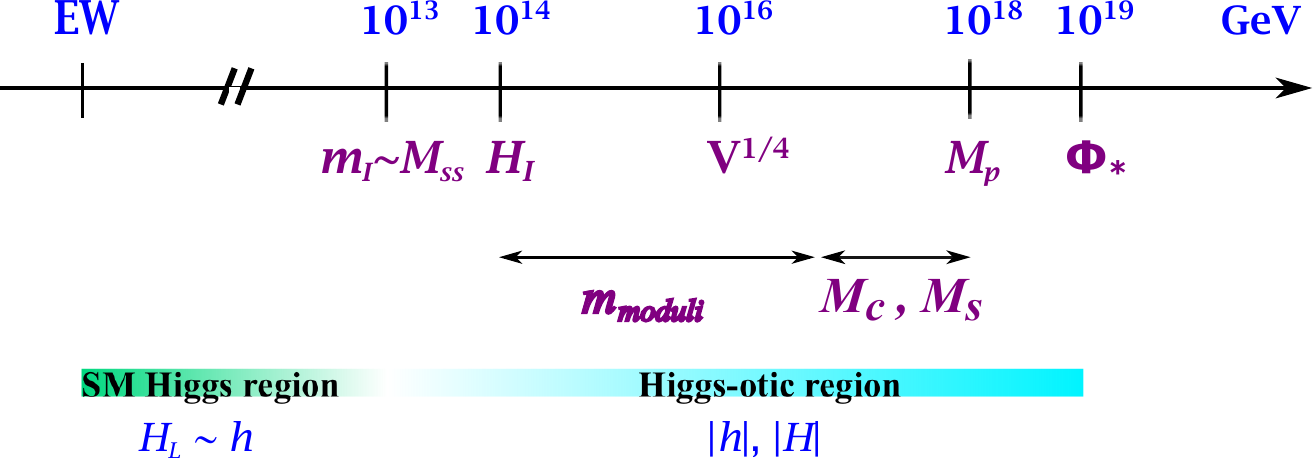}}
\caption{\small{Energy scales in the Higgs-otic Inflation scenario. Below $10^{13}$ GeV the light degrees of freedom in the Higgs sector are given by the $SU(2)$ doublet $H_L$. Above this scale $SU(2)$ is broken and they lie within the neutral components of $h$ and $H$.}}
\label{scalesfig}
\end{figure}

Before considering specific embeddings of our scheme 
let us briefly discuss the scale structure of a large field inflation string  model, see figure \ref{scalesfig}. 
The fundamental scale is the string scale which is in the region
$M_s\simeq 10^{16}-10^{18}$ GeV.  The (reduced) Planck
scale is  $M_p\simeq 10^{18}$ GeV and the inflaton initial value $\Phi_*$ is typically of order 
10-15 $M_p$ to obtain the appropriate number of e-folds. Using field theory and a scalar potential makes sense only at
energies below the compactification/unification scale  $M_c$, which should be sufficiently below $M_s$ so that the
10d action we start with makes sense.  The Hubble scale at inflation is $H_I\simeq 10^{14}$ GeV
and the inflaton mass is $m_I\simeq 10^{13}$ GeV. In the Higgs-otic scenario the latter is also of the order of the SUSY breaking  scale $M_{SS}$. 

\section{String theory embeddings of an inflaton-Higgs}\label{embedding}

In order to allow for consistent large field inflaton/Higgs,   we will search for string constructions 
in which a MSSM Higgs  sector of doublets $H_u,H_d$ appear. We want the neutral components of these
doublets to be associated with either continuous Wilson lines or position D-brane moduli.
In this chapter we will provide examples of both possibilities. 
The first example  is a compact ${\bf Z}_4$ toroidal heterotic orbifold in which Higgs fields are
identified with certain scalars in the untwisted sector. In the second example we will identify the Higgs
scalars with the position moduli of a $D7$-brane in a IIB orientifold with ${\bf Z}_4$ singularities.
The subsequent analysis will focus on this second possibility since the addition of
ingredients that give rise to monodromy  is better understood.

\subsection{The MSSM Higgs system in heterotic orbifolds}
\label{heterotic}

As a first example we will consider a Heterotic compactification in which a
MSSM-like  Higgs sector appears.
We start with the  $Spin(32)$ heterotic string compactified on a ${\bf T}^2\times {\bf T}^2\times {\bf T}^2$
torus, with each 2-torus defined in terms of an $SO(4)$ lattice. The model is subject to a twist in the compact dimensions 
defined by a ${\bf Z}_4$  shift $v=1/4(1,1,-2)$ acting on the lattices as $\pi/2$ rotations in the first two tori and a reflection $z_3\rightarrow -z_3$
in the third torus. The embedding of this twist in the $ Spin(32)$ weight lattice is given by the 16-dimensional shift
(see e.g. \cite{Ibanez:2012zz} for notation and examples)
\beq
V\ =\ \frac {1}{4}\left(1,1,1,2,2,0\ ;\ 1,1,3,0,0,0,0,0,0,0\right) \ ,
\eeq
where the SM group $SU(3)\times SU(2)$ lives in the first five entries.
In addition we add discrete order-4 Wilson lines $a_1$ and $a_2$ around the first and second torus
respectively, with
\beq
a_1\ =\  \frac {1}{4} \left(1,1,1,1,1,-1\ ;\ 0,0,-1,1,0,0,0,0,0,0\right)
\eeq
\beq
a_2\ =\  \frac {1}{4}\left(-1,-1,-1,-1,-1,1\ ;\ 0,0,-1,1,2,0,0,0,0,0\right) \ .
\eeq
As required both $4V$ and $4a_1$,$4a_2$ belong to the  $Spin(32)$ weight lattice.  The shift and Wilson lines verify the 
modular invariance constraints (see e.g. \cite{Ibanez:2012zz})
\beq
4\times  \left( (V\ \pm a_1\ \pm a_2)^2\ -\ v^2\right) \ =\ 2s, \ \ s\in {\bf Z} \ ,
\eeq
which automatically guarantee anomaly cancellation.
The projections $P.V=n$, $P.a_1=m$, $P.a_2=q$, with $P_I\in \Lambda_{Spin(32)}$ and
$n,m,q\in {\bf Z}$,  give us the invariant gauge group which is 
\beq
SU(3)\times SU(2)\times U(1)\times (SO(10)\times SU(2)'\times U(1)^6) \ .
\eeq
The chiral matter fields in the untwisted sector are obtained from $P_I$ verifying  $P.V=-1/4$ (mod integer)  but $P.a_i \in {\bf Z}$ for the first
two complex planes and $P.V=1/2$ mod integer for the third.
One gets
\beq
2(3,2)\ +\ 2({\bar 3},1) \ +\ (1,2)\ +\ (1,{\bar 2}) \ + \ hidden
\eeq
under the SM gauge group $SU(3)\times SU(2)$. By hidden we denote matter fields  not transforming with respect to this SM group.
Note there is a minimal set of Higgs fields, which is vector like, and can be identified with the $H_u,H_d$ scalars of the 
MSSM.  They are associated to the third complex plane. 
In addition the untwisted sector contains two generations of left- and right-handed  quarks, associated to the first two complex planes. 
In addition to the above matter fields, there will be additional ones from the $\theta,\theta^2$ and $\theta^3$ twisted sectors. 
They will provide for the rest of the two MSSM generations plus additional stuff, cancelling all anomalies. 
 We will not display those since they are
not relevant for our purposes.

As discussed in refs.\cite{continuousWL} the vevs of  untwisted fields in an orbifold along D-flat directions
correspond to switching on continuous Wilson lines in the underlying torus, in this case along the third torus.
So this is an example of  a consistent global string construction in which MSSM-like Higgs vevs  are parametrised by 
continuous Wilson lines.

The inflation potential is however flat so far. In order to obtain a potential (and hence a mass) for the
Higgs/inflaton system we would need some source of monodromy. A natural source could be the presence of some sort of
fluxes, like those geometric fluxes present in the definition of massive Wilson lines given in \cite{Marchesano:2014mla}.
However our understanding  of fluxes in heterotic compactifications is still quite incomplete compared to that in
type IIB compactifications. This is why  in the next section we turn to the description of the Higgs/inflaton system in type IIB orientifolds.

Before turning to the IIB case let us recall what is the structure of the K\"ahler potential involving
untwisted matter and moduli fields in ${\bf Z}_{2N}$ orbifolds in which one complex plane (i.e., the third) suffers
only a twist of order 2.  In this case the untwisted matter fields associated to the third complex plane
are vector like, i.e., chiral matter  multiplets $A,B$ with opposite gauge quantum numbers, like is the case for $H_u, H_d$ in the MSSM.
This is what happens in the ${\bf Z}_4,{\bf Z}_6', {\bf Z}_8'$ and ${\bf Z}_{12}'$ heterotic orbifolds, (see e.g. \cite{Ibanez:2012zz}). 
Then the K\"ahler potential has a contribution of the form
\beq
K\ =\ -{\rm log} \left[(T_3+T_3^*)(U_3+U_3^*)\ -\ \frac {\alpha'}{2}\ (A+B^*)(A^*+B)\right]  ,
 \label{kahlershift}
 \eeq
 where $T_3$ and $U_3$ are the K\"ahler and complex structure modulus of the ${\bf T}^2$ in the
 third complex direction.  In the above ${\bf Z}_4$ example we will have that $A+B^*=H_u+H_d^*$.
 The consequences of this structure, which is also present in the type IIB orientifold model of next subsection, 
 will be discussed in sections \ref{sugra} and \ref{corr}.

\subsection{The MSSM Higgs system in type IIB orientifolds}
\label{HiggsIIB}

In this second example we  will concentrate on type IIB compactifications with $O3/O7$ orientifold planes, in which the addition of RR and NS 3-form fluxes is at present best understood. The addition of these fluxes will give rise to the desired monodromy for the
inflaton/Higgs. This is so for the position moduli of D7-branes which are directly sensitive to the presence of ISD closed string 3-fluxes.\footnote{The case of D3-branes (or rather anti-D3-branes)  would be more subtle  since they  may feel the presence of ISD fluxes only through the back-reaction of the geometry, see \cite{Camara:2003ku}.}  In what follows we will thus concentrate on the case in which one identifies the Higgs/inflaton field with a D7 position modulus in a IIB orientifold

In particular, we will consider a type IIB O3/O7 orientifold with a stack of D7-branes sitting on a ${\bf Z}_4$ singularity, with a local
geometry of the form $(X\times {\bf T}^2)/{\bf Z}_4$, with X some complex two-fold. The D7-branes are transverse to the ${\bf T}^2$ and are initially located at its origin, on top of the singularity. The D7-branes wrap the compact 4-cycle $X$ which may be taken  to be ${\bf T}^4$ for simplicity, but whose structure will not be crucial for the relevant Higgs sector. We will consider this setting as a local model and do not care much about global RR tadpoles.

Examples of D-brane models in the case where $X={\bf T}_4$  have been given in \cite{Camara:2004jj,Marchesano:2004yn,Ibanez:2014kia}. Such orbifold has a geometric action of the form 
\be
\th :\ (z_1,z_2,z_3)\ \mapsto\ (e^{-2\pi i/ 4} z_1, e^{-2\pi i/ 4} z_2, e^{2\pi i/ 2} z_3)\ = \ (-i z_1, -iz_2, -z_3)
\label{geomZ4}
\ee
encoded in the shift vector $v = \frac{1}{4} (1,1,-2)$, as in the previous heterotic example.
We then consider a stack of $N$ D7-branes extended over 
the first two complex coordinates, and such that the action of the orbifold generator $\theta$ on the Chan-Paton degrees of freedom is
\be
\g_{\th,7} = {\rm diag\ } \left(\id_{n_0}, i \id_{n_1}, - \id_{n_2}, -i \id_{n_3} \right) 
\label{ChanZ4}
\ee
with $\sum_{i=1}^4 n_i = N$. Implementing the standard procedure  (see e.g.\cite{Ibanez:2012zz})
one obtains the following spectrum for open strings in the 77 sector:
\be
\begin{array}{ll}\vspace*{.2cm}
{\rm \bf Vector\ Multiplets} & \prod_{i=1}^{4} U(n_i) \\
{\rm \bf Chiral\ Multiplets} & \sum_{r=1}^{3} \sum_{i=1}^4 (n_i, \bar{n}_{i+4 v_r})
\end{array}
\label{specD3}
\ee
where the index $i$ is to be understood mod 4. 

Let us now follow \cite{Ibanez:2014kia} and consider the case where $n_0 =1$, $n_1= 3$, $n_2 = 2$, $n_3 = 0$. The spectrum in the 77 sector is then given by a gauge group $U(3) \times U(2) \times U(1)$ and matter spectrum
\be
2 \times (\bar{3},1)_{+1} \, +\, 2\times (3, \bar{2})_0 \, +\, (1,\bar{2})_{+1} \, +\, (1,2)_{-1}
\ee
where the subscript stands for the charge under the U(1) of the 0$^{th}$ node.

What is more relevant for us is how these representations arise in terms of the original stack of D7-branes and its fields, which correspond to three adjoints $(A_{\bar{1}}, A_{\bar{2}}, \Phi)$ of U(6). After performing the orbifold projection we obtain that these matrices get projected down to off-diagonal entries that contain the above matter fields. More precisely
\be
A_{\bar{i}}\, =\, 
\left(
\begin{array}{ccc}
\ido_3 & Q_L^i & \\
& \ido_2 & \\
 U_R^i & & 0
\end{array}
\right) \quad \quad
\Phi\, =\, 
\left(
\begin{array}{ccc}
\ido_3 &  & \\
& \ido_2 &  H_u \\
 & H_d & 0
\end{array}
\right)
\label{mat}
\ee
where we used standard notation to label the matter fields.\footnote{In (\ref{mat}) we have made a change of basis so that (\ref{ChanZ4}) reads $\g_{\th,7} = {\rm diag\ } \left(i \id_3, -\id_2, 1\right)$.} In particular the hypercharge generator  is given by the non-anomalous U(1) combination
\be
Q_Y\, =\, -\frac{Q_3}{3} - \frac{Q_2}{2} - Q_1
\ee
where $Q_n$ is the generator for $U(1) \subset U(n)$. This justifies the following notation for the Higgs sector
\be
H_u\, =\, (1,2)_{-1} \quad \quad H_d\, =\, (1,\bar{2})_1
\ee
The other two $U(1)$'s within the local model  are anomalous and become massive through the GS mechanism. 
From (\ref{mat}) one can compute the Yukawa couplings of this system by using the D7-brane superpotential formula
\be
W\, =\, \tr ([A_{\bar{1}},A_{\bar{2}}]\Phi) \, \raw \,  Q_L^2 H_u U_R^1 - Q_L^1 H_u U_R^2 
\label{yukis}
\ee
or simply orbifold CFT techniques. Here superindices denote generations.
Notice that the representation $H_d$ does not enter in the superpotential, which is to be expected because the representation $D_R$ will only appear when we include fractional D3-branes that cancel the twisted tadpoles of the model. 
One can also compute the D-term potential of this model  from
$V_D \sim \tr D D^\dag$
with 
$D\, =\, [A_{\bar{1}}, A_1]$$ + [A_{\bar{2}}, A_2] $$+ [\Phi, \bar{\Phi}]$.  From here one obtains the D-term quartic potential described in section 2.

The twisted tadpole cancellation conditions  allow for sets of D7-branes with traceless contribution to quit the
singularity and to travel to the bulk. In particular if one of the two $U(2)$ branes combines with  the $U(1)$ brane,
they do not give net contribution to the tadpole and can travel through  the bulk, in particular they can  travel 
over through  ${\bf T}^2$ in the $z_3$ direction. They should do that in a way consistent with the ${\bf Z}_4$ symmetry, which acts on $z_3$ through a the reflection $z_3\rightarrow -z_3$, and so the two wandering D7-branes should travel at mirror locations $z_3$ and $-z_3$ respectively. 
When that happens,  the 4 D7-branes remaining on the singularity have gauge group $U(3)\times U(1)$ whereas the
wandering couple  carries a single U(1).  Taking into account that the GS mechanism gave masses to two U(1)'s, a single 
$U(1)_{em}$ remains unbroken, corresponding to electromagnetism. All in all there is a symmetry breaking process
\beq
U(3)\times U(2)\times U(1)\rightarrow SU(3)\times SU(2)\times U(1)_{Y}\rightarrow SU(3)\times U(1)_{em} \ ,
\eeq
whereas the first symmetry breaking is due to the GS mechanism, and the last one is due to the Higgs mechanism induced
by the wandering pair of branes.

\begin{figure}[h!]
\begin{center}
{\includegraphics[width=0.4\textwidth]{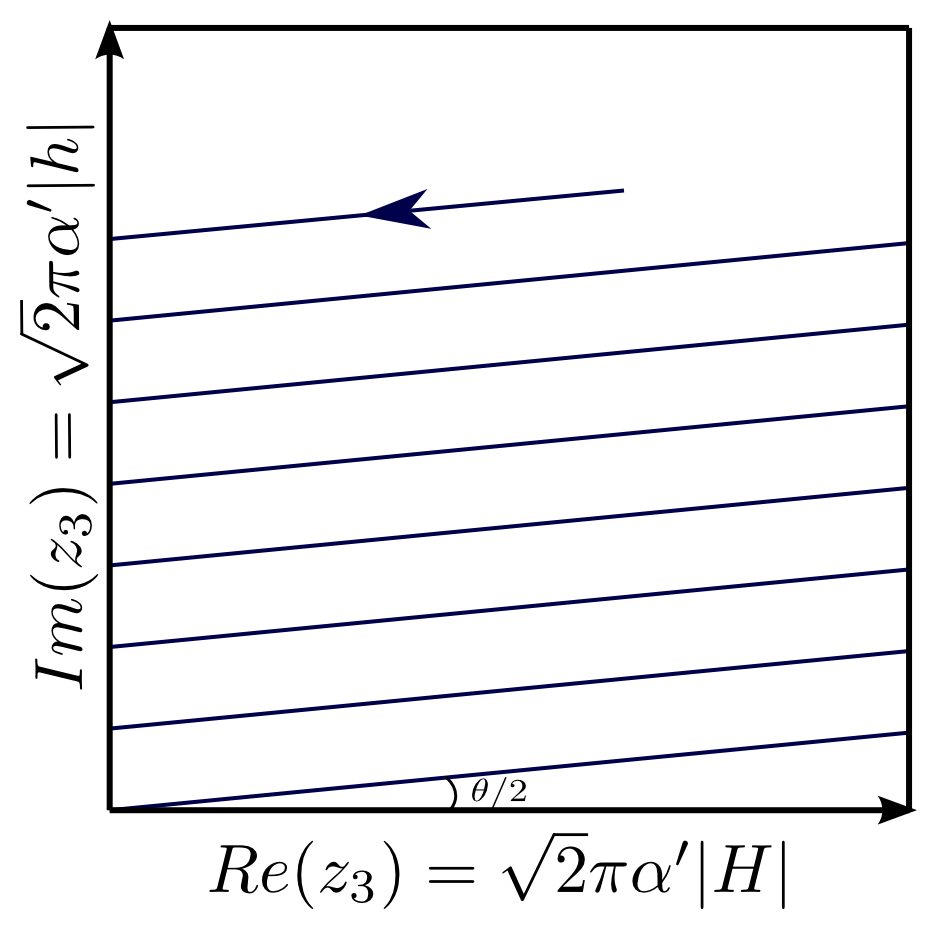}}
\end{center}
\vspace{-15pt}
\caption{A possible trajectory of the inflaton/Higgs D7-brane cycling around the ${\bf T}^2$ before
fluxes are turned on.}
\label{V}
\end{figure}

The fact that the wandering   D7's  can travel freely through $T^2$  corresponds to the existence of a flat direction
 $|\langle(1,{\overline 2})\rangle|=|\langle({\overline 1},2)\rangle|$, i.e., $|H_u|=|H_d^*|$.   The position of the D7-brane as it moves in the third
 ${\bf T}^2$ is parametrised by the vevs ($\sigma , \theta$). In particular one has for this coordinate\footnote{Note that it is $z_3^2$ , which is invariant under the ${\bf Z}_2$ reflection, which is well defined in the orbifold quotient space, rather than $z_3$ itself.}
 \beq
 z_3^2 \ =\ (2\pi \alpha ' ) ^2 \sigma^2 e^{i\theta} \ =\ (2\pi \alpha')^2  H_uH_d\ =\ (2\pi \alpha')^2 \frac {(|H|+ i |h|)^2}{2} e^{-i\g}
 \label{zetilla}
 \eeq
 Thus $2\pi\alpha' \sigma$ 
   corresponds to the distance of the wandering D7-branes to the branes remaining at  the ${\bf Z_4}$ singularity. 
   This separation corresponds to spontaneous gauge symmetry breaking.
      A possible trajectory  of the wandering-D7/Higgs/inflaton  branes over ${\bf T}^2$ is illustrated in figure \ref{V}, where we assume $\g=0$.
   The open strings going from the $D7$ to the singularity will give rise  to 
   massive $W^\pm,Z^0$ gauge bosons and their SUSY partners.
 In particular, consider a D7-brane at the point $z_3 = x + iU_3 y$, where $iU_3$ is the complex structure of the third ${\bf T}^2$. 
 Then the mass formula for the open string states between the singularity and the D7-brane is given by 
 \beq
 M^2 \ = \frac{1}{(2\pi  \alpha')^2}  \left| z_3 - (w_1 +  iU_3 w_2)2\pi R\right|^2,
 \label{winding}
 \eeq
 where $w_{1,2}$ are the winding numbers around the two cycles of the transverse ${\bf T}^2$, whose radius along $x$ is given by $R$. We thus obtain $M^2  = \sigma^2$ for $w=0$ and small $x$, so that the mass is controlled by $\langle\sigma\rangle$.\footnote{The  familiar factor proportional to the square of the gauge coupling appears upon normalising the fields canonically.}

The massive states include not only $W^\pm, Z^0$, but also three massive scalars $H^{\pm}, h^0$, which are  the scalars 
included  in the $N=1$ SUSY massive vector multiplets. The counting of degrees of freedoms is as follows: We start with 8 real scalars
from $H_u,H_d$. Three of them become goldstone bosons, whereas other three ($H^\pm,h^0$), complete a massive vector multiplet.
The two remaining scalars are massless at this level, and correspond to the two neutral scalars from $\sigma,\theta$, which parametrise the position of the D7 wandering branes through the third ${\bf T}^2$. In the model the 2 families of quarks become also massive due to the Yukawa couplings in eq.(\ref{yukis}).

Note that  the Higgs vev $\sigma$ may be arbitrarily large, even larger than the Planck scale. This however {\it does not lead to new states with masses larger than 
$M_p$}.  In particular this applies to the massive $W^\pm,Z^0$ boson and their partners, which can never get masses larger than the 
KK scale of the ${\bf T}^2$. Indeed, as shown in eq.(\ref{winding}), for $|z_3|>\pi R$, the lightest states to be identified with these bosons correspond to  winding numbers $w_{1,2}\not=0$, and no longer to the initial states with $w_{1,2}=0$. 
In this sense the {\it effect of the inflaton/Higgs vev in this string context is very mild}, not deforming the structure of the  KK/string spectra in a substantial manner. This is to be contrasted to a purely 4d field theory model of the MSSM  in which the gauge boson masses {\it are } proportional to the vev of the scalar and hence would produce masses larger than $M_p$, with physics difficult to control, if at all.

Note that an interesting property of the wandering D7-branes is that, as the position varies and the inflaton vev decreases, the masses
of $W^\pm$, $Z^0$ etc. decrease in an oscillating manner, since the distance of the brane to the singularity also oscillates. 
In some particular limits in which the brane travel along one of the axis or the diagonal, these fields become periodically massless
as the vev of the inflaton decreases.  This is however not generically the case, and it will not be the case in the 
relevant Higgs-otic model.

\section{Fluxes and the Higgs/inflaton potential}\label{fluxpotential}

In the previous section we have discussed how a vev within the MSSM Higgs sector may be understood in terms of the
motion  of a  D7-brane on a ${\bf T}^2$. However, up to now the full scalar potential is flat along such
D-term flat direction. We will now induce mass terms for the inflaton/Higgs as required in oder to obtain a
chaotic-like potential. To do that we will consider the case in which there are imaginary self-dual (ISD) 
3-form fluxes  $G_{3}$ acting as a background. As is well known, such classes of ISD fluxes are solutions of the type IIB 10d 
equations of motion in warped Calabi-Yau backgrounds \cite{Giddings:2001yu}.
In such type IIB compactifications there are two types of ISD fluxes, with tensor structure $G_{(0,3)}$ and $G_{(2,1)}$ respectively.
The first class breaks SUSY and induces SUSY-breaking soft-terms: scalar and gaugino masses. The second class preserves SUSY and may induce
supersymmetric F-term masses to the chiral multiplets. These  flux-induced  terms were analysed  in 
\cite{Camara:2003ku,Grana:2003ek,Camara:2004jj,Lust:2004fi,Aparicio:2008wh,Camara:2014tba}.
In our discussion below we will consider the generic case in which both classes of fluxes are turned on 
simultaneously. More precisely, we will consider the following closed string background
\begin{align}
ds^2 &= Z(x^m)^{-1/2} \eta_{\mu\nu} d\hat{x}^\mu d\hat{x}^\nu + Z(x^m)^{1/2} ds^2_{\rm CY} \label{back1}\\
\tau &= \tau(x^m) \nonumber\\
G_3 &= \frac{1}{3!}G_{lmn}dx^l\wedge dx^m\wedge dx^n \nonumber\\
\chi_4 &= \chi(x^m)d\hat{x}^0\wedge d\hat{x}^1\wedge d\hat{x}^2 \wedge d\hat{x}^3 \nonumber \\
F_5 &= d\chi_4 + *_{10}d\chi_4 \nonumber
\end{align}
with $\tau = C_0 +ie^{-\phi}$  the 10d axio-dilaton, $Z$ a warp factor that depends 
on the internal coordinates $x^m$, and $ds_{\rm CY}^2$ the Ricci-flat metric of the internal covering space, namely ${\bf T}^4\times {\bf T}^2$. Finally, $G_3=F_3-\tau H_3$ is the complexified three-form flux, with $F_3$ and $H_3$ the RR and NSNS fluxes respectively. As mentioned before we take this flux to be of the form $G_3 = G_{(0,3)} + G_{(2,1)}$, and in particular we choose $G_{(0,3)}  = G_{\bar 1\bar 2\bar 3}\, d\bar z_1 \wedge d\bar z_2 \wedge d\bar z_3$ and $G_{(2,1)} = G_{\bar 1\bar 2 3}\, dz_1 \wedge dz_2 \wedge d\bar z_3$, as these are the two fluxes that are invariant under the ${\bf Z}_4$ action (\ref{geomZ4}). Since we are considering only ISD 3-form fluxes, the background dilaton $\tau$ must be holomorphic in order to satisfy the IIB supergravity equations of motion. For simplicity we will consider $\tau$ to be constant, although our results can easily be generalised for a non-constant profile. 

The potential for the fields living in the D7-brane worldvolume can be obtained by evaluating the D7-brane DBI+CS action in the above background, as we now describe.

\subsection{Flux induced scalar potential from DBI+CS}
\label{DBICS}

We will consider again the toroidal setting and compute the effect of the $G_3$ fluxes 
on the  $U(6)$ adjoint complex scalar existing in the model in the previous section before orbifolding.
This adjoint contains off-diagonal components containing the $H_{u,d}$ fields of interest, which we will
display at the end.

The effective action for the microscopic fields of a system of D7-branes in the 10d Einstein frame is given by the Dirac-Born-Infeld (DBI) + Chern-Simons (CS) actions
\beq
S_{DBI}=-\mu_7 g_s^{-1}\text{STr}\left(\int d^8\xi\sqrt{-\text{det}(P[E_{MN}]+\sigma F_{MN})\text{det}(Q_{mn})}\right)
\label{DBID7}
\eeq
\beq
S_{CS}=\mu_7g_s\text{STr}\left(\int d^8\xi P\left[-C_6\wedge B_2+C_8\right]\right)
\label{CSD7}
\eeq
where
\beqa
E_{MN}=g_s^{1/2}G_{MN}-B_{MN}\quad \quad Q^m_n=\delta^m_n+i\sigma[\phi^m,\phi^\rho]E_{\rho n}\quad \quad \mu_7=(2\pi)^{-3}\sigma^{-4}g_s^{-1}
\eeqa
and $\sigma=2\pi\alpha'$. Here $M,N$ are D7-brane worldvolume indices and $P[\cdot]$ denotes the pullback of the 10d background onto such worldvolume, while $m,n$ are indices transverse to the D7-brane. Finally, `STr' stands for the symmetrised trace over gauge indices.\footnote{The  parameter $\sigma$ in here should not be confused with the inflaton field $\sigma$ defined in eq.(\ref{sigthetadef}).}

The D7 world volume spectrum  compactified to 4d contains before orbifolding  two 
adjoints $A_{1,2}$ which come from 8d vectors and an adjoint $\Phi$ which parametrises 
the D7-position and that will be the subject of our interest.
The determinant in the DBI action  can be factorised between Minkowski and the internal space (labelled by $\mu,\nu$ and $a,b$ indices respectively) and after some calculations we obtain
%
%
\beq
\textrm{det}(P\left[E_{MN}\right] + \sig F_{MN}) = - g_s^4 f(B)^2 \left[ 1 + 2Z\sigma^2D_\mu\Phi D^\mu\bar\Phi+\frac{1}{2g_s}\sigma^2 Z F_{\mu\nu}F^{\mu\nu}\right]
\label{detP}
\eeq
and
\beq
\textrm{det}(Q_{mn})=1-\frac{Zg_s\sigma^2}{2}[\Phi_m,\Phi_n]^2
\label{Qmn}
\eeq
where  
\beq
f(B)^2=1+\frac12 Z^{-1}g_s^{-1}B_{ab}B^{ab}-\frac{g_s^{-2}}{4}Z^{-2}B_{ab}B^{bc}B_{cd}B^{da}
+\frac{g_s^{-2}}{8}Z^{-2}\left[B_{ab}B^{ab}\right]^2
\label{fB0}
\eeq
The details of the computation can be found in Appendix \ref{DBI}. Recall that $Z$ is a possible warp factor which we will often set to unity
when doing explicit computations. Nevertheless, a non-constant warp factor  might have interesting phenomenological consequences, as we will briefly discuss later on.  The contribution coming from \eqref{Qmn} will give rise to the usual D-term potential. Since this term does not change formally when including the $\alpha'$ corrections, we will skip it in the computation below and restore it only at the end of the section, to avoid clutter.

For simplicity we are not considering neither Wilson lines nor  magnetic fluxes on the branes worldvolume, that is,  we are setting $\langle A_a\rangle =0$. In our configuration only the adjoint field $\Phi$ will take a non-zero vacuum expectation value, which will parametrise the position of the D7-branes in their transverse space $z_3$ via the equation
\beq
{\rm det}\left(\langle \Phi\rangle-\sigma^{-1}z_3I\right)=0\ .
\eeq 
For this reason, in \eqref{detP} we have already neglected all the terms that are not relevant for the scalar potential (like $BF$, $FF$ and $[A,\Phi]$ couplings), since they vanish for $\langle F \rangle = \langle A\rangle=0$.  Notice however that we have kept all those depending only on $B$ to all orders. The reason is that, in the presence of a background three-form flux $H_3$, changing the vev of $\Phi$ induces a B-field on the D7-brane worldvolume. Hence, since our model of inflation the vev $\langle \Phi \rangle$ is going to take large values, we cannot neglect the dependence on $B$ to any order in the DBI expansion.

Let us for now ignore the contribution coming from ${\rm det} (Q_{mn})$, which gives the D-term scalar potential.
Then, plugging (\ref{detP}) into the DBI action (\ref{DBID7}) we obtain
\beq
S_{DBI}=-\mu_7 g_s\text{STr}\int d^8\xi\,  \sqrt{f(B)^2\left[ 1+ 2Z\sigma^2  D_\mu\Phi D^\mu\bar\Phi+\frac12 Zg_s^{-1}\sigma^2 F_{\mu\nu}F^{\mu\nu}\right]}
\eeq
with $f(B)$ the same as in (\ref{fB0}). One can check that whenever the B-field is a $(2,0)+(0,2)$-form on the D7-brane internal worldvolume $f(B)$ can be written as
\beq
f(B)\, =\, 1+\frac12 Z^{-1}g_s^{-1}B^2
\label{fB}
\eeq
where we have denoted $B^2\equiv B_{ab}B^{ab}/2$ and used that $4B_{ab}B^{bc}B_{cd}B^{da}=\left[B_{ab}B^{ab}\right]^2$. This implies that all corrections in $\alpha '$, which appear as powers of the B-field in $f(B)^2$, can be completed into a perfect square. The reason is the underlying supersymmetry of the system, which imposes that for a worldvolume flux $\CF$ which is a self-dual two-form on the D7-brane internal dimensions the D7-brane gauge kinetic function must be holomorphic on the axio-dilaton $\tau$, while for an anti-self-dual two-form it must be anti-holomorphic. In both cases (ours being the second) no square roots should appear multiplying $F_{\mu\nu}F^{\mu\nu}$, because there are none multiplying $F_{\mu\nu}\tilde F^{\mu\nu}$. We refer to Appendix \ref{DBI} for further details.

Even if $\Phi$ is supposed to take large vacuum expectation values their derivatives must remain small, since we are interested in slow-roll dynamics. We can then expand the square root neglecting higher orders in $\partial_\mu\Phi$, obtaining
\beq
S_{DBI}=-\mu_7 g_s\text{STr}\int d^8\xi  f(B)\left[1+Z\sigma^2 D_\mu\Phi D^\mu\bar\Phi+\frac14 Zg_s^{-1}\sigma^2F_{\mu\nu}F^{\mu\nu} +\mathcal{O}(\partial^4)\right]
\label{DBIfin}
\eeq
where we have taken the same approximation for $A_\mu$ and its derivatives.

In order to proceed further we have to express the B-field in terms of the fluctuations of the 8d field $\Phi$. Recalling that $G_3=F_3-\tau H_3$ (with $F_3$($H_3$) being the RR(NSNS) 3-form flux), we can integrate
\beq
dB_2=\frac{\text{Im}G_3}{\text{Im}\tau}
\eeq
to obtain the B-field induced on the brane due to the presence of a constant $G_3$ background flux. The result for the B-field components is given by
\beq
B_{12}=\frac{g_s\sigma}{2i}(G_{(0,3)}^*\Phi-G_{(2,1)}\bar\Phi)\quad ;\quad B_{\bar 1\bar 2}=-\frac{g_s\sigma}{2i}(G_{(0,3)}\bar \Phi-G_{(2,1)}^*\Phi)
\label{Bcomp}
\eeq
where recall that, in tensorial notation the (0,3)-form flux corresponds to components $G_{\bar 1\bar 2\bar 3}$ while the (2,1)-form flux to $G_{\bar 1\bar 2 3}$. 
From now on we will denote the fluxes as $G\equiv G_{\bar 1\bar 2\bar 3}$ and $S\equiv \epsilon_{3jk}G_{3\bar j\bar k}$ for simplicity in the notation. Plugging this in \eqref{fB} we get that $f(B)$ becomes
\beq
f(\Phi)=1+\frac{Z^{-1}g_s\sigma^2}{4} |G^*\Phi-S\bar\Phi|^2 \ ,
\label{fphi}
\eeq
%

Let us now consider the Chern-Simons piece. 
From the equations of motion of type IIB supergravity one can derive the following relations between the RR fields and the 3-form fluxes
\begin{align}
&dC_6=H_3\wedge (C_4+\frac12 B_2\wedge C_2)-*\text{Re}\, G_3\\
&dC_8=H_3\wedge C_6-*\text{Re}\,d\tau
\end{align}
Integrating these equations and using that the background for the dilaton is constant, we obtain the following RR 6-form and 8-form potentials 
\beq
(C_6)_{12}=-\frac{Z^{-1}\sigma}{2i}(G^*\Phi-S\bar\Phi)
\eeq
\beq
(C_8)_{1\bar 1 2\bar 2}=\frac{Z^{-1}g_s\sigma^2}{4}\left((|G|^2+|S|^2)|\Phi|^2-4G^*S^*\Phi^2+\text{c.c.}\right)
\eeq
Plugging these expressions in the Chern-Simons action of the D7-branes we get
\beq
S_{CS}=\mu_7g_s\text{STr}\int d^8 \xi \left(-\frac{Z^{-1}g_s\sigma^2}{4}|G^*\Phi-S\bar\Phi|^2\right)
\eeq
which combined with the DBI part results in the following 8d action
\beq
S_{8d}=-\mu_7 g_s\text{STr}\int d^8\xi \left( f(\Phi)\left(Z\sigma^2 D_\mu\Phi D^\mu\bar\Phi+\frac14 Zg_s^{-1}\sigma^2F_{\mu\nu}F^{\mu\nu}\right)-\tilde{V}(\Phi)\right)
\eeq
where the scalar potential is given by
\beq
\tilde{V}(\Phi)\,=\,2(f(\Phi)-1)\,=\,\frac{Z^{-1}g_s\sigma^2}{2}|G^*\Phi-S\bar\Phi|^2
\eeq
In this last step we have also subtracted the D7-brane tension (which is cancelled by the contribution of the orientifold planes). Notice that once done so the DBI and the CS parts of the action contribute the same amount to the scalar potential, so we cannot neglect the contribution from the CS action, as is oftentimes done in the literature.

Finally, integrating over the internal ${\bf T}^4$ wrapped by the
D7-branes (using that the internal profile of the wavefunctions for $\Phi$ is constant, see \cite{mms}) and rescaling the fields such that
\beq
\Phi\rightarrow \Phi (V_4\mu_7g_sZ\sigma^2)^{-1/2}\quad ;\quad A_\mu\rightarrow A_\mu(V_4\mu_7Z^{-1}\sigma^2)^{-1/2}
\label{fredef}
\eeq
we obtain the following 4d effective Lagrangian 
\beq
\mathcal{L}_{4d}=\text{STr} \left(f(\Phi)D_\mu\Phi D^\mu\bar\Phi+\frac{1}{4g_{YM}^2}F_{\mu\nu}F^{\mu\nu}-V(\Phi)-\frac12 g_{YM}^2[\Phi,\bar \Phi]^2\right)
\label{L4d}
\eeq
where we have restored the D-term. Notice that all the dependence of the D-term on the higher order corrections is absorbed in $g_{YM}^{-2}=V_4 \mu_7 Z^{-1}\sigma^2f(\Phi)$, with $V_4$ being the volume of the internal ${\bf T}^4$. The rescaled scalar (F-term) potential and $f(\Phi)$ become
\beq
V(\Phi)=\frac{Z^{-2}g_s}{2} |G^*\Phi-S\bar\Phi|^2\ ,
\eeq
\beq
f(\Phi)=1+\frac{Z^{-2}(V_4\mu_7)^{-1}}{4} |G^*\Phi-S\bar\Phi|^2 \ .
\eeq

As expected, this potential looks like a quadratic potential for the adjoint scalars. However, one has to take into account  the field redefinition required 
 to have canonical kinetic terms in eq.(\ref{L4d}), which becomes important for large values of $\langle \Phi\rangle$. As we will describe in section \ref{slow}, this redefinition modifies the large $\Phi$ behaviour of the system, which turns close to a linear potential. Note that this {\it flattening} effect is similar to that obtained in previous examples of monodromy inflation models \cite{McAllister:2008hb,Dong:2010in,Gur-Ari:2013sba}.
  It is however important to realise that in the present case the flattening effect is purely due to the field redefinition, and not to the square root of the DBI action. In fact notice that the CS piece suffers the same flattening effect with no square root involved whatsoever.

\subsection{Kaloper-Sorbo Lagrangian}

While it may not be obvious from the above discussion,  the system of D7-branes described above is an example of F-term axion-monodromy inflation model \cite{Marchesano:2014mla}, in the sense that for small values of $\langle \Phi \rangle$ the scalar potential can be understood as a standard F-term potential. This has already been shown for the case of D7-branes in smooth Calabi-Yau geometries, see for instance \cite{Camara:2004jj,Gomis:2005wc,Martucci:2006ij}. For the orbifold model of interest to this paper the connection with $N=1$ supergravity turns out to be more involved, but as we will show in section \ref{sugra} a similar result applies. Hence, we can also consider this model as an example of F-term monodromy inflation.

Now, as pointed out in \cite{Marchesano:2014mla}, in general models based on F-term axion monodromy have a direct connection with the 4d effective framework developed in \cite{Kaloper:2008fb,Kaloper:2011jz,Kaloper:2014zba}, which features a Lagrangian of the form (\ref{KSLag}). Following \cite{Marchesano:2014mla}, it is for instance straightforward to obtain the Kaloper-Sorbo Lagrangian from a heterotic or type I model where the inflaton is a massive Wilson line in a twisted torus, this being the most direct way to give a mass to the Higgs system of the model of section \ref{heterotic}. Nevertheless, a similar derivation for F-term monodromy models where the inflaton is a D-brane position has so far not been worked out. 

In order to see how to derive the 4d Lagrangian (\ref{KSLag}) from a model of wandering D7-branes, let us consider a single D7-brane transverse to $z_3$ and in the presence of the ISD three-form fluxes $G \sim G_{\bar{1}\bar{2}\bar{3}}$ and $S \sim G_{12\bar{3}}$. Now, looking at the DBI action  in the Yang-Mills approximation we have that
\be
 \mu_7 \int  \oh (\sig F_2 + B_2) \wedge *_8 (\sig F_2 + B_2)\, =\,\mu_7  \int  \oh \sig^2 F_6 \wedge *_8 F_6 + \sig B_2 \wedge F_6 + \dots
\label{D7YM}
\ee
where we have only kept terms that depend on $F_6= dA_5$, the magnetic dual of $F = dA$. If we assume that the D7-brane has a position modulus $\phi$, then it means that the four-cycle $S_4$ wrapped by the D7-brane contains a (2,0)-form $\omega_2$ \cite{Jockers:2004yj}, in which we can expand the magnetic potential $A_5$ as
\be
A_5\, =\, i C_3 \wedge \bar{\omega}_2 - i \bar{C}_3 \wedge \omega_2
\ee
where $C_3$ is a complex three-form in 4d. For instance, if $S_4 = {\bf T}^4$ such (2,0)-form will be given by $\omega = dz_1 \wedge dz_2$. Plugging this decomposition into the kinetic term for $A_5$ in (\ref{D7YM}) and performing dimensional reduction we obtain
\be
\mu_7 \sig^2 \oh \int_{\IR^{1,3} \times S_4}  \hspace*{-.5cm}   F_6 \wedge *_8 F_6 \quad \raw \quad \rho \int_{\IR^{1,3}} \hspace*{-.25cm} d^4 x \, |dC_3|^2\, ,  \quad \quad \rho\,=\, \mu_7 \sig^2 \int_{S_2} \omega_2 \wedge *_4 \bar{\omega}_2 
\ee
which is nothing but the complex generalisation of the term $\int |F_4|^2$ in (\ref{KSLag}), in the sense that $F_4 = dC_3$ is now a complex four-form in 4d. 

Let us now dimensionally reduce the second term in the rhs of (\ref{D7YM}). By taking into account that 
\be
B_2 \, =\, \frac{g_s\sigma}{2i}(G^*\phi-S\bar\phi)\, \omega_2 + \text{c.c.} 
\ee
as derived in the previous section we obtain
\be
\mu_7 \sig  \int_{\IR^{1,3} \times S_4}  \hspace*{-.5cm}B_2 \wedge F_6 \quad \raw \quad - g_s \rho \int_{\IR^{1,3}} \hspace*{-.25cm} \phi (G^* dC_3 - S^* d\bar{C}_3) + \text{c.c.}
\ee
where we have used that $*_4 \omega_2 = - \omega_2$. Again, we obtain a generalisation of the axion-four-form term $\int \phi F_4$ in (\ref{KSLag}), where a complex scalar $\phi$ couples to the four-form $F_4 = dC_3$ and its complex conjugate via the presence of fluxes. Notice that a similar expression was found in \cite{Dudas:2014pva} for the coupling of a complex scalar to a complex four-form. In our case we find a more general expression, in the sense that $\phi$ can couple to both $\bar{F}_4$ and $F_4$ due to the respective presence of supersymmetric ($S$) and non-supersymmetric ($G$) background fluxes respectively. 

From this Lagrangian and following the general philosophy of \cite{Kaloper:2008fb,Kaloper:2011jz,Kaloper:2014zba} one finds that after integrating out $F_4$ the potential generated for the scalar field $\phi$ is given by
\be
V(\phi)\, =\, \frac{g_s}{2} |G^* \phi -  S\phi^*|^2
\ee
just as found in the previous section when setting $Z=1$, as we have done here. Of course this will only be the potential in the small field regime, receiving corrections for large values of $\langle \phi \rangle$. Nevertheless, due to the symmetry properties of the Kaloper-Sorbo Lagrangian such corrections can only arise in powers of the initial scalar potential $V(\phi)$ and not of the field $\phi$ itself, see \cite{Kaloper:2008fb,Kaloper:2011jz,Kaloper:2014zba} and also \cite{Dvali:2005ws,Dvali:2005an,Dvali:2013cpa}. In our analysis of the previous section we have seen that this is the case, occurring in the form of a redefinition for the kinetic term of $\phi$, and giving rise to flattening effect for the potential. In section \ref{corr} we will discuss from an independent, string theoretical viewpoint why the Planck suppressed corrections to the inflaton potential should be of this form. 

Finally, in this section we have only discussed the appearance of the Kaloper-Sorbo Lagrangian for the case of a single D7-brane with an Abelian gauge group. This is indeed the case of interest in our Higgs-otic D7-brane model, since away from the orbifold singularity we have a single wandering D7-brane. We nevertheless expect that a similar result applies to the non-Abelian case, given that the results of the previous section involving the large field corrections, flattening etc. are valid for any U(N) gauge group or even orbifolds thereof. Such non-Abelian analysis is however beyond the scope of this paper and we hope to return to this problem in the near future.

\subsection{Estimation of the scales of the model}\label{scales}

The coefficient of the quadratic term in the inflation potential, and hence the inflaton mass, is 
determined by the size of the fluxes. We can try to estimate the size of the fluxes in terms of the
energy scales in the theory, assuming an approximate isotropic compactification.

Since the 3-form fluxes  have to be quantised over the internal 3-cycles $\gamma_j$ that they wrap, 
they  are expected to scale as
\beq
\frac{1}{2\pi\alpha'}\int_{\gamma_j}G_3=2\pi n_j\rightarrow G_3\simeq \frac{4\pi^2\alpha'n}{V_6^{1/2}}
\eeq
where $V_6$ is the volume of the internal dimensions and $n$ are integer quanta.
Using the following identities from type IIB   compactifications for the Planck mass and the 
compactification/unification scale \cite{Ibanez:2012zz}
\beq
m_p^2=(8\pi)M_p^2=\frac{8M_s^8V_6}{(2\pi)^6g_s}\label{Mp} \ \ ,\ \ 
M_c=M_s\left(\frac{2\alpha_G}{g_s}\right)^{1/4} \ ,
\eeq
where we have defined the compactification/unification scale as $M_c=1/R_c$ with $V_4=(2\pi R_c)^4$, we find
\beq
G_3=\frac{n}{\pi}\frac{M_c^2}{\alpha_G^{1/2}m_p} \ .
\label{G3}
\eeq
One can then estimate the scale of SUSY breaking which is given by
\beq
M_{SS}=\frac{Z^{-1}g_s^{1/2}}{\sqrt{2}}G_3=\frac{Z^{-1}n}{\pi}\frac{M_s^2}{g_s^{1/2}m_p}
\eeq
For $n\sim O(1)$ one gets $M_{SS}\sim 10^{12}-10^{13}$ GeV if $M_s\simeq 10^{16}$ GeV.
Thus the above simple dimensional argument implies a SUSY breaking scale of the 
required order so that the SM Higgs potential is saved from its instability, see \cite{imrv} for further details.

We have seen that the effect of considering higher order corrections on $\Phi$ is the presence of a function $f(\Phi)$ multiplying the kinetic terms given by 
\beq
f(\Phi)=1+\frac{Z^{-2}(V_4\mu_7)^{-1}}{4} |G^*\Phi-S\bar\Phi|^2 \ .
\eeq
For small field this function is approximately 1 and we recover canonically normalised kinetic terms. To estimate how important is the effect for large field we define the parameter $\hat G\equiv Z^{-1}V_4^{-1/2}\mu_7^{-1/2} G_3$ and using \eqref{G3} we get
\beq
[\hat G]=[Z^{-1}V_4^{-1/2}\mu_7^{-1/2} G_3]\simeq 0.3Z^{-1}g_s^{-1/2}n\frac{1}{M_p}
\label{warpping} 
\eeq
For $n\sim O(1)$  one obtains $\hat G\sim 0.3\frac{1}{M_p}$, so this effect becomes appreciable approximately for $\langle\Phi\rangle>7M_p$. We can also write the SUSY breaking scale in terms of $\hat G$ such that
\beq
M_{SS}^2=V_4\mu_7 g_s|\hat G|^2\sim 0.05 M_s^4 |\hat G|^2
\eeq
so $\hat G$ gives us the relation between the SUSY breaking scale and the string scale. This relation will be useful later on when checking that the potential energy never becomes bigger than the string scale.

\subsection{The Higgs/inflaton scalar potential}\label{potential}

Even if the analysis in section \ref{DBICS} is done for an adjoint field of a U(N) gauge theory, it also applies after we have made an orbifold projection that converts the adjoint into a set of bifundamental fields charged under the orbifolded gauge group. In particular, we may consider the ${\bf Z}_4$ orbifold projection of section \ref{HiggsIIB} and hence take $\Phi$ to be the $6 \times 6$ matrix containing the Higgs system of the model 
\begin{equation}
\Phi=\left(\begin{array}{ccc}{\bf 0}_3&&\\& {\bf 0}_2&H_u\\&H_d&0\end{array}\right)
\label{phi}
\end{equation}
as in (\ref{mat}). Then applying the results from section \ref{DBICS} we obtain the standard D-term contribution to the scalar potential and the F-term contribution which is given by
\beq
V(\Phi)=\text{STr}\left(\frac{Z^{-2}g_s}{2} |G^*\Phi-S\bar\Phi|^2\right)\ ,
\eeq
which in terms of the bifundamental fields $H_u$, $H_d$ gives rise to
\begin{equation}
V=\frac{Z^{-2}g_s}{2}\left[(|G|^2+|S|^2)(|H_u|^2+|H_d|^2)-4{\rm Re} (G^* S^*H_uH_d)\right]
\label{VHud}
\end{equation}
once we trace over the gauge indices.
This potential can be rewritten in terms of the combinations
\begin{gather}
h\, =\, \frac {e^{i\g/2}H_u - e^{-i\g/2}H_d^*}{\sqrt{2}}  \quad {\rm and} \quad H\, =\, \frac {e^{i\g/2}H_u+e^{-i\g/2}H_d^*}{\sqrt{2}} 
\label{hH}
\end{gather}
where $\g = \pi - {\rm Arg} (GS)$ as
\begin{equation}
V=\frac{Z^{-2}g_s}{2}\left[(|G| -|S|)^2|h|^2+(|G|+|S|)^2|H|^2\right]
\label{VhH2}
\end{equation}
Note that, at this level, before field rescaling to canonical kinetic terms, the potential has the
structure of double chaotic inflation.
Note also  that for  $|S|=|G|$, $h$  becomes massless. Thus, if eventually we want to
 fine-tune a massless SM Higgs, we would need to be close to a situation where $|S|=|G|$. 
 The subsequent running from $M_c$ down to the scale $M_{SS}$ of soft parameters will give rise 
to a massless SM Higgs.


We may now write this potential in terms of the real scalars $(\sigma,\theta)$ which we defined in eq.(\ref{sigthetadef}).
They parametrise the neutral Higgs along the D-flat direction.
One finds
\begin{equation}
V(\sigma , \theta) =Z^{-2}g_s(|G|^2+|S|^2)\left(1- A\, {\rm cos}\,{\tilde \theta }\right)\sigma^2
\label{Vst}
\end{equation}
where we have defined 
\begin{equation}
A=\frac{2|SG|}{|G|^2+|S|^2} \quad \quad \text{and}\quad  \quad  {\tilde \theta} \ =  \theta - {\rm Arg} (GS) \ .
\label{defA}
\end{equation}
Note that $0\leq A\leq 1$ and one also has
\beq
A\ =\ \frac {m_H^2-m_h^2}{m_H^2+m_h^2} \ =\ |cos2\beta|  \ ;\   \frac{m_H}{m_h} \ =\ \sqrt{\frac {1+A}{1-A}} \ ,
\eeq
with $tan\beta=m_H/m_h$.
The potential in eq.(\ref{Vst}) will be our inflation potential.  It is essentially a quadratic potential in $\sigma$ modulated by the dependence on
${\tilde \theta }$. Note however that  we still have to include the effect that the kinetic terms are non-canonical and field dependent, as we will
discuss later.  However, the qualitative structure of the scalar potential can already be discussed at this point.

\begin{figure}[h!]
\begin{center}
{\includegraphics[width=0.3\textwidth]{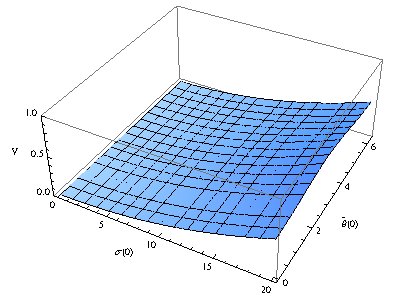}}
{\includegraphics[width=0.3\textwidth]{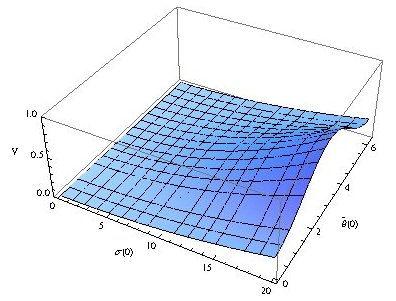}}
{\includegraphics[width=0.3\textwidth]{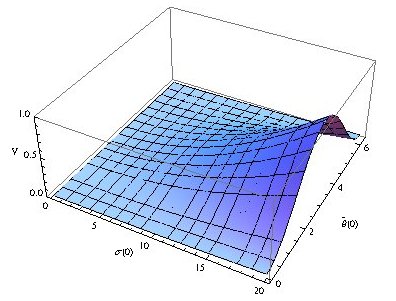}}
\end{center}
\vspace{-15pt}
\caption{Scalar potential for three different values of $A$, $A=0.1$ (left), $A=0.5$ (centre) and $A=0.9$ (right).}
\label{3pot}
\end{figure}

Roughly speaking, the shape of the potential  depends on the value of the parameter $A$ which parametrises the relative size of both 
types of ISD fluxes. In figure \ref{3pot} we show the structure of the scalar potential for three characteristic values $A=0.1, 0.5,0.95$.
For $A\simeq 0$, which can happen  if either $G$ or $S$ vanish,
the potential is simply given by
\begin{equation}
V=\frac{Z^{-2}g_s}{2}|G|^2(|H|^2+|h|^2)=Z^{-2}g_s|G|^2\sigma^2
\label{VA0}
\end{equation}
which is ${\tilde \theta }$-independent. This case will be essentially identical to a single inflaton case with a chaotic, quadratic potential for $\sigma$ (before flattening). This case  with $A\simeq 0$ is depicted in the left plot in figure \ref{3pot}. 
Getting the same result with either $G=0$ or $S=0$ is expected by symmetry arguments, since a D7-brane which is point-like in the third complex plane cannot locally distinguish between the real and imaginary parts of $z_3$, and both choices of fluxes are related by interchanging $z_3$ by $\bar z_3$.

For the case $A=1$ one has the fluxes related as $|G|=|S|$, and $h$ is massless. The potential is then given by
\begin{equation}
V\, =\, 4Z^{-2}g_s|G|^2 {\rm cos}^2 (\tilde\theta/2) \sigma^2\, =\, 2Z^{-2}g_s|G|^2|H|^2
\label{VA1}
\end{equation}
This corresponds to the right plot in figure \ref{3pot}.
This choice of fluxes corresponds to a non-supersymmetric situation in which the NSNS 3-form flux $H_3$ only has a leg in one of the real directions of the transverse space, so the other direction is a flat direction for the D7-branes. In the 4d effective theory this is reflected by the presence of a massless real scalar which is given by $|h|$. 

These two cases $A=0,1$ are limiting cases in which the potential reduces to a single field inflation model. For a generic choice of fluxes, one expects a situation in between, with both scalars playing an important role in inflation. In section \ref{slow} we will compute the slow-roll parameters first for the cases $A=0,1$ and then for the general 2-field inflation case. 

\begin{figure}[h!]
\begin{center}
{\includegraphics[width=0.6\textwidth]{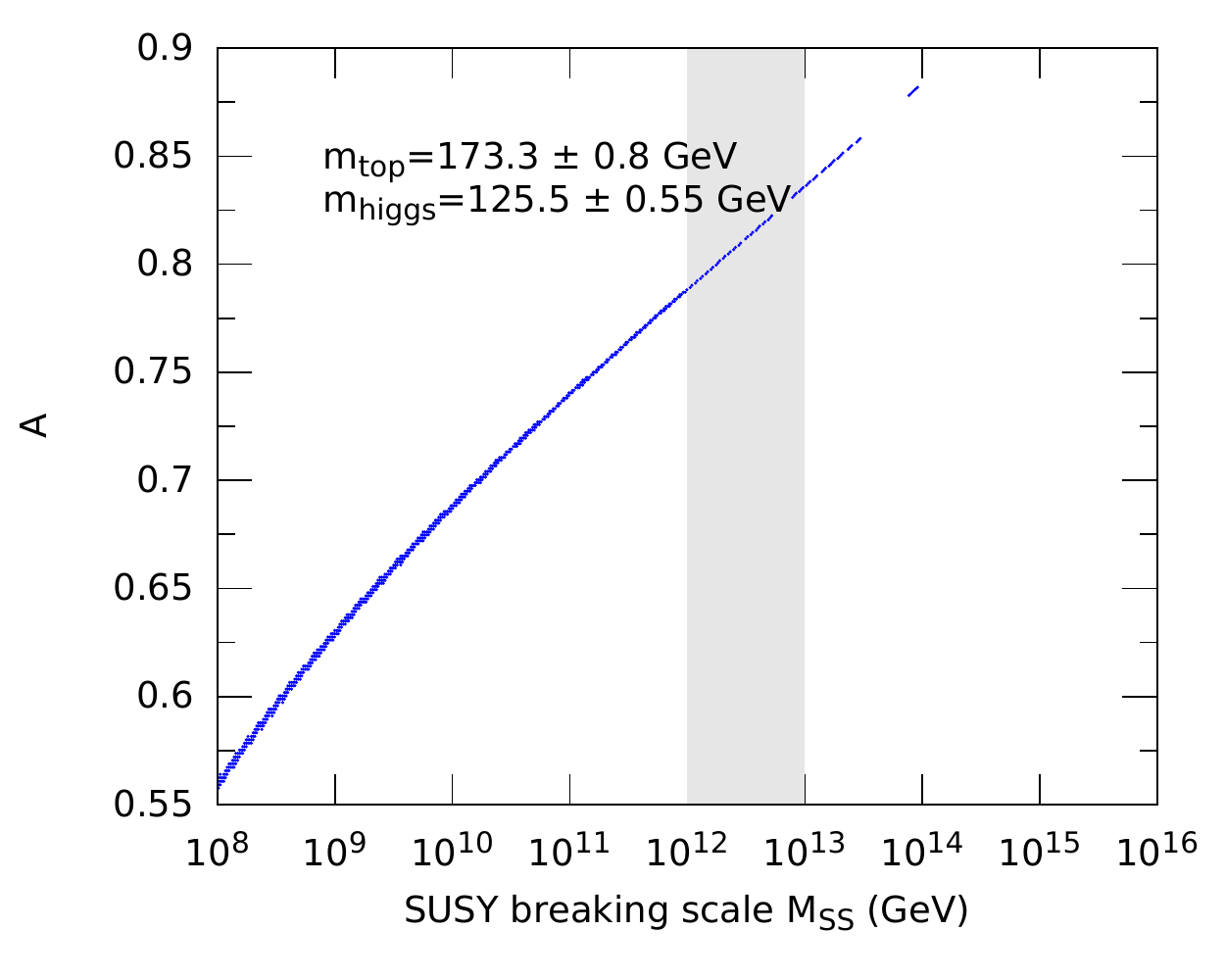}}
\end{center}
\vspace{-15pt}
\caption{The required value of A in order to have a massless eigenstate at $M_{SS}$ as a function of the SUSY breaking scale.}
\label{AMss}
\end{figure}

Notice however that if we want to have a massless eigenstate at the SUSY breaking scale (in order to get a light SM Higgs), $A$ is not a free parameter anymore. In terms of the mass parameters of the Higgs mass matrix in \eqref{matrizmasas}, $A$ parametrises the ratio between the off-diagonal entries $|m_3|$ and the diagonal ones $m^2_{H_u}=m^2_{H_d}$ at $M_c$. Thus a massless eigenstate implies $|m_3|^2=m^2_{H_u}m^2_{H_d}$ which corresponds indeed to $A=1$ as we already commented. However, as we discussed in section \ref{sechiggs}, we need the eigenstate to become massless at $M_{SS}\sim 10^{12}-10^{13}\text{ Gev}$ and not at the inflation scale $\sim 10^{16}$ GeV, so $A$ needs to be slightly lower than 1. We have computed the running between both scales and obtained that the optimal value to have a zero eigenvalue at $M_{SS}$ is $A\simeq 0.83$, corresponding to $m_H/m_h=3.28$. We take here the unification scale $M_c$ as the scale at which $\a_2 = \a_3$, see \cite{Ibanez:2014zsa}. Of course this result depends on the exact value of $M_{SS}$ which is in turn parametrised by the global factor in the potential, whose size was estimated in section \ref{scales} obtaining $M_{SS}\sim 10^{12}-10^{13}\text{ Gev}$. In figure \ref{AMss} we plot the value of $A$ that we need to start with in order to have a light SM Higgs boson, as a function of the SUSY breaking scale.  We have also imposed to get the experimental value of the top and Higgs mass at the EW scale. We can see that for $M_{SS}\sim 10^{12}-10^{13}$, we have $0.8<A<0.85$, so in any case, we will be in a situation quite close to the single field case $A=1$, in which the heavy Higgs $H$ is the scalar which plays the role of the inflaton.

\subsection{$N=1$ supergravity description}
\label{sugra}

Before turning to the computation of the slow roll parameters, let us compare the scalar potential of the previous section with the one that we would have obtained from a $N=1$ supergravity computation. As we will see, upon introducing the appropriate K\"ahler potential and superpotential one recovers an F-term scalar potential with the same structure as the one found microscopically via the D7-brane action. The exact matching does however only occur for small values of the inflaton vev. For large field values there will be $\alpha'$ corrections that the supergravity approach fails to capture, and can only be seen by means of our previous DBI+CS analysis.

In eq.(\ref{kahlershift}) we showed the K\"ahler potential for the Higgs fields in a ${\bf Z_4}$ heterotic orbifold. It is easy to
convince oneself  (e.g. by application of S-duality and T-duality along the third complex plane) that the corresponding K\"ahler potential for the type IIB model with a stack of D7's is given by
\beq
K_H =\ -\text{log}[(S+S^*)(U_3+U_3^*)\ -\ \frac {\alpha'}{2}\ |H_u+H_d^*|^2]  \  \ -\  3\text{log}(T+T^*)
\label{KH}
\eeq
where $S$ is the complex type IIB dilaton.  We have also added the well known  K\"ahler moduli dependent piece 
in terms of a diagonal K\"ahler moduli field $T$ (i.e. we are taking $(T_i+T_i^*)=(T+T^*), \, \forall i$).
We have also set the other matter fields $A_{1,2}=0$ since they do not play any role in the discussion
and also the complex structure moduli to $U_1=U_2=1$. These simplifications are not important and the general case can be 
easily included in the discussion. The important point is that this dependence of the K\"ahler potential on $T$ yields a no-scale structure 
for the F-term scalar potential, typical of type IIB compactifications with ISD fluxes \cite{Giddings:2001yu}. 

In fact, it is well known that the effect of ISD fluxes on D7-brane fields can be understood macroscopically in terms of an $N=1$ supergravity description in which the SUSY-breaking effects are induced by the auxiliary fields of the K\"ahler moduli,see \cite{Camara:2003ku,Grana:2003ek,Camara:2004jj,Lust:2004fi,Aparicio:2008wh,Camara:2014tba}.
 In our case the relevant superpotential in this effective description includes a constant term $W_0$ and a $\mu$-term
\beq
W\ =\ W_0 \ +\ \mu H_uH_d  \ .
\eeq
Due to the no-scale structure of the K\"ahler potential, the scalar potential is simply given by
\begin{equation}
V=e^K(K^{i \bar j}D_iW D_{\bar j}\bar W)
\label{potF}
\end{equation}
where the indices run over the dilaton and complex structure moduli. Let us assume that the above potential is minimised when
\begin{equation}
D_S W=0\quad ;\quad D_U W=0
\label{FSU}
\end{equation}
which  implies $V_0=0$. Moreover, as mentioned before, we assume that supersymmetry breaking comes from the K\"ahler moduli sector, namely
\begin{equation}
F^t=e^{K/2} K^{\bar T T}D_T W\ =\ -\frac {W_0}{\sqrt{st} } \ \neq 0\ ,
\label{Ft}
\end{equation}
where $s=(S+S^*), t=(T+T^*)$. This is nothing but the assumption of modulus dominance  SUSY breaking in type IIB  which was studied in detail in
  \cite{Camara:2003ku,Grana:2003ek,Camara:2004jj,Lust:2004fi,Aparicio:2008wh,Camara:2014tba}.
   Plugging all these data in standard $N=1$ sugra formulae \cite{Brignole:1997dp} leads to a bilinear scalar potential of the form
\beq
V=(m^2_{H_u}+\hat\mu^2)|H_u|^2+(m^2_{H_d}+\hat\mu^2)|H_d|^2+B\hat\mu H_uH_d+\text{h.c.}
\eeq
where ${\hat \mu}$ is the Higgsino mass with fields canonically normalised, and
\beq 
m^2_{H_u}=m^2_{H_d}\ =\ |M|^2 \ ,\ {\hat \mu } \ =\ \frac {W_0\ +\mu s}{t^{3/2}\sqrt{s}}  \ ,\ B\ =\ -2M \ ,
\eeq
where 
\beq 
M\ =\ - \frac {W_0^*}{t^{3/2}\sqrt{s}} \ ,
\eeq
is a universal gaugino mass. Note that the physical $\mu$-term ${\hat \mu }$ has two 
contributions, one coming from the original $\mu$-term of the superpotential, and the other arising after SUSY breaking from
the K\"ahler potential via a Giudice-Masiero mechanism, which is implicit in the form of the K\"ahler potential.
All in all the scalar potential is given by 
\beq
V\ =\ (|M|^2+|\hat\mu|^2)(|H_u|^2+|H_d|^2)-2M\hat\mu H_uH_d+\text{h.c.}
\eeq
This scalar potential is identical to the one we derived from explicit fluxes eq.(\ref{VHud})
upon the identifications
\begin{equation}
G^*=\left(\frac{g_s}{2}\right)^{-1/2}\frac{W_0^*}{\sqrt{s}t^{3/2}}\quad ,\quad S^*=-\left(\frac{g_s}{2}\right)^{-1/2}\frac{W_0+\mu s}{\sqrt{s}t^{3/2}}
\label{idenGS}
\end{equation}
which implies $M=-\frac{g_s}{2}G^*$ and $\hat\mu=-\frac{g_s}{2}S^*$, in agreement with the results of \cite{Camara:2004jj}. 

Finally we can write the scalar potential in terms of the fields $H,h$ obtaining\footnote{In terms of $H$ and $h$ (\ref{KH}) reads $-\text{log}[(S+S^*)(U_3+U_3^*) - \alpha' ({\rm cos}^2 (\g/2) |H|^2  + {\rm sin}^2 (\g/2) |h|^2)]$. It is then quite remarkable that the scalar potential is independent of which combination of $H$ and $h$ appears in the K\"ahler potential.} 
\begin{equation}
V=  \left[(|\hat\mu|+|M|)^2|H|^2+(|\hat\mu|-|M|)^2|h|^2\right]
\label{VhH}
\end{equation}
Note that in the absence of an explicit $\mu$-term  one has ${\hat \mu }=-M^*$ so that the  $h$ doublet is massless.
So from the $N=1$ sugra point of view, the desired situation with $m_h^2\ll m_H^2$  would correspond to a suppressed
explicit $\mu$-term in the superpotential. 
This limit with a massless $h$ field corresponds  in terms of fluxes to a situation with $G=-S^*$.  
It is interesting to have this $N=1$ sugra description for this equality which could be unmotivated from a 
microscopic point of view. Note finally that we will also have a similar situation whenever $|W_0| = |W_0 +\mu s|$.

Since the $N=1$ sugra formalism is quite familiar one may be tempted to discuss inflation only in terms
of the above formulae (see e.g. \cite{Ellis:2014opa} for a recent two-field analysis in no-scale supergravity). The structure 
would be just the one of double chaotic inflation. 
However this $N=1$ sugra formulation misses important $\alpha'$ stringy corrections.
On the other hand, the  DBI+CS D7-brane action on which we have based our analysis contains corrections to 
all orders in $\alpha'$, and so include all higher order terms in the expansion on the Higgs field vevs. These 
higher order terms are missed by the sugra formulation. In particular, the flattening of the inflation potential due 
to the kinetic field redefinitions is such an $\alpha'$ correction, and the sugra scalar potential would only capture 
the first term in the $\alpha'$ expansion. 

\section{Computing slow roll parameters for large inflaton}
\label{slow}

In this section we compute the slow-roll dynamics of our inflation model and the resulting cosmological observables. We first review the generalisation of the slow roll parameters to multiple field inflationary models in which the kinetic terms are not canonically normalised. Then we will solve the slow roll equations of motion and show the results for different values of $A$, distinguishing between the single field and two-field cases.

\subsection{Slow roll equations of motion}\label{slowformulae}

In the previous section we derived the effective action for the Higgs/inflaton sector obtaining for a general choice of fluxes a two-field inflation model. The 4d effective Lagrangian in terms of the neutral Higgs scalars $H_u,H_d$ is given by
\beq
\mathcal{L}_{4d}=f(H_u,H_d)(|D_\mu H_u|^2  +|D_\mu H_d|^2 )-V_F(H_u,H_d)-V_D(H_u,H_d)
\eeq
where we have explicitly separated the F-term \eqref{VHud} and D-term \eqref{Dterm}  contribution of the potential. The function multiplying the kinetic terms is given also in terms of the F-term potential such that
\beq
f=1+\frac{(V_4\mu_7g_s)^{-1}}{2} V_F\ .
\eeq
We saw that the D-term potential is minimised for
\beq
H_u=H_d^*e^{i\theta}\ ,\ |H_u|=|H_d|=\sigma
\eeq
with $\theta=\theta_u+\theta_d$. Thus in terms of the remaining scalar degrees of freedom $\sigma,\theta$ the potential becomes
\begin{equation}
V_F=Z^{-2}g_s(|G|^2+|S|^2)(1-A\, {\rm cos}\,\tilde\theta)\sigma^2
\end{equation}
as we derived in \eqref{VA1}. Recall that $\tilde \theta=\theta -{\rm Arg}(GS)$ and $A$ gives the relative size of the moduli of the fluxes (see \eqref{defA}). The kinetic terms read
\beq
|D_\mu H_u|^2  +|D_\mu H_d|^2\rightarrow 2(D_\mu \sigma)^2  +\frac{\sigma^2}{2}(D_\mu \theta)^2
\eeq
implying the following 4d effective Lagrangian for the fields $\sigma,\theta$,
\beq
\mathcal{L}_{4d}=f(\sigma,\theta)\left(2|D_\mu \sigma|^2  +\frac{\sigma^2}{2}(D_\mu \theta)^2 \right)-Z^{-2}g_s(|G|^2+|S|^2)(1-A\, {\rm cos}\,\tilde\theta)\sigma^2
\eeq
One could think that the first step is to absorb the prefactor $f(\sigma,\theta)$ in a redefinition of the fields in order to have canonically normalised kinetic terms. Comparing with the general form of a Lagrangian of multiple fields
\beq
\mathcal{L}_{4d}=\frac12 G_{ab}(\phi)D_\mu \phi^a D^\mu\phi^b- V(\phi) 
\label{def} 
\eeq
this is equivalent to ask if there  exists an appropriate field redefinition such that $G_{ab}=\delta_{ab}$, where in our case the metric is given by
\beq
G_{ab}=\left(\begin{array}{cc}4f(\sigma,\theta)&0\\0&\sigma^2 f(\sigma,\theta)\end{array}\right)
\label{metric}
\eeq
This is always possible for a  single field, making a field redefinition of the form
\beq
\phi'=\int d\phi f^{1/2}(\phi)
\eeq
where we have assumed $G_{\phi\phi}=f(\phi)$. However, in general this can not be done globally (i.e. for all values of $\phi$) for two or more fields simultaneously.
Notice that $G_{ab}$ transforms as a rank two tensor under field redefinitions of the form $\phi\rightarrow f(\phi)$ and is positive definite, so it can be interpreted as a metric on the moduli space parametrised by the fields. Therefore a change of variables which brings  the metric to the flat metric $G_{ab}=\delta_{ab}$ can only be done globally if the  curvature scalar vanishes everywhere. In fact, the metric \eqref{metric} is conformal to the flat metric, so the Ricci scalar of curvature will be proportional to the Hessian of the function $f$. It can be checked that this scalar vanishes
\beq
R\propto\frac{1}{f}\Delta({\rm Ln}\, f)=0
\eeq
if the function $f$ can be written as $f=|h(z_3)|^2$ where $h(z_3)$ is a holomorphic function on $z_3$. By absorbing all the global factors in the potential into a single overall parameter given by 
\be
|\hat G|^2=Z^{-2}(V_4\mu_7)^{-1} (|G|^2+|S|^2) 
\label{hatG}
\ee
as in section \ref{scales} we obtain the function 
\beq
f=1+\frac{|\hat G|^2}{2} (1-A\, {\rm cos}\, \tilde\theta)\sigma^2
\eeq
Then, recalling that $z_3=(2\pi\alpha') \sigma e^{i\theta/2}$, we see that $f$ is not a holomorphic function in general so it does not exist any field redefinition that canonically normalises simultaneously both fields $\sigma$ and $\theta$.  Therefore for the general 2-field case we will have to keep track of the non-flat metric all over the computation of the slow roll parameters.

The scalar equations of motion for several inflaton fields are given by
\begin{equation}
\ddot{\phi}^a+\Gamma^a_{bc}(\phi)\dot{\phi}^b\dot{\phi}^c+3H\dot\phi^a=-G^{ab}\frac{\partial V(\phi)}{\partial\phi^b} \ .
\label{eqsfirst}
\end{equation}
with $H$ being the Hubble constant. 
The slow roll condition for inflation implies that the potential energy has to be dominant with respect to the kinetic energy over the whole inflationary trajectory, so we can drop the first two terms in \eqref{eqsfirst} leading to the well known slow roll equations of motion 
\begin{equation}
3H\dot\phi^a=-G^{ab}\frac{\partial V(\phi)}{\partial\phi^b} \ .
\label{eqmotion}
\end{equation}
This is a good approximation whenever the slow roll parameters $\epsilon$, $\eta$ remain smaller than one. The generalisation of the $\epsilon$ parameter for multiple field inflation is given by  (see e.g. \cite{Burgess:2007pz})
\beq
\epsilon=\frac{M_p^2}{2}G^{ab}\frac{V'_aV'_b}{V^2}
\label{e}
\eeq
where the primes denotes derivatives with respect to the fields $V'_a=\frac{\partial V}{\partial\phi^a}$. The $\eta$ parameter would correspond though to the smallest eigenvalue of the matrix of second derivatives of the potential given by
\beq
N^a_b=M_p^2\frac{G^{ac}V''_{cb}}{V}
\label{n}
\eeq
where $V''_{cb}=\frac{\partial V'_a}{\partial\phi^b}-\Gamma^a_{bc}V'_a$ is the covariant derivative.

The $\epsilon$-parameter can also be defined in the multi-field case in terms of the number of efolds as
\beq
\epsilon=\frac12 G_{ab}\frac{d\phi^a}{dN_{\rm efolds}}\frac{d\phi^b}{dN_{\rm efolds}}\ .
\eeq
This implies the following formula that  we will use to compute $N_{\rm efolds}$ in terms of $\epsilon$,
\beq
N_{*}=\int_{\phi_0^2}^{\phi_{end}^2}\frac{1}{\sqrt{2\epsilon}}\sqrt{G_{11}\left(\frac{d\phi^1}{d\phi^2}\right)^2+G_{22}}\ d\phi^2
\label{Nefo}
\eeq
in the two field case.\footnote{Note that here $\phi^2$ stands for $\phi^b$ with b=2, so it is not an exponent but an index.} Finally, the scalar spectral index and the tensor to scalar ratio are defined as in section \ref{inflation} (for single field) but using the multi-field generalisation of $\epsilon$ and $\eta$ explained here.

Below we show the results first for the single field limit cases ($A=0$ and $A=1$) and then for a general two field case with arbitrary $A$, but with special focus on the case of
special  interest $A\simeq 0.83$.

\subsection{Single field limit cases}

We showed in section \ref{potential} that for specific choices of fluxes the potential reduces to a single field inflationary potential where the inflaton has a clear geometric interpretation.
In particular, we get the potential
\begin{equation}
V=Z^{-2}g_s(|G|^2+|S|^2)\phi^2
\end{equation}
with $\phi\equiv\sigma$ for $A=0$ ($G=0$ or $S=0$) or $\phi\equiv |H|$ for $A=1$ ($|G|=|S|$).
Recall that the position of the D7-branes in the transverse torus is parametrised by
\beq
z_3=2\pi \alpha'\ \sigma e^{i\theta/2}=2\pi \alpha'\frac{1}{\sqrt{2}}(|H|+i|h|)e^{-i\gamma/2}
\eeq
If $A=0$ the inflaton $\sigma$ parametrises the distance of the travelling D7-branes to the singularity ${\bf Z}_4$, while if $A=1$ the inflaton corresponds to the distance along one of the 1-cycles of the torus, the orthogonal 1-cycle being a flat direction.

\begin{figure}[h!]
\begin{center}
{\includegraphics[width=0.6\textwidth]{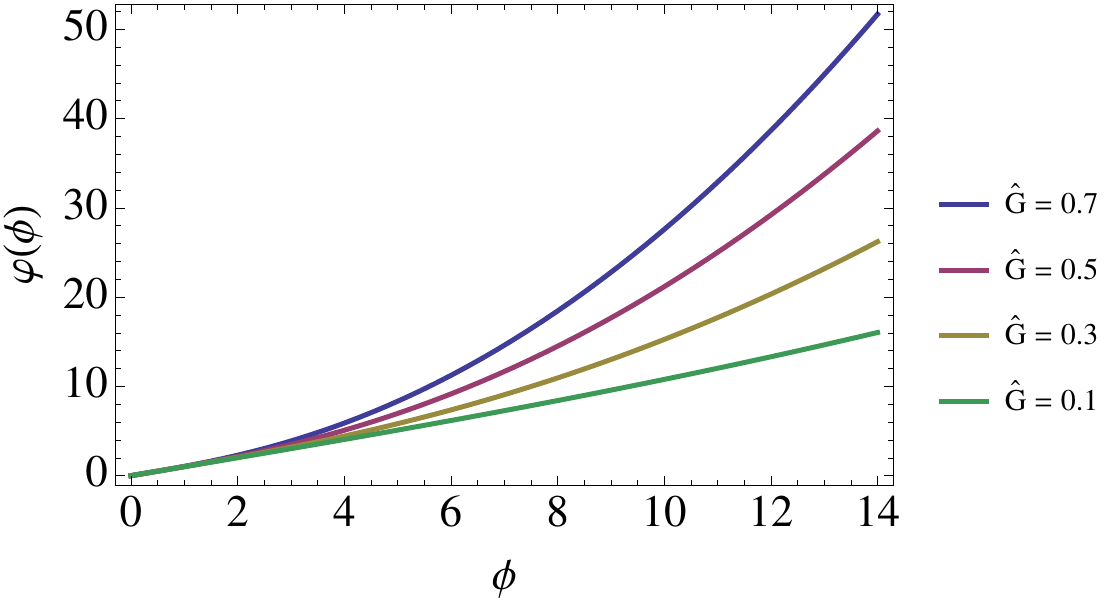}}
\end{center}
\vspace{-15pt}
\caption{Field redefinition (new field $\varphi$ vs old field $\phi$) for different values of $\hat G$.}
\label{figphi}
\end{figure}

\begin{figure}[h!]
\begin{center}
{\includegraphics[width=0.6\textwidth]{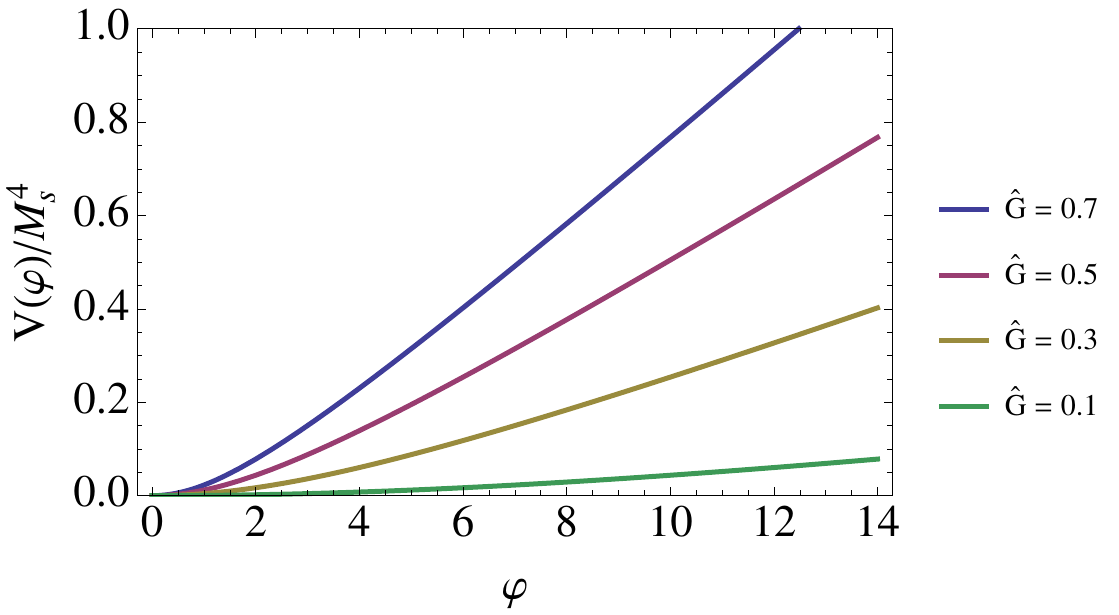}}
\end{center}
\vspace{-15pt}
\caption{Scalar potential in terms of the canonically normalised field $\varphi$ for different values of $\hat G$.}
\label{pot}
\end{figure}

Before taking into account the field redefinition the potential is quadratic on the fields, corresponding to a soft mass induced by breaking SUSY with the closed string fluxes. However, since we are interested in large field values, higher order corrections to the potential become important and can not be neglected. These corrections were computed from the DBI+CS action of the D7-brane and their effect is to induce non-canonical kinetic terms, with a prefactor
\beq
f=1+\frac{(V_4\mu_7g_s)^{-1}}{2}V=1+\frac{|\hat G|^2}{2}\phi^2
\label{f}
\eeq
where we have again defined $|\hat{G}|$ by (\ref{hatG}). 
In the single field case, the kinetic term can always be canonically normalised by an appropriate redefinition of the field.
Therefore the effect of the higher order corrections can be encoded on a field redefinition given by
\beq
\varphi=\int d\phi f^{1/2}(\phi)
\label{redefi}
\eeq
which becomes important for large field. Inserting \eqref{f} in \eqref{redefi} we get
\beq
\varphi=\frac{1}{2\sqrt{2}} |\phi|\sqrt{2+|\hat G|^2|\phi|^2}+\frac{1}{\sqrt{2}}|\hat G|^{-1}\text{sinh}^{-1}[|\hat G||\phi|/\sqrt{2}]
\label{redef}
\eeq
In fig.\ref{figphi} we plot the new normalised field $\varphi$ in terms of the old one $\phi$. Notice that for large field this yields
\beq
\varphi\simeq \frac{1}{2\sqrt{2}}|\hat G|\phi^2
\eeq
and the potential becomes linear in the new normalised field $\varphi$. Hence the effect of the higher order corrections is indeed a flattening of the potential. In fig.\ref{pot} we plot the scalar potential in terms of the new canonically normalised field, for different values of $\hat G$. The bigger $\hat G$ is, the sooner the flattening effect takes place. 
To work  this plot out  we have used the fact that the overall factor in the potential (which parametrises the SUSY breaking scale) is related to $|\hat G|$ by
\beq
M_{SS}^2=Z^{-2}g_s(|G|^2+|S|^2)=V_4\mu_7g_s|\hat G|^2\simeq 0.05g_s M_s^4 |\hat G|^2
\eeq
where $M_s$ is the string scale. Hence the scalar potential interpolates between quadratic and linear depending on the SUSY breaking scale (through $\hat G$). 
For $|\hat G|>1/M_p$ the potential becomes bigger than the string scale during inflation (i.e. $V^{1/4}>M_s^4$)
and the computation is inconsistent, since new KK and string modes should be taken  into account.

%
%
\begin{figure}[h!]
\begin{center}
{\includegraphics[width=0.8\textwidth]{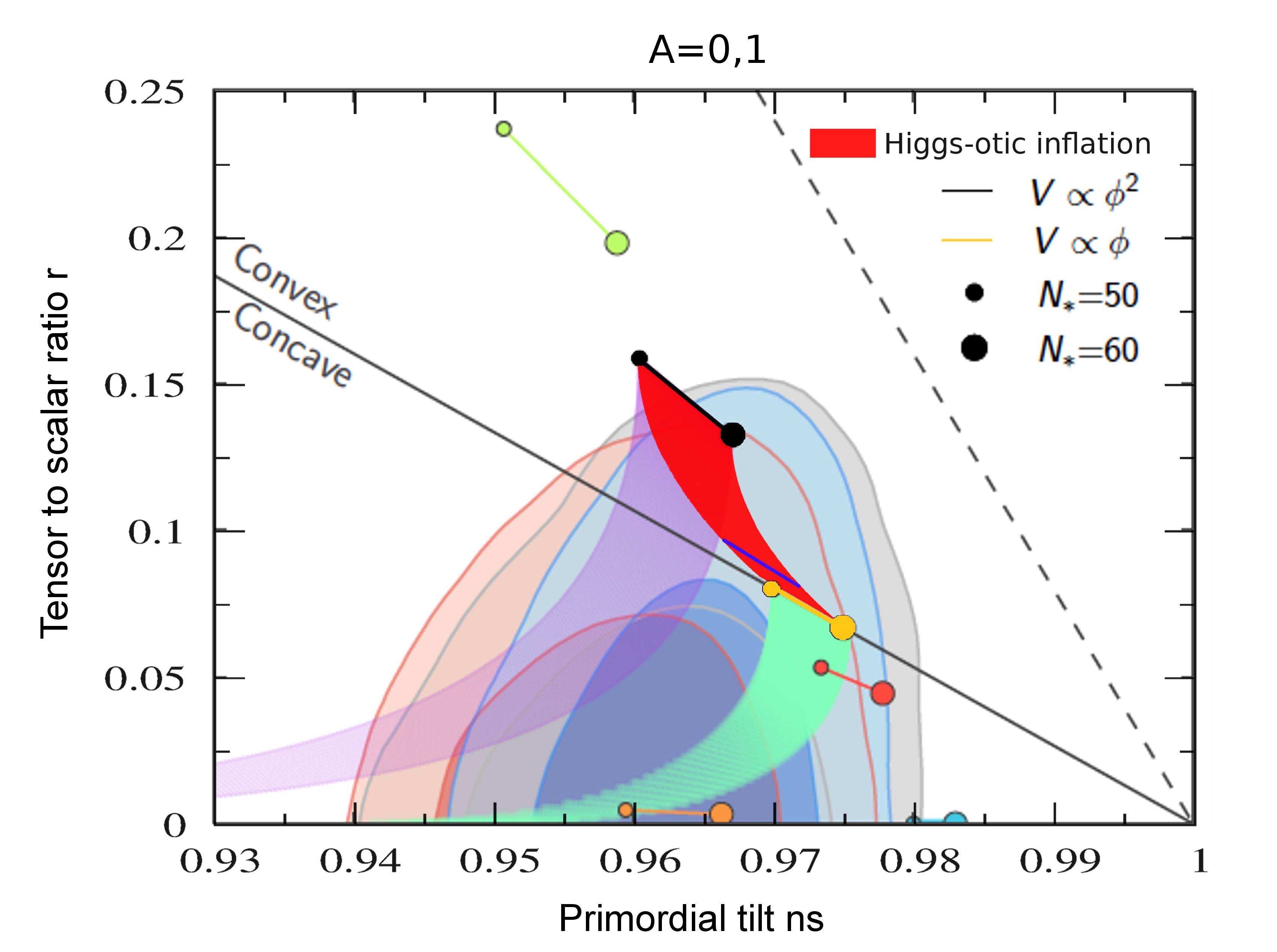}}
\end{center}
\vspace{-15pt}
\caption{Tensor to scalar ratio vs scalar spectral index for $A=0,1$ in Higgs-otic inflation (red band).}
\label{rnsonefield}
\end{figure}

Let us compute now the tensor-to-scalar ratio $r$ and the scalar spectral index $n_s$. We compute the field value $\phi_0$ at which inflation starts by imposing to get between 50 and 60 efolds before inflation ends. Notice that inflation ends when $\epsilon(\phi_{end})=1$. Once we know the initial value $\phi_0$, we can compute $r$ and $n_s$ by using eqs.(\ref{en}-\ref{nsr}). We plot the result in fig.\ref{rnsonefield}. The result for Higgs-otic inflation (red band) has been superimposed over the figure with the 
Planck experimental exclusion limits and some  inflationary models in the literature. Remark those corresponding to quadratic and linear potentials, given respectively by black and yellow points. Our model interpolates precisely between both of them, recovering a quadratic potential in the small $\hat G$ limit, and a linear potential in the large $\hat G$ limit.
 There is a special value for $\hat G$ (corresponding to the blue line inside the red band) given by considering generic fluxes in an isotropic compactification, 
 as estimated in section \ref{scales}.  It  corresponds to $\hat G\simeq 0.3/M_p$, implying a SUSY breaking scale around $10^{12}-10^{13}$ GeV (depending on the exact value of the string scale). The numerical results for  $\hat G\simeq 0.3/M_p$ are shown in table \ref{table}. Notice that the field range is given in units of the reduced Planck mass $M_p$. We can see that the prediction for the tensor to scalar ratio is around $r\simeq 0.09$.
%

\begin{table}
\begin{center}
\begin{tabular}{|c|cccc|}
\hline
$N_{\rm efolds}$&$\varphi_{end}$&$\varphi_{0}$&$r$&$n_s$\\
\hline
60&1.38&13.38&0.080&0.972\\
50&1.38&12.33&0.098&0.966\\
\hline
\end{tabular}
\caption{Results for $\hat G=0.3/M_p$ in isotropic compactifications.\label{table}}
\end{center}
\end{table}

Finally one could also wonder about the density of scalar perturbations. These have been measured experimentally by Planck obtaining an order of magnitude of
\beq
P_s=\frac{V}{24\pi^2M_p^4\epsilon}\sim \left(\frac{\delta \rho}{\rho}\right)^2\sim (10^{-5})^2
\eeq
Using that $V=M_{SS}\phi(\varphi)^2$ where $M_{SS}$ is the SUSY breaking scale and taking into account the field redefinition $\phi(\varphi)$, we can use the experimental result for the density scalar perturbations to estimate the SUSY breaking scale. The result is $M_{SS}\simeq 10^{12}-10^{13}$ GeV depending on the exact value of the string scale, in agreement with the assumption of closed string fluxes as the main source of SUSY breaking. More precisely, for $\hat G\simeq 0.3/M_p$ fixed, we obtain $M_{SS}\simeq 3\cdot 10^{12}$ GeV.

We can also estimate the number of times that the inflaton has to travel  along the torus. For simplicity let us assume that the overall internal space is a direct product of the internal 4d space wrapped by the D7-branes and the transverse torus such that
\beq
{\rm Vol}(B_3)={\rm Vol}(X_4){\rm Vol}({\bf T}^2)
\label{vol}
\eeq
where also $X_4={\bf T}^4$. Then ${\rm Vol}(X_4)=(2\pi R_c)^4$ and ${\rm Vol}({\bf T}^2)=(2\pi r)^2$.
The position of the branes is parametrised by
\beq
z_3=2\pi \alpha'\langle \varphi\rangle
\eeq
and the inflaton completes a period when $\langle \varphi\rangle_0=\frac{r}{\alpha'}$.  Using eq.(\ref{vol}) and the identities (\ref{Mp}) one period along the transverse torus is given by
\beq
\langle \varphi\rangle _0=\frac{1}{2\pi\alpha'}\left(\frac{{\rm Vol}(B_3)}{{\rm Vol}(X_4)}\right)^{1/2}=\frac{g_s^{1/2}m_p}{2\alpha_G^{-1/2}}\sim 0.5g_s^{1/2}M_p
\eeq
Hence if we need $\Delta \varphi\simeq 10 M_p$, we will need about 20 periods. Of course this is the worst case in which we are assuming the same radius for both cycles of the torus and that the inflaton is circling only around one of them. In general 
\beq
\Delta \varphi=\frac{R}{\alpha'} |m + iU_3 n|
\eeq
with $m,n$ the number of periods along both 1-cycles, so the effective number of periods can be considerably smaller (although always bigger than 1).

Note that all these  $A=0,1$ results are independent on whether the inflatons have the quantum numbers of the MSSM Higgs bosons. If they were describing any other scalar field, but still corresponding to the position of a D7-brane in such closed string flux background, then their potential would be described by the analysis of section \ref{DBICS} or and orbifold thereof and the same results would apply.
However, the case in which the inflaton is a Higgs field is further constrained by known Higgs physics. In particular, 
for Higgs-otic inflation we are interested in obtaining a massless eigenstate at the SUSY breaking scale that could play the role of SM Higgs boson, so we need a specific choice of fluxes satisfying $A\simeq 0.83$. This leads us to the two-field inflation case. However, if we start with initial conditions such that $<H> \gg <h>$   (implying $H_u=H_d^*$) the inflaton is mostly $H$ and the analysis here described is a good approximation. 
This is essentially the initial proposal in ref.\cite{Ibanez:2014kia}.
For generic initial conditions, however, both fields are relevant for inflation and a more general analysis is needed. We turn now to describe the more general case of two fields.

\subsection{The general 2-field Higgs/inflaton case}

In this section we deal with the more general and interesting case of the two field inflationary potential.

\subsubsection{Results for small field}

As a first approximation we assume that the fields take only small values such that the function $f$ is approximately $f(\sigma,\theta)\approx 1+\dots$ and we do not have to worry about the field redefinition. Notice that this is not consistent for our inflationary model in which the fields necessarily have to take large trans-planckian values in order to obtain of the order of 60 efolds during inflation. But this simplification allows us to solve analytically the equations of motion making easier the presentation of the new features that arise in a 2-field inflationary model with respect to the previous single field case. It is also a good approximation for very small values of $\hat G$. In the next subsection we will deal with the more general case including the field redefinition and obtaining a flattening of the potential. This will imply a reduction in the tensor to scalar ratio obtained in this subsection.

Neglecting the field redefinition coming from higher order corrections on $\alpha'$ in the DBI+CS action, the metric is simply given by $G_{ab}={\rm diag}(4,\sigma^2)$. This leads the following slow roll equations of motion
\begin{align}
&\frac{d\sigma (t)}{dt}=-c\ \sigma(t)(1-A\, {\rm cos}\tilde\theta(t))\\
&\frac{ d\tilde\theta (t)}{dt}=-c\ 2A\, {\rm sin} \tilde\theta(t)  \ .
\label{eqs}
\end{align}
where $c=Z^{-2}g_s(|G|^2+|S|^2)/6H$. These equations can be solved analytically, obtaining
\beq
\sigma (t)=\sigma(0)e^{-c(1+A)t}
\left(       \frac {1+e^{4{\rm Act}}{\rm cot}\left(\frac{\tilde\theta(0)}{2}\right)^2} 
   {1+{\rm cot}\left(\frac{\tilde\theta(0)}{2}    \right)^2} \right)^{1/2}
   \eeq
   \beq
{\rm tan}\left(\frac{\tilde\theta (t)}{2}\right)=e^{-2{\rm Act}}{\rm tan}\left(\frac{\tilde\theta(0)}{2}\right)
\eeq
which can be combined to obtain the slow roll trajectory $\sigma(\tilde\theta)$. 
This trajectory will be independent of the parameter $c$, recovering the well known result that the observables $r,n_s,N_{\rm efolds}$ are independent of the global factor of the potential in chaotic-like inflation models. Instead, these observables will depend only on the relative size of the fluxes parametrised by $A$.

By looking at the above equations, we can see that the phase remains unchanged $\tilde\theta(t)=\tilde\theta(0)$ for the case $A=0$, while 
$\sigma(t)=\sigma(0)e^{- c t}$. This is the typical exponentially decreasing behaviour of single field inflation and we recover the results described in the previous section.
The case $A=1$ is a bit special since the minimum of the potential is at $\tilde\theta=0$ for any value of $\sigma$, 
including $\sigma\neq 0$,  which  implies that at the end of inflation the gauge group $SU(2)\times U(1)$  remains broken. This is an unwanted
situation, since we want to maintain the SM gauge symmetry unbroken after inflation. So this particular limit would not be viable generically.
This case can also be reduced to single field inflation as we explained in the previous section. Here we are going to focus on an intermediate situation in which $A$ takes a value in between 0 and 1, so both fields may be important for inflation.

There is a novel feature of the 2-field case comparing with single field inflation: the dependence of the results on the initial conditions $\sigma(0)$ and $\theta(0)$. Depending on which initial point on the field space inflation starts, the slow roll trajectory  will be different giving rise to different values of the cosmological observables. Although one of the initial conditions can be fixed by imposing a specific number of efolds (as in single field inflation) the other one remains as a free parameter. This extends the range of possibilities but in principle also makes the model  less predictive.

\begin{figure}[h!]
\begin{center}
{\includegraphics[width=0.6\textwidth]{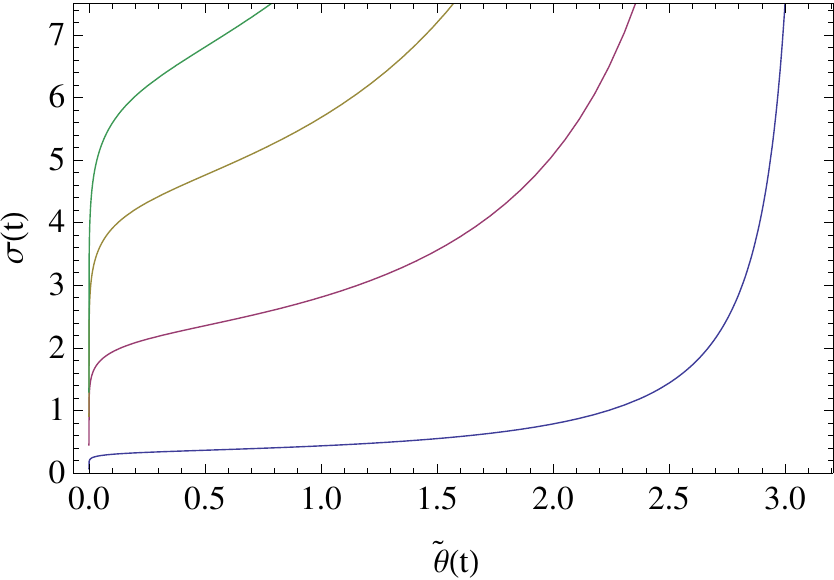}}
\end{center}
\vspace{-15pt}
\caption{Trajectory $\sigma(\tilde\theta(t))$ described by the slow roll eqs. of motion for $A=0.83$ and the different initial values $\tilde\theta(0)=3,3\pi/4,\pi/2,\pi/4$.}
\label{trajectory}
\end{figure}

As we argued above, for the SM Higgs to be fine-tuned and (approximately) corresponding to the $h$ linear combination we need to have $m_h^2\ll m_H^2$ at the string scale. This corresponds to a value of $A\simeq 1$.
In fact in section \ref{potential} we estimated the  required value of  $A$ in order to have a vanishing SM Higgs eigenvalue at a scale $\simeq 10^{13}$ GeV, obtaining a value around $A=0.83$. 
For this case of interest ($A=0.83$) we have plotted the trajectory followed in the $(\sigma,\tilde\theta)$-plane in fig.\ref{trajectory}, for different initial values $\tilde\theta(0)$.
We see that for an initial value at the top of the hill ($\tilde\theta(0)\simeq \pi$, $\sigma\simeq 7$) the inflaton goes downhill in the $\sigma$ direction keeping $\tilde\theta$ almost constant. Eventually the opposite happens and the phase goes fast to zero. 
For initial values at large $\sigma(0)$  but smaller $\tilde\theta(0)$ both $\sigma$ and $\tilde\theta$ decrease simultaneously. For small values of $\tilde\th(0)$ the inflaton goes fast to $\tilde\th = 0$ and then goes downhill in $\sigma$.

%

\begin{figure}[h!]
\begin{center}
{\includegraphics[width=0.4\textwidth]{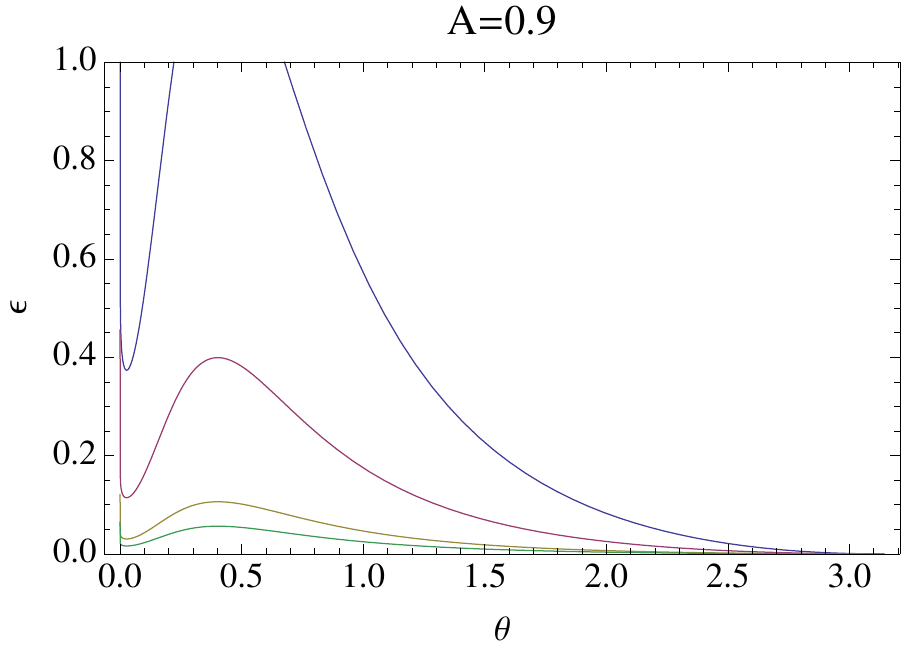}}
{\includegraphics[width=0.4\textwidth]{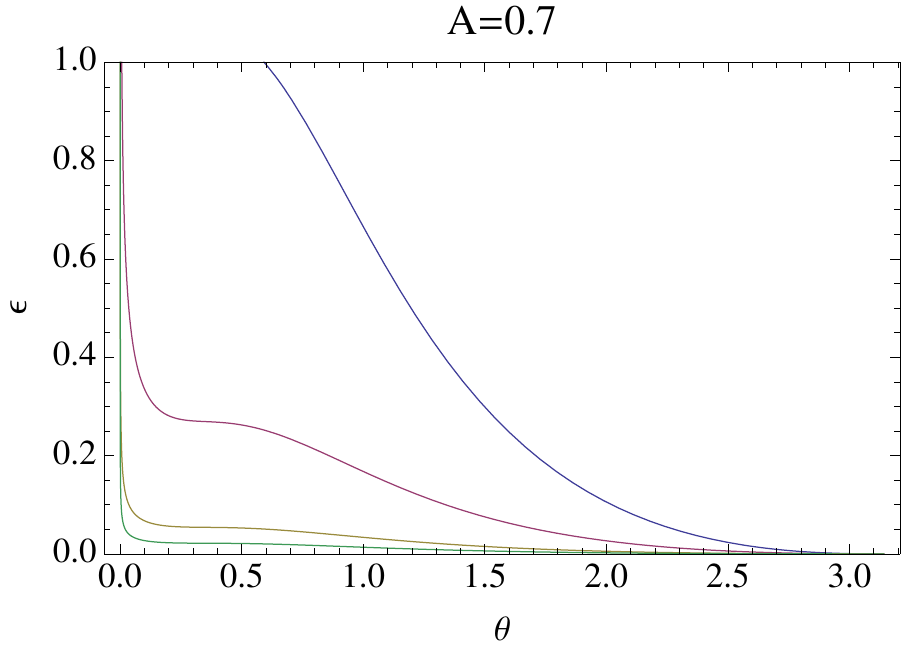}}\\
{\includegraphics[width=0.4\textwidth]{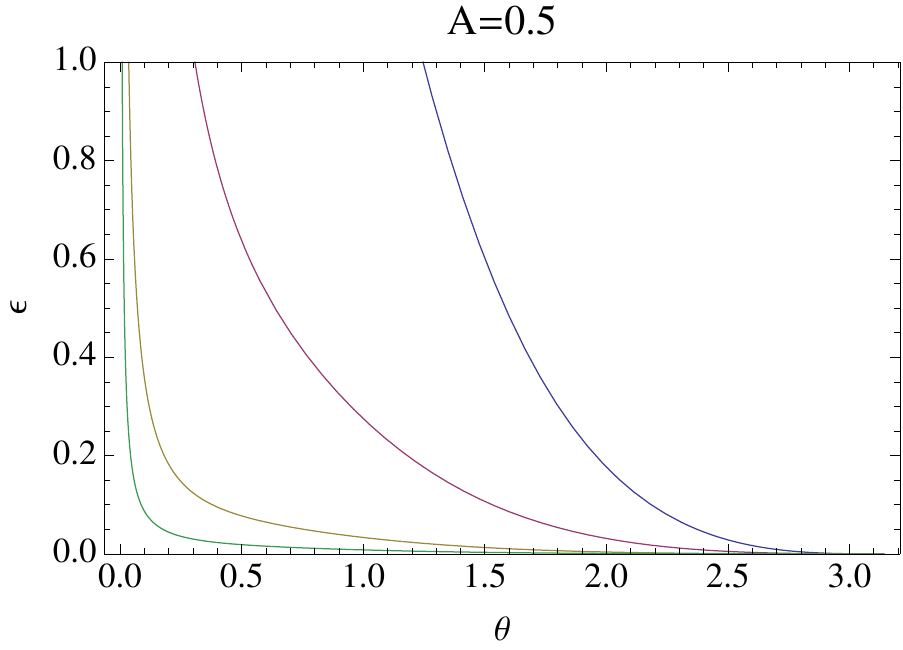}}
{\includegraphics[width=0.4\textwidth]{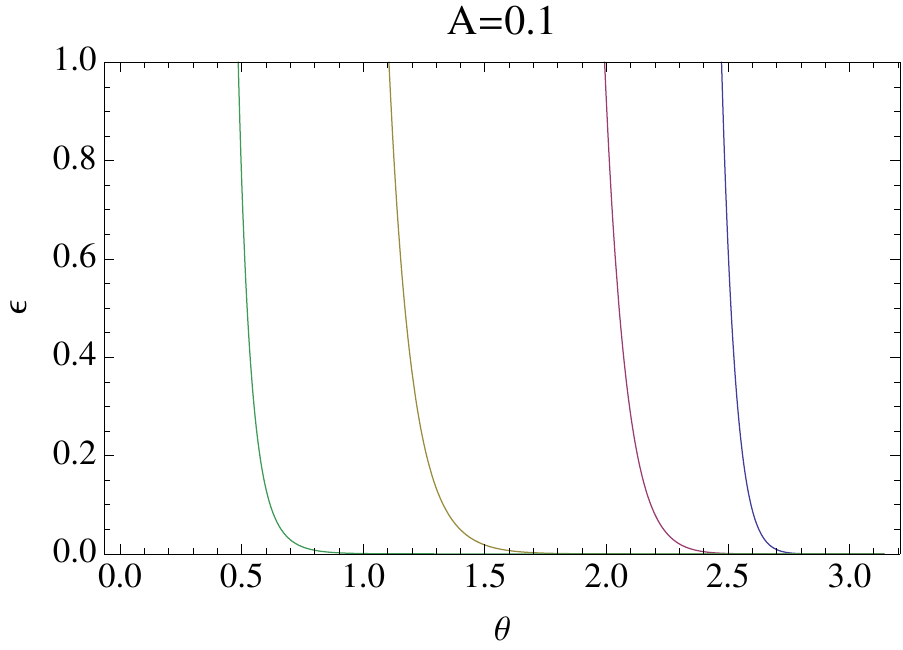}}
\end{center}
\vspace{-15pt}
\caption{The slow roll parameter $\epsilon$ as a function of $\tilde\theta$ for different values of $A$ and different possible trajectories. }
\label{epsilon}
\end{figure}

By using \eqref{e} we get the following formula for the slow roll $\epsilon$ parameter,
\beq
\epsilon=\frac{M_p^2}{2\sigma^2}\left(1+A^2\frac{{\rm sin}^2\tilde\theta}{(1-A\, {\rm cos}\, \tilde\theta)^2}\right)
\eeq
Given a value for $A$ and for the initial conditions $\sigma(0),\tilde\theta(0)$, we can compute the $\epsilon$-parameter along the inflationary trajectory $\sigma(\tilde\theta)$. The result is shown in fig.\ref{epsilon} for the same choices of trajectories depicted in fig.\ref{trajectory}, and this time also for different values of $A$. Inflation ends when this parameter becomes order 1, or alternatively when both fields reach their minima. 

\begin{figure}[h!]
\begin{center}
{\includegraphics[width=0.49\textwidth]{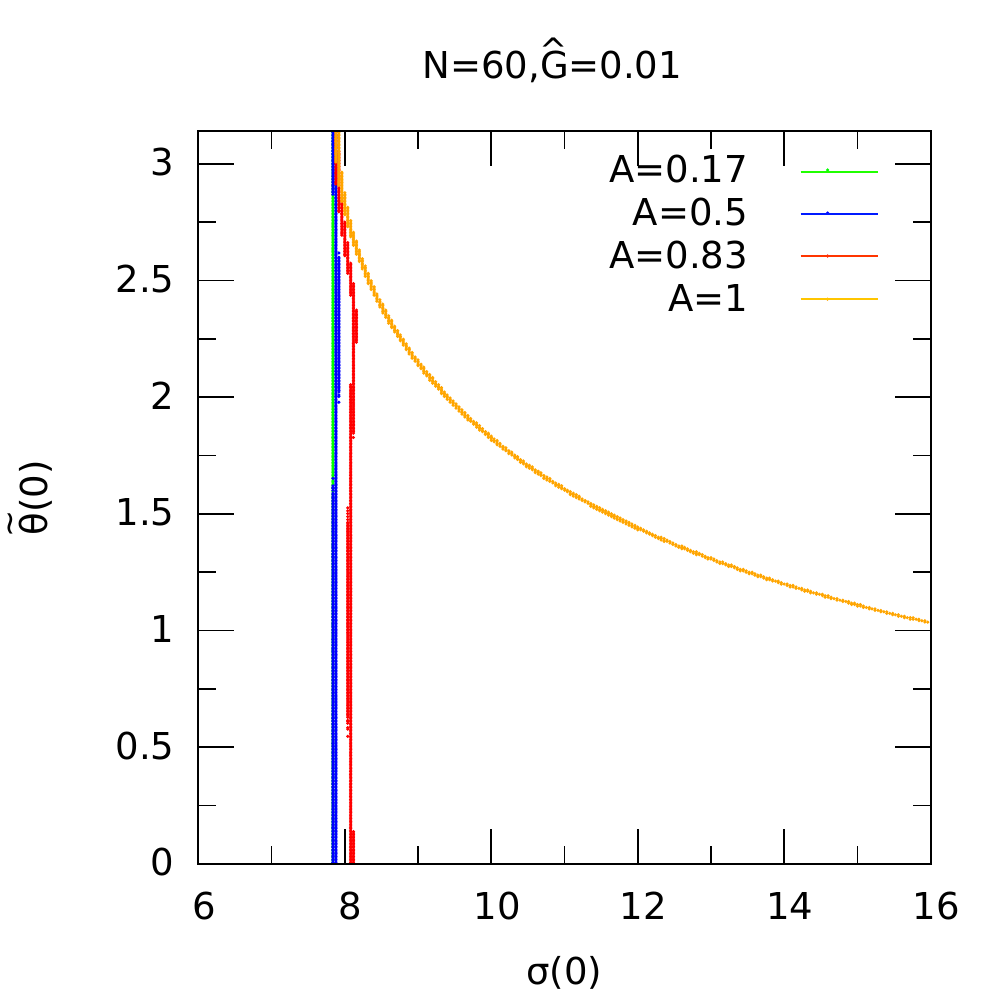}}
{\includegraphics[width=0.49\textwidth]{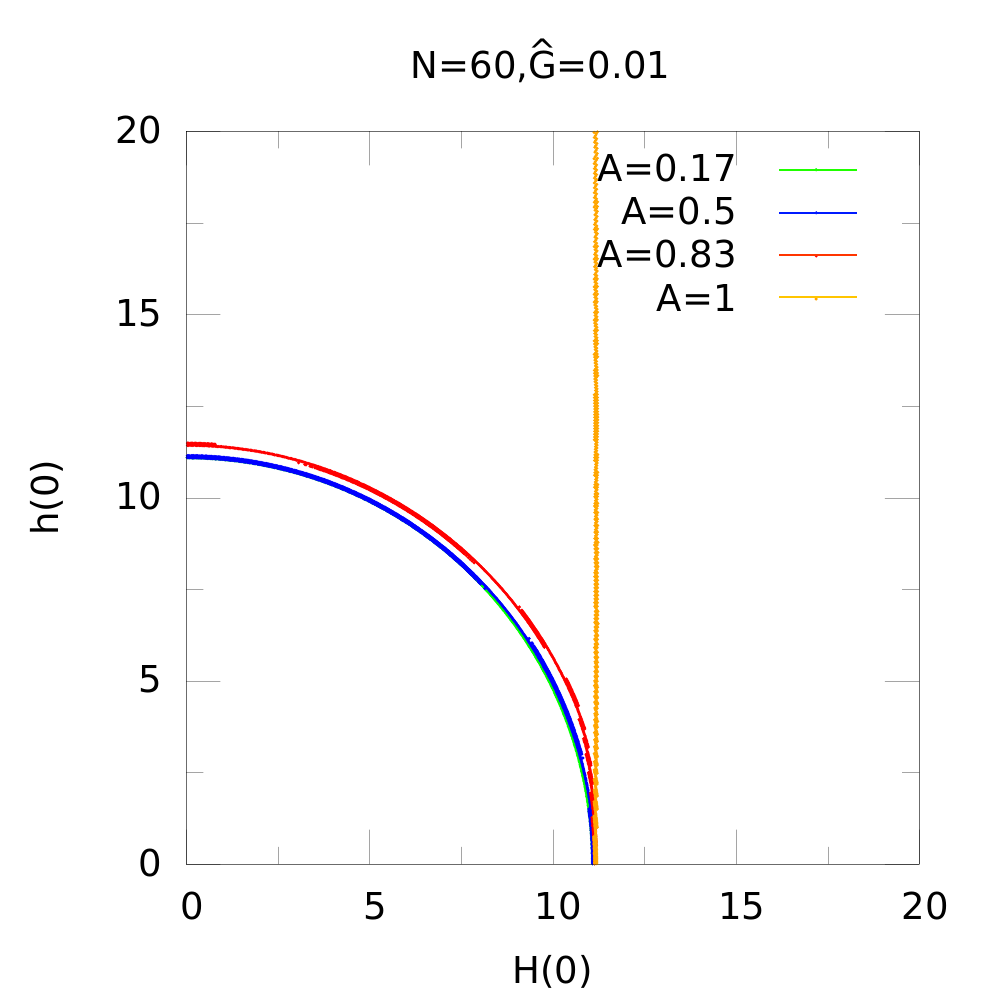}}
\end{center}
\vspace{-15pt}
\caption{Possible initial values that will give rise to $N_{\rm efolds}=60$. Each curve corresponds to a different value for $A=1,0.83,0.5,0.17$. For simplicity in the right plot we have assumed ${\rm Arg} (GS) =0$.}
\label{Nefolds}
\end{figure}

Replacing the metric in \eqref{Nefo} we get the following formula for the number of efolds,
\beq
N_{\rm efolds}=\int_{\tilde	\theta(0)}^{\tilde\theta_{end}}\frac{1}{\sqrt{2\epsilon(\tilde\theta,A,\sigma(0),\tilde\theta(0))}}\sqrt{4\left(\frac{d\sigma(\tilde\theta)}{d\tilde\theta}\right)^2+\sigma(\tilde\theta)^2}\ d\tilde\theta
\eeq
The value $\tilde\theta_{end}$ is the one at which $\epsilon=1$ and inflation ends. For some choices of initial conditions, we can see that $\epsilon$ remains $\epsilon<1$ until the fields almost reach the minimum of the potential, so $\tilde\theta_{end}\simeq 0$. Finally the tensor-to-scalar ratio is proportional to the $\epsilon$-parameter evaluated at the beginning of inflation,
\beq
r=16\epsilon|_{\tilde\theta(0),\sigma(0)}
\eeq
and the same for the primordial tilt,
\beq
n_s=1+2\eta|_{\tilde\theta(0),\sigma(0)}-6\epsilon|_{\tilde\theta(0),\sigma(0)}
\eeq
We have studied the possible trajectories in our parameter space that give rise to $N_{\rm efolds}=50-60$ before inflation ends. This constraint implies a curve in the parameter space of initial conditions $(\tilde\theta(0),\sigma(0))$ for each value of $A$ (fig.\ref{Nefolds}). Note  that the number of efolds (for $A<1$) is almost independent of $\tilde\theta(0)$. All the dependence comes from the fact that $\epsilon,\eta$ do depend on $\tilde\theta(0)$, and thus $\tilde\theta_{end}$ may be different for different initial values $\tilde\theta(0)$. The dependence of $N_{\rm efolds}$ on $A$ also comes from the slight dependence of $\tilde\theta_{end}$ on $A$. Therefore, the behaviour for $A<1$ is quite similar to that of $A=0$, in which $\sigma$ is the only inflaton. For $A=1$ the situation changes drastically and $N_{\rm efolds}$ only depends on $H(0)$, being this field the inflaton. Notice that in this case 60 efolds are obtained if $H(0)=11M_p$. Taking into account the definition of canonically normalised fields \eqref{def} for which the physical field would actually be $\sqrt{2}H$, this implies a physical field range of $15.5M_p$, as usual in chaotic inflation. Therefore we recover the results of chaotic inflation in the cases $A=0,1$.

\begin{figure}[h!]
\begin{center}
{\includegraphics[width=1\textwidth]{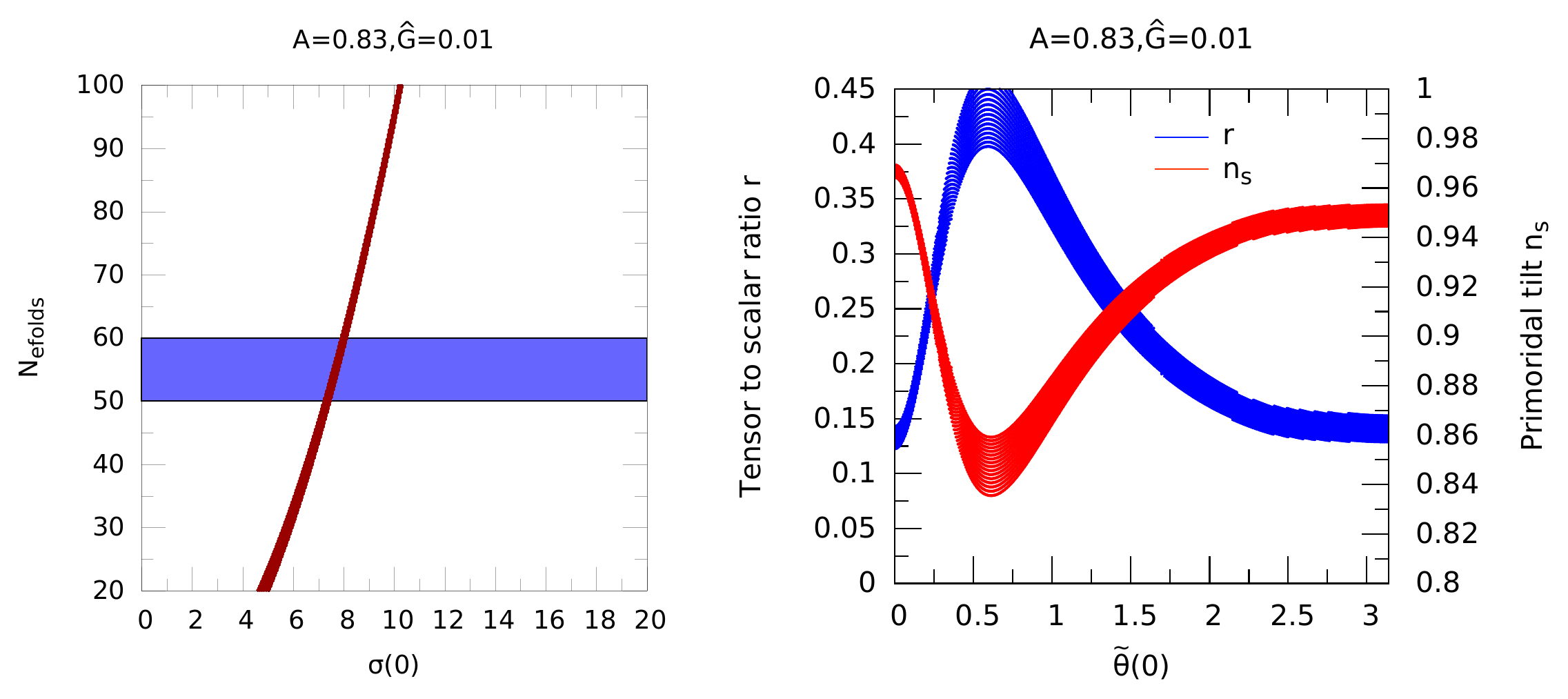}} 
\end{center}
\vspace{-15pt}
\caption{Left: Number of efolds vs the initial point $\sigma(0)$ for $A=0.83$ before flattening.
 Right: Tensor-to-scalar ratio (blue curve) and scalar spectral index (red curve) as functions of the initial point $\tilde\theta(0)$ for $A=0.83$.}
\label{refolds}
\end{figure}

Although the behaviour of $N_{\rm efolds}$ does not differ much from the single field cases, the results for $r$ and $n_s$ do. Let us explain the reason. We can use the constraint of getting 50-60 efolds to fix one of the initial conditions $(\sigma(0))$, as we can see in fig.\ref{refolds} (left). In the single field cases this determines completely $r$ and $n_s$, but here we have another free parameter, the other initial condition $\tilde\theta(0)$. We can then plot $r$ and $n_s$ in terms of $\tilde\theta(0)$ obtaining the functions depicted in fig.\ref{refolds} (right) for the case $A=0.83$. It is clear that these observables do depend on $\tilde\theta(0)$. The minimum value for $r$ that we can get corresponds to the result of chaotic inflation ($r\simeq 0.13$), while the freedom of choosing $\tilde\theta(0)$ allows us to get bigger values for the tensor to scalar ratio. However if we impose the experimental constraint for the primordial tilt $n_s$ only the region $\tilde\theta(0)> 1.7$ survives, implying $0.13 < r < 0.15$ again.

\begin{figure}[h!]
\begin{center}
{\includegraphics[width=0.49\textwidth]{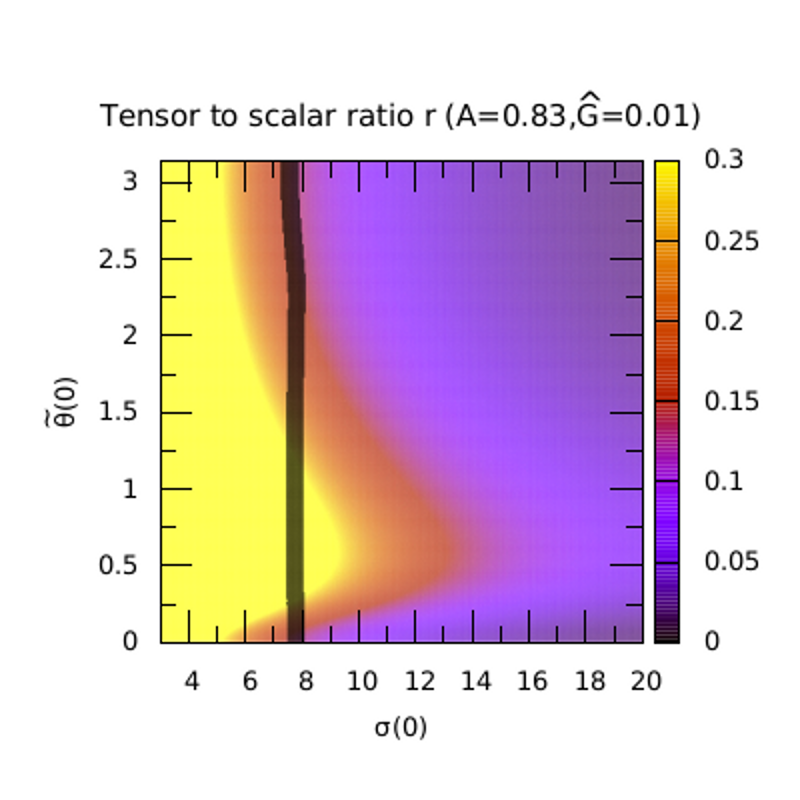}} 
{\includegraphics[width=0.49\textwidth]{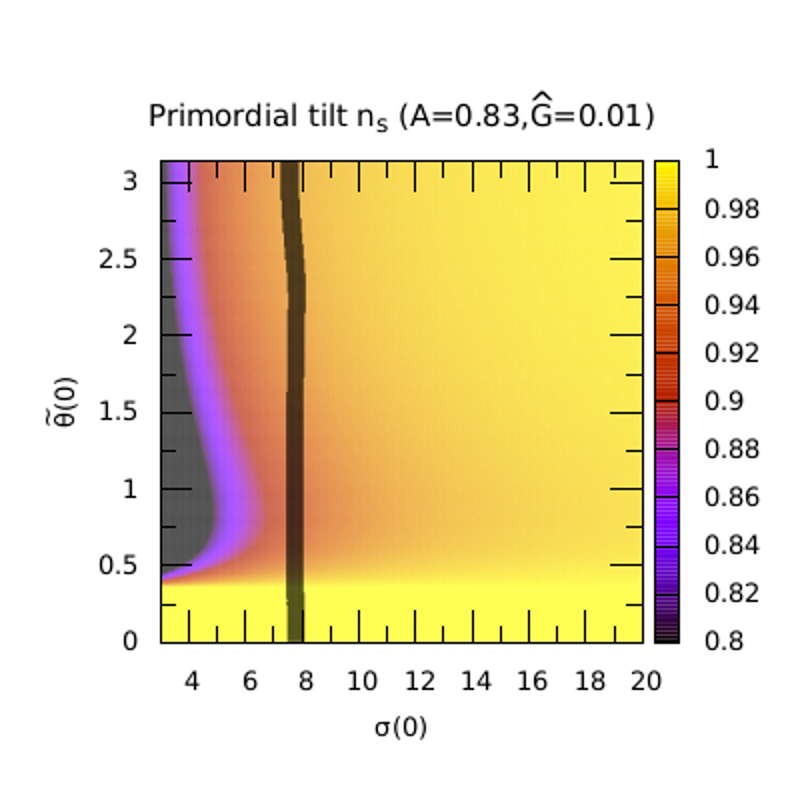}} 
\end{center}
\vspace{-15pt}
\caption{Left: Tensor-to-scalar ratio (contour plot) in the parameter space of initial conditions for $A=0.83$
before flattening. Only plotted those points which imply a potential $V^{1/4}<M_{s}$. The black band corresponds to those points which lead to 50-60 efolds. Right: The same for the primordial tilt.}
\label{rnscontour}
\end{figure}

In fig.\ref{rnscontour} we plot the value of the tensor-to-scalar ratio (left) and the scalar spectral index (right) without imposing a specific $N_{\rm efolds}$ in the parameter space of initial conditions for the relevant case $A=0.83$. It has been imposed that the potential energy remains lower that the string scale ($V^{1/4}<M_s$). This implies a lower bound in $r$ and an upper bound in $n_s$. Therefore although we allowed for more than 60 efolds, we could not get parametrically smaller values for $r$. It has also been superimposed a black band corresponding to the set of initial points which gives rise to $50<N_{\rm efolds}<60$, to guide the eye. Notice however that the region from the black band to the right part of the plot is also allowed (whenever the potential remains subplanckian) corresponding to $N_{\rm efolds}>60$ and a smaller $r$. These values for $r$ will decrease in the next section when including the flattening of the potential.

\subsubsection{Results for large field}\label{general2field}

Once we take into account higher order corrections in he DBI+CS action, the kinetic terms turn out to be non-canonically normalised. The metric in the field space is given by \eqref{metric}
\beq
G_{ab}=\left(\begin{array}{cc}4f(\sigma,\tilde\theta)&0\\0&\sigma^2 f(\sigma,\tilde\theta)\end{array}\right)
\eeq
with
\beq
f=1+\frac{|\hat G|^2}{2} (1-A\, {\rm cos}\,\tilde\theta)\sigma^2
\eeq
In the previous section we neglected this effect assuming $\hat G$ small. There the results did not depend on $\hat G$ because this parameter entered only as a global factor in the scalar potential. In this section we consider the most general case in which both $\hat G$ and $A$ can take arbitrary values. Hence, we have to deal with a two field inflationary model in which the kinetic terms are not canonically normalised, so we will use the generalisation for the slow roll parameters derived in \ref{slowformulae}. Now the results will also depend on $\hat G$ (and so in the SUSY breaking scale) as it enters in the field redefinition above.
As we explained in the single field cases, the effect of the field redefinition will be a flattening of the potential giving rise to a decrease in the tensor to scalar ratio (more important as $\hat G$ increases). The structure will no longer be that of double chaotic inflation.

\begin{figure}[h!]
\begin{center}
{\includegraphics[width=0.49\textwidth]{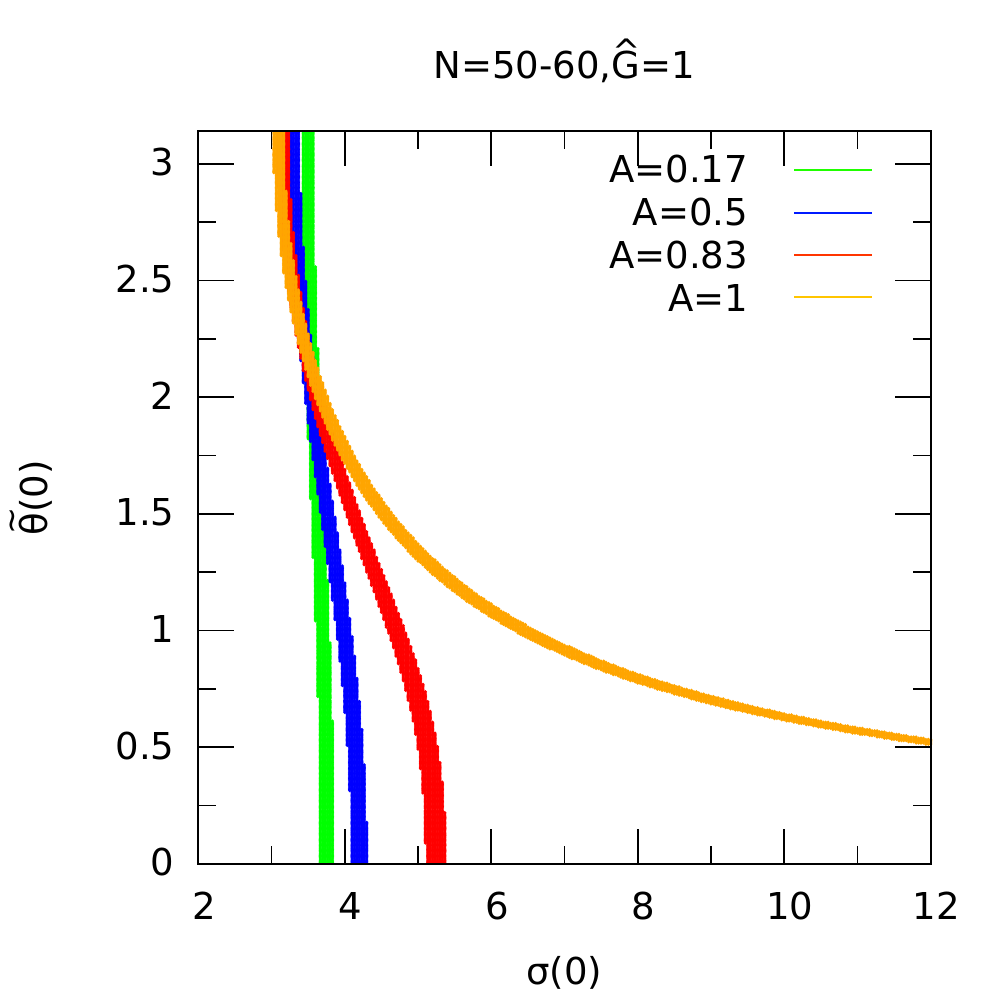}}
{\includegraphics[width=0.49\textwidth]{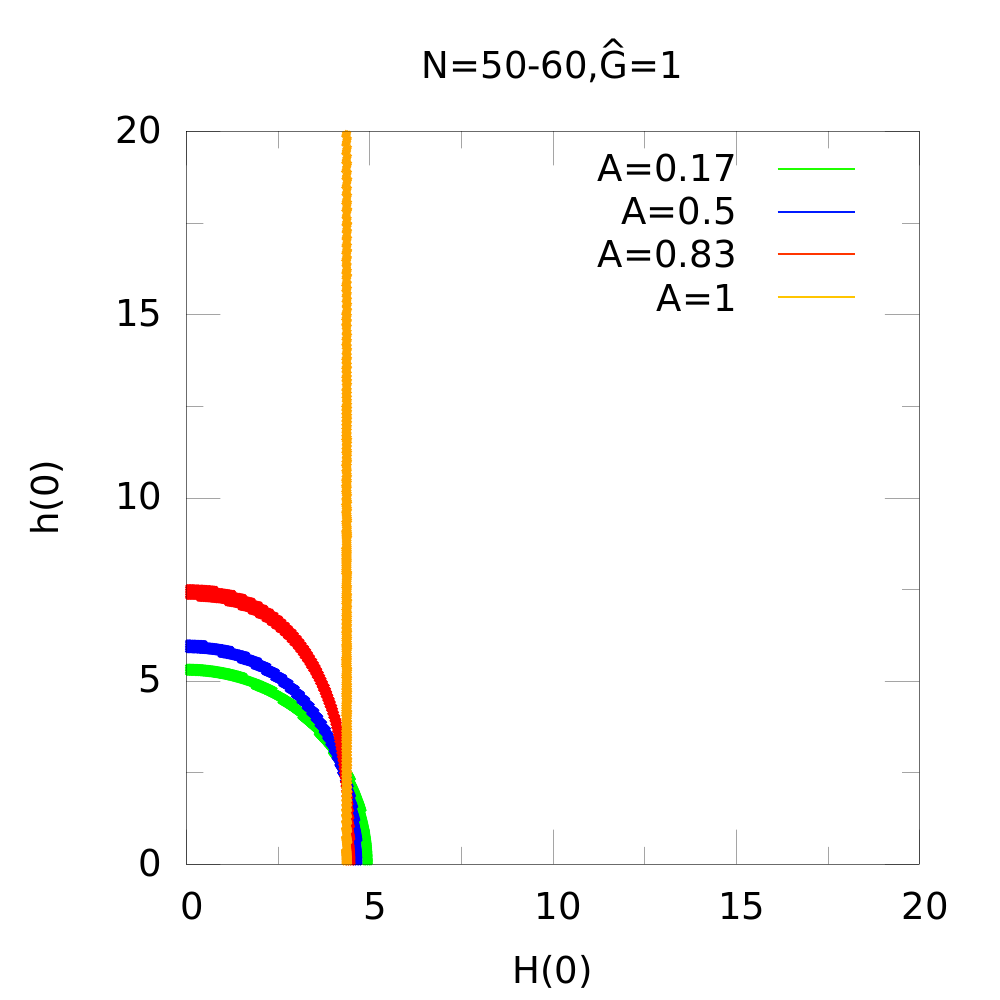}}
\end{center}
\vspace{-15pt}
\caption{Possible initial values that will give rise to $N_{\rm efolds}=50-60$. Each curve corresponds to a different value for $A=1,0.83,0.5,0.17$.}
\label{Nefolds_large}
\end{figure}

In the following we show the results for $\hat G=1/M_p$, corresponding to the biggest value for $\hat G$ that still implies a potential energy lower than the string scale. For this value, in the single field cases the potential was almost linear, so here we expect to recover the results of linear inflation for $A=0,1$. 
We show the same plots than in the previous section but now for  $\hat G=1/M_p$, to highlight the  flattening effect. 
Notice in fig.\ref{Nefolds_large} that 60 efolds are achieved now when $H(0)\simeq 4.7M_p$ for $A=1$. This field is not canonically normalised, so in order to compare with the physical field we have to compute the field redefinition, possible in this single field case. In fact, the result is  $H'(0)\simeq 11M_p$, recovering the result for linear inflation. 

\begin{figure}[h!]
\begin{center}
{\includegraphics[width=1\textwidth]{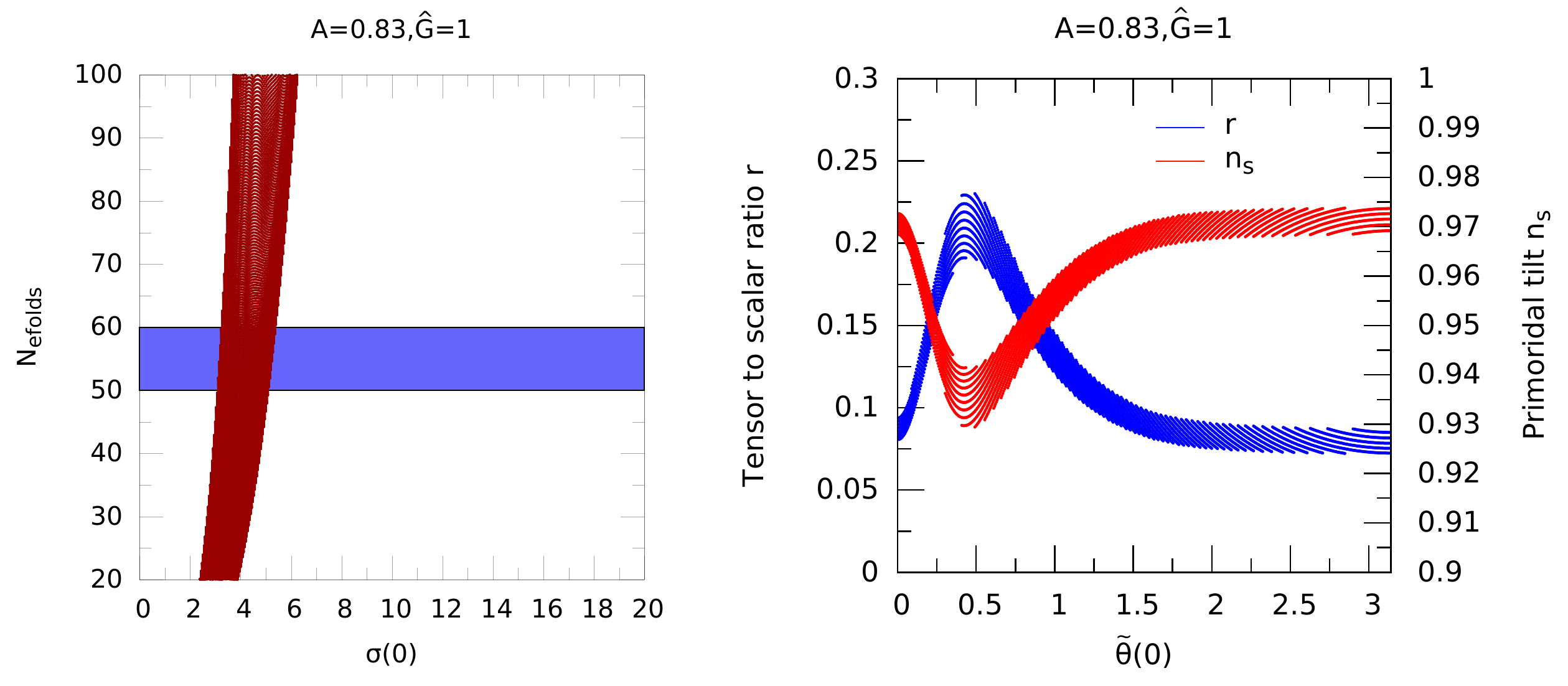}} 
\end{center}
\vspace{-15pt}
\caption{Left: Number of efolds vs the initial point $\sigma(0)$ for $A=0.83$. Right: Tensor-to-scalar ratio and scalar spectral index as functions of the initial point $\tilde\theta(0)$ for $A=0.83$.}
\label{refolds_large}
\end{figure}

In fig.\ref{refolds_large} we plot $N_{\rm efolds},r$ and $n_s$ for $A=0.83$. The tensor to scalar ratio is smaller than in the previous section for a bigger range of $\tilde\theta(0)$. In fact, after imposing the experimental bound on $n_s$, the value for $r$ is constrained to  the range $0.07$ - $0.1$, corresponding again to the result of a single field with a linear potential. Therefore, after imposing the experimental constraints, the results look quite similar to the single field case, as in the previous section.

\begin{figure}[h!]
\begin{center}
{\includegraphics[width=0.49\textwidth]{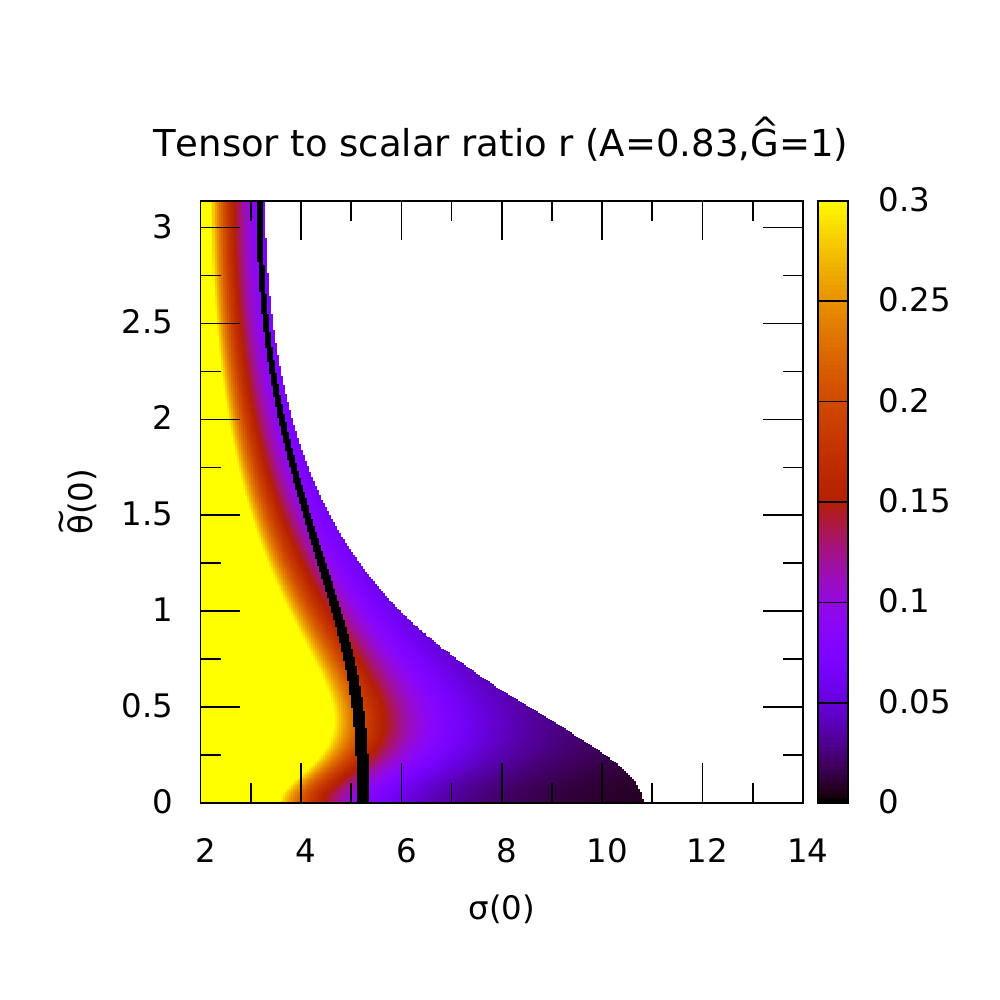}} 
{\includegraphics[width=0.49\textwidth]{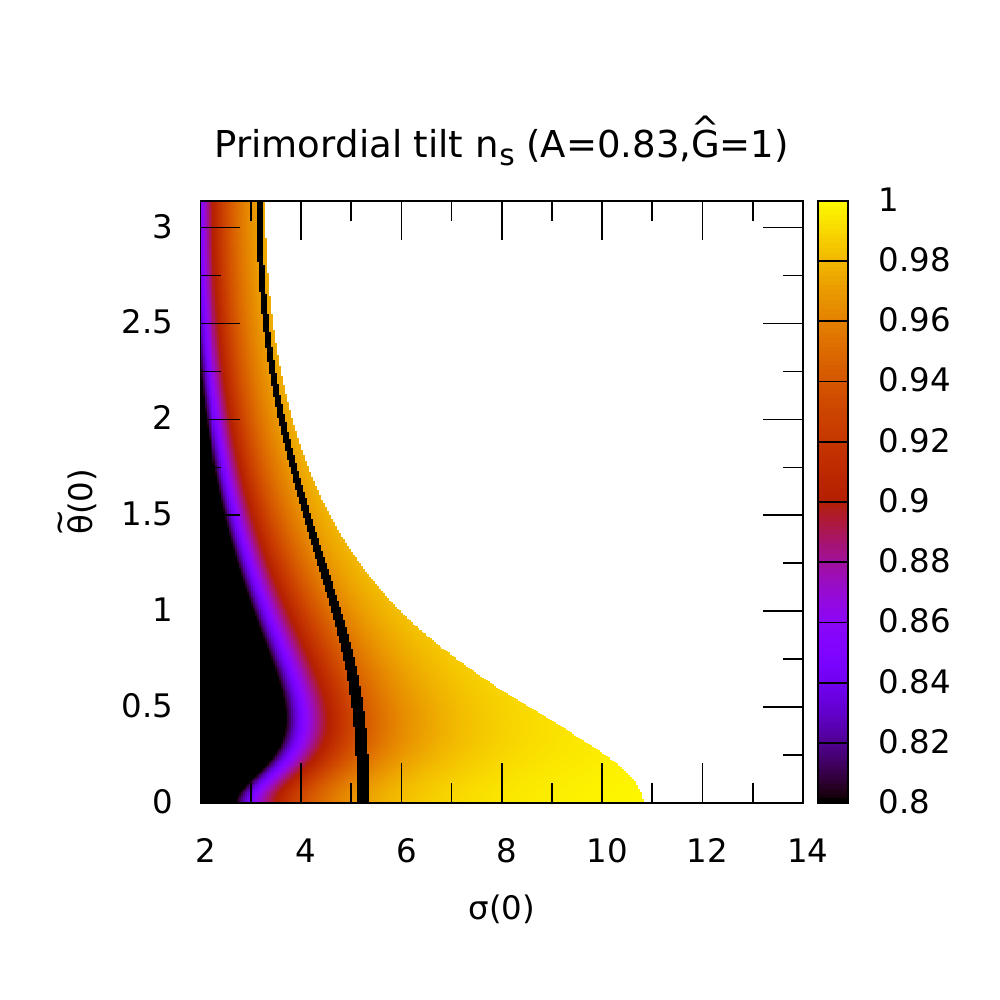}} 
\end{center}
\vspace{-15pt}
\caption{Left: Tensor-to-scalar ratio (contour plot) in the parameter space of initial conditions for $A=0.83$. Blank regions in the plots correspond to those points where $V^{1/4}>M_{s}$. The black band corresponds to those points which lead to 50-60 efolds. Right: The same for the primordial tilt.}
\label{rnscontour_large}
\end{figure}

In fig.\ref{rnscontour_large} we illustrate  the decrease on the tensor to scalar ratio due to the flattening of the potential. Notice also that the bound of getting $V^{1/4}<M_{s}$ is stronger, and the value $\hat G=1/M_p$ corresponds to the limit case in which this bound is still satisfied.

\begin{figure}[h!]
\begin{center}
{\includegraphics[width=0.39\textwidth]{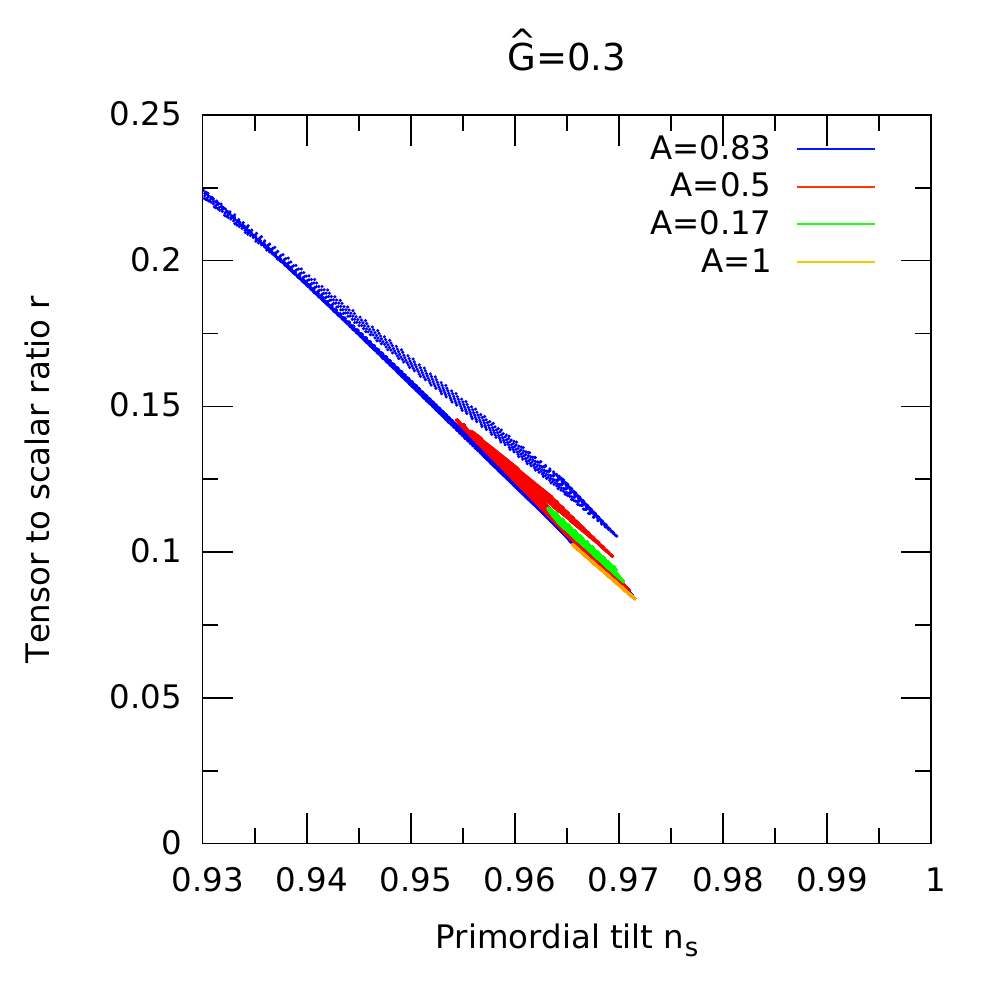}} 
{\includegraphics[width=0.6\textwidth]{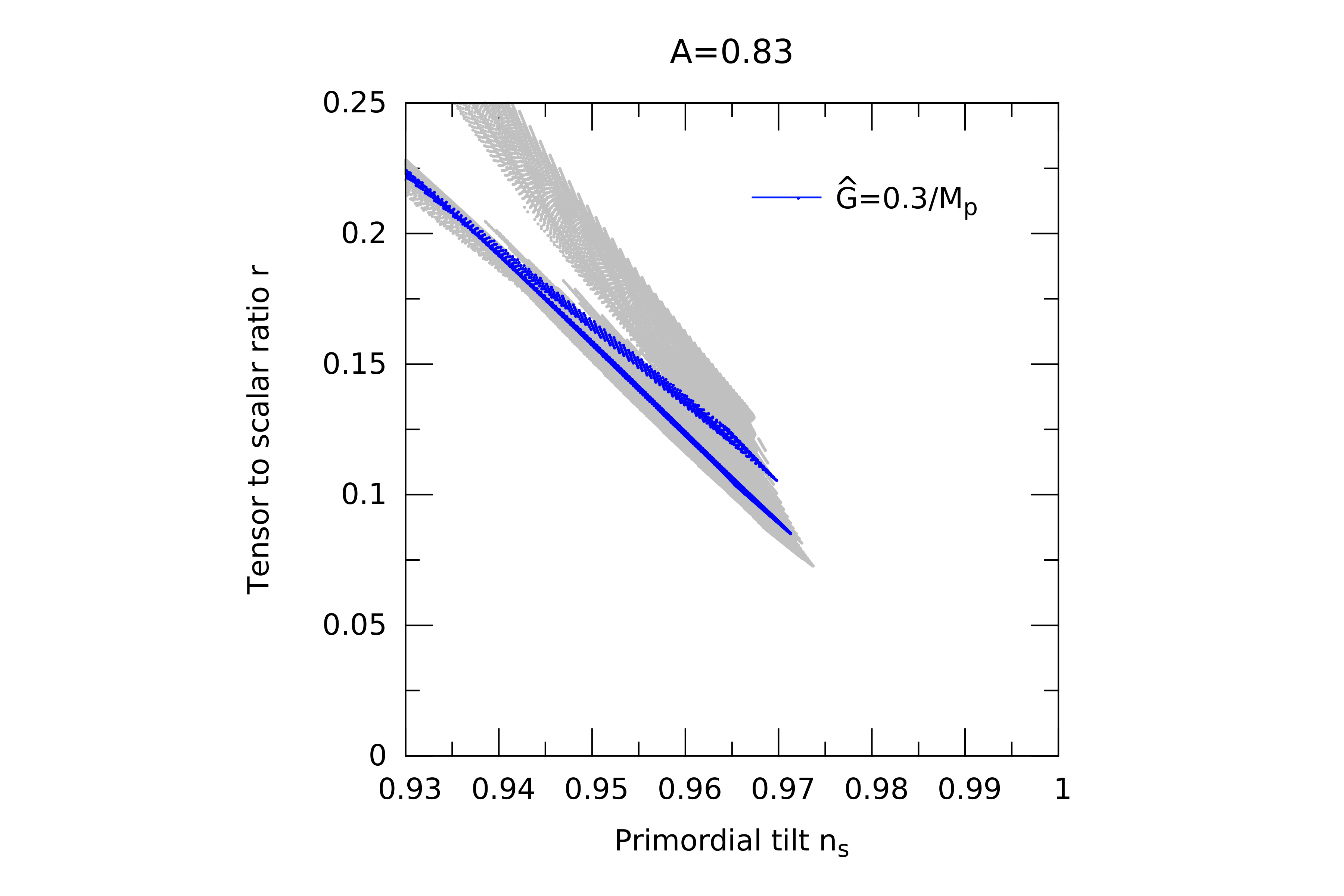}} 
\end{center}
\vspace{-15pt}
\caption{Allowed region of the parameter space ($r$ vs $n_s$) that gives rise to 50-60 efolds. Left: For different values of $A=1,0.83,0.5,0.17$ and $\hat G=0.3/M_p$. Right: For $A=0.83$ and any $\hat G$ (grey points). The blue points corresponds to $\hat G=0.3/M_p$.}
\label{rns_large}
\end{figure}
All these figures show the results for $\hat G=1/M_p$. The figures of the previous section can be recovered in this general analysis by fixing $\hat G$ small, around $\hat G\simeq 0.01/M_p$. For intermediate values of $\hat G$ we would have an intermediate situation between both sections. Recall that we have four free parameters in the model, two of them giving the absolute and relative size of the fluxes ($\hat G$ and $A$), and the other two parametrising the initial conditions of the two fields. We have seen that the initial conditions can be highly constrained by imposing a specific number of efolds and the experimental bound on the primordial tilt $n_s$. The relative size of the fluxes (given by $A$) is fixed by imposing the condition of getting a massless eigenstate at $M_{SS}$ which could play the role of the SM Higgs boson. Hence, only $\hat G$ remains as a free parameter. Although we are assuming an intermediate scale of SUSY breaking around $M_{SS}=10^{11}-10^{13}$ GeV (consistent with the density scalar perturbations), this flexibility still has a big impact in the results of the cosmological observables. 

In fig.\ref{rns_large} (right) we plot all the values for $r$ and $n_s$ that we can get for any possible value of $\hat G$. We only require to get between 50 and 60 efolds during inflation, and a light SM Higgs (so $A=0.83$). The minimum value for the tensor to scalar ratio that we can get is that one of linear inflation, around $r\simeq 0.07$. We have marked in blue those points that corresponds to an isotropic compactification with generic fluxes, ie. $\hat G\simeq 0.3/M_p$. 

For completeness, we also show the results for $r$ and $n_s$ for different values of $A$ and $\hat G=0.3/M_p$ (fig.\ref{rns_large} (left)). Although for arbitrary values of $A$ the inflaton could not be identified with a Higgs boson, the results still apply for a generic D7-brane position modulus playing the role of the inflaton in such a closed string background.

\begin{figure}[h!]
\begin{center}
{\includegraphics[width=0.8\textwidth]{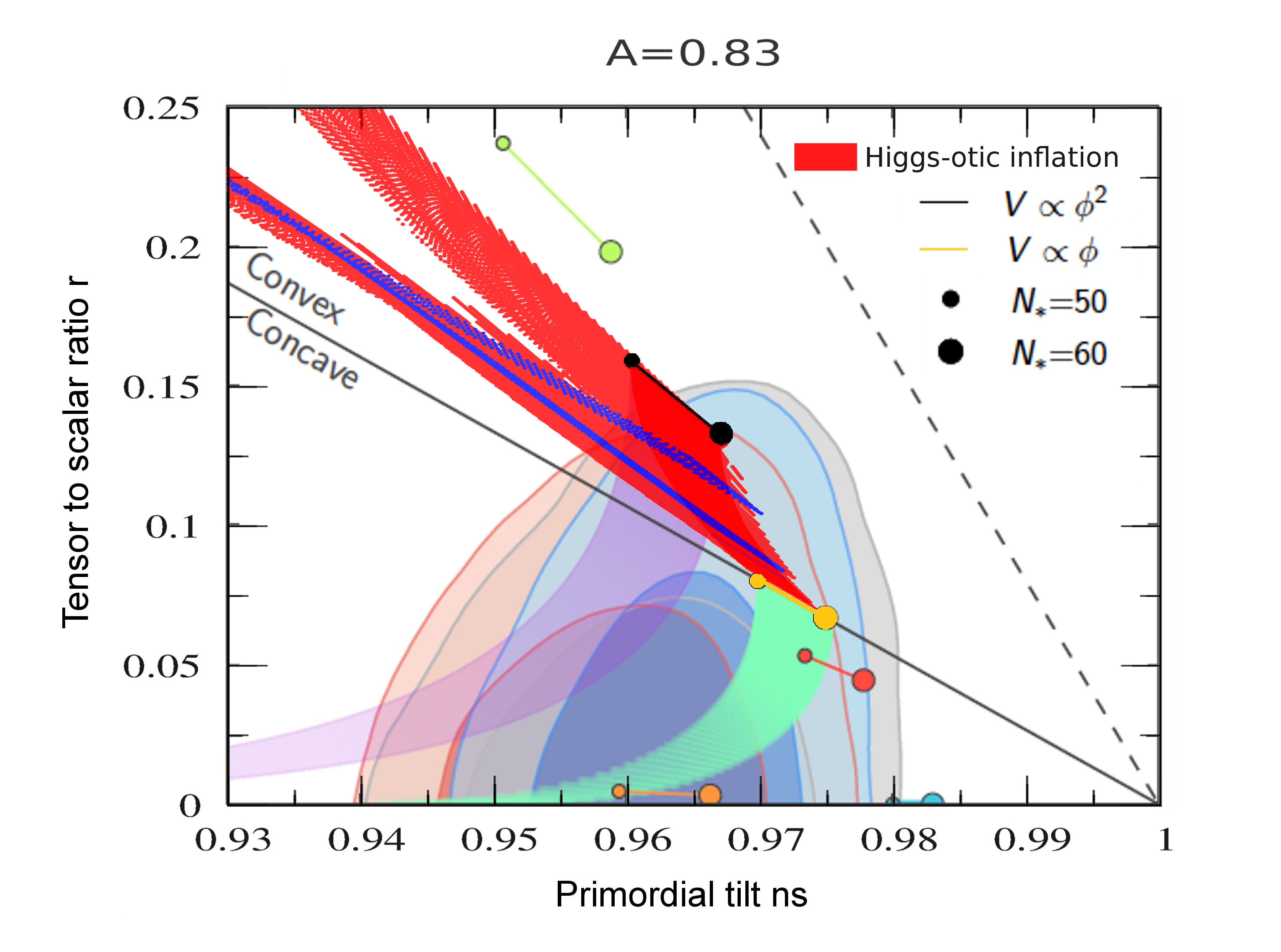}} 
\end{center}
\vspace{-15pt}
\caption{Higgs-otic inflation results for all possible values of $\hat G$ after imposing 50-60 efolds. They are superimposed over experimental Planck exclusion limits.
Region in blue corresponds to ${\hat G}=0.3/M_p$.}
\label{rns083}
\end{figure}

The general results in the $r$-$n_s$ plane for Higgs-otic inflation are shown in fig.18, superimposed over the Planck exclusion limits. The red band corresponds to results with $A\simeq 0.83$, 50-60 efolds and general initial conditions and $\hat{G}$. The blue region corresponds to $\hat{G} = 0.3/M_p$. Although not visible in the plot, the regions at smaller $r$ are more densely populated. This can be noticed in fig.15 (right plot) in which $0.07 < r < 0.1$ except for the region $0.1 \lesssim \tilde{\th}(0) \lesssim 1.2$.

\section{Inflaton potential corrections, backreaction and moduli fixing}\label{others}

We will consider here in turn several properties of our inflaton system concerning corrections 
and the possible back-reaction of the closed string sector on the potential. 
We first discuss the Planck suppressed corrections to the inflaton potential,
which are under control and fully given by the DBI+CS action.
We then study the possible induction of D3-brane RR-tadpoles for non-vanishing values of  the Higgs/inflaton. 
We show how there is a delicate cancellation coming from the closed string 
10d action which sets to zero such tadpoles. Finally, we briefly discuss the issue of
the moduli fixing potential and how one could hope to separate their dynamics from that of
the inflaton sector.

\subsection{Planck mass suppressed corrections}
\label{corr}

Higher  dimensional   Planck-supressed operators, i.e. terms of the form
$(\phi^{4+2n}/M_p^{2n})$ correcting the inflaton potential are a potential danger
for the slow-roll conditions.  The simplest such corrections to a an inflation potential $V_0$ 
are possible terms of the form
\beq
V_n \, \simeq \,  V_0 \times  \left(\frac {\Phi^2}{M_p^2}\right)^n 
\label{jodienda}
\eeq
with $n>0$. 
Such terms can give large contributions to the slow-roll parameters
driving  $\epsilon,\eta\simeq 1$ for  transplanckian excursions of the inflaton. 
To avoid the presence of such terms it is customary to assume the existence of a
shift symmetry under which $\phi \rightarrow \phi+c$  and the K\"ahler potential remains invariant.

The presence of such a symmetry helps also in trying to solve a second related problem,
the $\eta$ problem in $N=1$ supergravity. The idea is that the pre-factor $e^K$ appearing in
the supergravity potential will tend to give a large  (of order $H$) mass to the inflaton, once one
expands $K$ to leading order in the inflaton field. In the case of chaotic inflation this problem
is not severe because $m_I$ needs to be only one order of magnitude smaller than $H$, which can easily
be done by  a modest fine-tuning. If the inflaton does  not appear explicitly in the K\"ahler potential, as happens in the presence
of a shift symmetry, such mass term for the inflaton does not appear to leading order.

The effective action of string axions are known to possess shift symmetries which 
could protect the inflation potentials against these effects.
In fact such shift symmetries  are typically part of  larger non-compact groups 
leaving invariant the $N=1$ supergravity effective action. These large continuous groups
are broken by instanton effects down to discrete (infinite) groups which are 4d duality groups in general.
These shift symmetries are particularly welcome in models with large inflaton excursions, in which
one expects that the above Planck-supressed corrections could be very important.

In the case we are considering here, the  $N=1$ supergravity $\eta$ problem  is mixed up with the fine-tuning required to have 
a massless SM doublet left below the SUSY-breaking scale $M_{SS}$. There is a fine-tuning in the flux parameters such that
both a SM doublet survives and  the inflaton mass is slightly below $H$.  Both fine-tunings cannot be disentangled.

Concerning the first problem, 
in the case we studied above in which the inflaton is a D7-brane modulus
corrections to the inflation potential of the type (\ref{jodienda}) do indeed
appear. The important point however is that those corrections are computable 
to all orders in inverse Planck masses and are under control.  Indeed, in our case the inflaton/Higgs fields
are open string fluctuations and their action, including their interaction with closed string moduli
are given to all orders in $\alpha '$ by the DBI+CS action.\footnote{Corrections in $\a'$ to the non-Abelian DBI action which describes our MSSM system are to date not fully understood. However, notice that for large values of the inflaton, the inflationary system is described by a single D7-brane plus orbifold images. Thus, all $\a'$ corrections relevant to inflation should be captured by those of the Abelian DBI action.} 
For illustrative purposes let us look at the DBI+CS action for the $U(N)$ adjoint that we discussed in section \ref{DBICS}.
There we see that the full effect of those corrections is just a field redefinition. In particular one gets a structure of the form
\beq
{\cal L}_{4D}\ =\ \text{STr} \left(  \left[1+\frac {\kappa}{2} V_0(\Phi)\right]\ D_\mu \Phi  D^\mu {\bar \Phi}  \ - \ V_0(\Phi) \ +...\right) \ .
\eeq
with $\kappa = (V_4\mu_7g_s)^{-1}$ a constant. 
After a field redefinition one sees that corrections will always appear in powers  of  the initial fiducial potential $V_0$.
Thus indeed large corrections to the potential of the form in (\ref{jodienda})  do appear but
 in the D-brane case  here considered these  corrections are under control and
lead to a flattening of the potential, i.e., the potential becomes of linear type rather than quadratic,
leading to a new potential consistent with slow-roll.

It is interesting to consider the $N=1$ sugra avatar of this property.
We have seen how the K\"ahler potential involving the Higgs/inflaton fields is
\beq
K_H =\ -\text{log}[(S+S^*)(U_3+U_3^*)\ -\ \frac {\alpha'}{2}\ |H_u+H_d^*|^2]  \  \ -\  3\text{log}(T+T^*) \ .
\eeq
As shown in refs.\cite{shiftsymm} for the S-dual heterotic case, the Lagrangian described by this 
K\"ahler potential is invariant under a $SL(2,{\bf Z})_{U_3}$  geometric symmetry associated to 
reparametrisation of the corresponding ${\bf T}^2$. In particular it is easy to check that under the 
continuous transformations
\beqa
U_3\  &\longrightarrow \ &\frac {a U_3-ib }{ic U_3 +d} \\
S\  &\longrightarrow \      &S\ -\ \frac {ic}{2}    \frac {H_uH_d}{ic U_3+d} \\
H_u\  & \longrightarrow \  &\frac{H_u}{ic U_3+d } \\
H_d\  & \longrightarrow \  &\frac{H_d}{ic U_3+d}  \ , \ a, b, c, d \in{\bf R}
\eeqa
with $ac-bd=1$, the K\"ahler potential transforms like
\beq
K_H\ \longrightarrow \ K_H\ +\ \text{log} |ic U_3 + d|^2  \ .
\eeq
The latter is a K\"ahler transformation, so the Lagrangian will be invariant under it even if the K\"ahler potential is not.  
One can also easily check that the addition of a $\mu$-term does not spoil this symmetry.
 In fact the low-energy effective action is invariant under 
 these continuous symmetries, while
 the discrete subgroup with 
$a,b,c,d \in {\bf Z}$ is preserved to all orders in
perturbation theory or sigma model expansions and it is only broken spontaneously 
once the moduli are fixed. 
Note that these transformations act both on the moduli and the 
Higgs/inflaton fields so that e.g. the particular dependence on the
combination $H_u+H_d^*$ is dictated by the symmetries. 
In particular  the linear combinations of Higgs fields transform as
\beq
H_u \pm H_d^*\ \longrightarrow \
\frac {d (H_u\pm H_d^*)\ +\ i c(U_3H_d^*\mp U_3^*H_u) }
{| i c U_3\ +\ d |^2} \ .
\eeq
In the case with $a=d=0$, $b=1,c=-1$, one has $U_3\rightarrow 1/U_3$ and 
\beq
e^{i\g/2} H_u \pm e^{-i\g/2} H_d^*\ \ \longrightarrow \ -i\ \left(\frac{e^{-i\g/2}H_d^*}{U_3^*} \mp \frac{e^{i\g/2}H_u}{U_3} \right) \ .
\eeq
For a square torus $U_3=1$ and the above transformation just exchanges the fields $h$ and $H$.
This is somehow expected because in this case the transformation $U_3\rightarrow 1/U_3$
 corresponds to the exchange of the two cycles of the torus. The transformation also implies a shift of the 
 complex dilaton $S\rightarrow S-\frac {1}{2}H_uH_d$, as expected from the fact that $H_uH_d$ parametrises
 the wandering D7-brane position. 
 Finally, an analogous symmetry  $SL(2,{\bf Z})_{S}$ 
acting on the complex dilaton $S$ exists. The transformations look the same
as the ones above replacing $U_3\leftrightarrow S$.  However this $S$-duality symmetry
is in general broken by quantum corrections.

A direct  consequence of the modular  symmetries is that, since the K\"ahler potential is {\it not} 
invariant and only the Lagrangian and the potential are, one expects corrections of higher
order in $\alpha'$ to appear as powers of the potential itself, that is
\beq
\delta V_H \simeq \sum_{n>1} \    \frac {(V_0)^{n}}{(M_p)^{4(n-1)}} 
\eeq
Indeed this is consistent with the higher order corrections in $\alpha'$ given by the DBI+CS
action that we studied, and with the fact that such action is related to a Kaloper-Sorbo 4d effective Lagrangian.  
As stressed, in our case these Planck suppressed corrections are known and give rise to the 
flattening of the inflaton potential at large field. 

Note that a corollary of this discussion is that using the  (two-derivative) $N=1$  sugra standard formalism
does not capture these corrections leading to the flattening of the scalar potential for large fields. This is particularly the case for
any model in which the inflaton is an open string mode. 

Let us finally comment that above the inflaton mass, where SUSY couplings are recovered, 
 the loop corrections to the potential  are only expected to lead to
logarithmic corrections with small coefficients, and should not modify in any important way
the shape of the potential. Among these loop corrections one expects the presence of minute 
modulations for  the overall linear potential at large field. They would arise from the fact that, as we 
mentioned at the end of section (\ref{HiggsIIB}), as the  D7-brane position varies over the torus, the
masses of the massive fields $W^\pm,Z^0, H^\pm$ etc. oscillate. This oscillation should induce  in turn one-loop minute 
field dependent oscillations on the inflaton mass parameters.

\subsection{Backreaction and induced RR-tadpoles}

In general one expects that wandering D7-branes may lead to some level of backreaction in the surrounding geometry. 
This is a well known fact present in all perturbative IIB orientifolds with  D7-branes. However our setting is initially
supersymmetric, with SUSY broken spontaneously, and in this sense is more stable that settings  in which there
are both branes and antibranes and SUSY is broken at the string scale. In any event,  taking into account this back reaction would
require to go to a F-theory setting. We will have nothing to add concerning this issue other than pointing out that it would be interesting the
embedding of this type of models into an F-theory background. For previous proposals of large field inflation models within F-theory see \cite{Palti:2014kza,Hebecker:2014eua,Grimm:2014vva,Arends:2014qca,Hebecker:2014kva}.

Other than that, the presence of a non-vanishing vev for the inflaton may have also an impact on the surrounding geometry.
In particular as we have seen in section \ref{DBICS} the inflaton vev induces a background for the B-field in the D-brane worldvolume,
which in turn leads to induced lower dimensional D-brane RR charges. This fact has already appeared in previous monodromy inflation models,
leading to the introduction of brane-antibrane pairs to cancel the tadpoles. We show here that this is not the case in our setting, and therefore there is no need to introduce anti-branes.

The different sources of D3-brane tadpole in a type IIB compactification with O3/O7-planes are captured by the 5-form Bianchi identity as follows
\be
d\tilde{F}_5\, =\, F_3 \wedge H_3 + \sum_i \delta_6(p^i_{D3}) + \sum_j \delta_4(\pi^j_{D5}) \wedge B + \sum_k \delta_2(S^k_{D7}) \wedge \oh B^2 + \dots
\label{dF5}
\ee
ignoring factors of $\alpha'$, etc. Here $p_{D3}^i$ runs over all points where D3-branes are located, $\pi_{D5}^j$ over the 2-cycles where D5-branes are located, and $S^k_{D7}$ over the divisors wrapped by the D7-branes. The $\d_{2n}$'s are $2n$-form bump functions localised on their respective worldvolumes.\footnote{In general the D7 and D5-branes will be magnetised by an open string worldvolume flux $F$, so one should replace $B \raw \CF = B + F$ everywhere in (\ref{dF5}). For the sake of simplicity we will stick to the above notation, the generalisation being straightforward.} Finally, the dots represent similar delta function contributions of opposite sign that come from the O3 and O7-planes. Typically, it is this negative contribution that allows for the integral of the r.h.s of (\ref{dF5}) over the compact manifold $X_6$ to vanish, in agreement with the fact that $\tilde{F}_5$ should be globally well-defined. If this integral over $X_6$ does not vanish we say that we have a D3-brane tadpole. 

The problem arises when we have a non-trivial $H_3$ in our compactification, because then the contribution from D5-branes and D7-branes, which depends on the pull-back of the B-field in their worldvolume, is position-dependent. Hence it is not clear if, given that we can solve the tadpole condition in one particular point in open string moduli space $\{p^i_{D3}, \pi_{D5}^j,  S_{D7}^k\}$, we can solve it for a different point $\{p^{i\, \prime}_{D3}, \pi_{D5}^{j \, \prime},  S_{D7}^{k \, \prime}\}$. In other words, when we move a D7-brane from $S_{D7}$ to $S_{D7}'$ the pull-back of $B^2$ on its worldvolume changes, and so does its induced D3-brane charge. It then seems that, during inflation, we will generate a D3-brane tadpole as soon as we move the D7-brane from its initial location.

In the following we will show that this is not the case. Basically, when we move D5 and/or D7-branes their induced D3-brane  charge changes, but the contribution to the D3-brane tadpole coming from $F_3 \wedge H_3$ changes by a similar amount. Both effects cancel each other upon integration over $X_6$, and so $\tilde F_5$ is always well-defined and there is no tadpole. We will show this first for the case where we only have D5-branes in our model (which is unrealistic in SUSY compactifications) and then for the more interesting case of models with D7-branes. 

\subsubsection*{Magnetised D5-branes}

Let us consider the case where in our compactification there are only space-time filling D3-branes, D5-branes and fluxes $(F_3, H_3)$. Take a D5-brane in a 2-cycle $\pi_2$ and move it to a new location $\pi_2'$ within the same homology class. The difference in the contribution to the D3-brane tadpole can be measured by the integral
\be
\int_{X_6} \d_4(\pi_2') \wedge B - \int_{X_6} \d_4(\pi_2) \wedge B \, =\, \int_{\pi_2'} B -  \int_{\pi_2} B\, =\, \int_{\Sigma_3} H_3
\label{intB}
\ee
where $\Sigma_3$ is a 3-chain such that $\p \Sigma_3 = \pi_2' - \pi_2$. So in general we see that the contribution to the D3-brane tadpole changes when we move one or several D5-branes.\footnote{Together with this D5 we should move its orientifold image on $\Omega\CR (\pi_2)$. Taking this into account will not change much the discussion, so we will ignore the effect of orientifold images in the following.}

We should however take into account that, in the presence of D5-branes, $F_3$ is not a harmonic form, which is the case when we only have D3-branes. On the contrary, it satisfies the equation
\be
dF_3\, =\, \sum_j \delta_4(\pi^j_{D5})
\label{dF3}
\ee
which we assume corresponds to a globally well-defined but non-closed $F_3$. As a result, when we move the D5-brane from $\pi_2$ to $\pi_2'$ the field strength $F_3$ will change because (\ref{dF3}) changes. Let us represent by $F_3$ the background flux with the D5 located at $\pi_2$, and by $F_3'$ the flux with the D5 located at $\pi_2'$ and $\Delta F_3 = F_3' - F_3$. Then it is easy to see that
\be
d \Delta F_3 \, =\, \delta_4(\pi_{2}') - \delta_4(\pi_2)
\label{dDF3}
\ee
Moreover notice that, even if non-closed, $F_3$ and $F_3'$ are quantised 3-forms on $X_6$. Hence so is $\Delta F_3$, and this fact together with (\ref{dDF3}) can be used to show that \cite{Marchesano:2014bia}
\be
\int_{X_6} \Delta F_3 \wedge \omega_3\,=\, - \int_{\Sigma_3} \omega_3
\label{chain}
\ee
for any closed 3-form $\omega_3$, and where again $\p \Sigma_3 = \pi_2' - \pi_2$ is a 3-chain describing the deformation of the D5-brane location.

We can now use (\ref{chain}) to prove that the D3-brane tadpole induced by the background fluxes changes. Indeed, assuming that there are no NS5-branes in our compactification $H_3$ is a harmonic form and we can apply (\ref{chain}). Hence
\be
\int_{X_6} F_3' \wedge H_3 - \int_{X_6} F_3 \wedge H_3 \, =\,  \int_{X_6} \Delta F_3 \wedge H_3\, =\, - \int_{\Sigma_3} H_3
\ee
This is precisely the opposite as the previous change (\ref{intB}), so tadpoles still cancel when we change the D5-brane position.

\subsubsection*{Magnetised D7-branes}

Let us now consider the case where we have D3-branes and D7-branes, as in the inflationary model of section \ref{HiggsIIB}, and that we move one of the latter as $S_4 \raw S_4'$. The change in D3-brane tadpole is given by 
\be
\oh \left[ \int_{X_6} \d_2(S_4') \wedge B^2 - \int_{X_6} \d_2(S_4) \wedge B^2\right] \, =\, \oh \left[ \int_{S_4'}  B^2 -  \int_{S_4}  B^2 \right]\, =\, \int_{\Sigma_5} H_3 \wedge B
\label{intB2}
\ee
with $\Sigma_5$ a 5-chain with $\p \Sigma_5 = S_4' - S_4$ and describing the above deformation. 

Because the D7-branes are magnetised by the B-field they carry a D5-brane charge, and so again $F_3$ is not a closed 3-form. Instead it must satisfy the equation
\be
dF_3\, =\, \sum_k \delta_2(S^k_{D7}) \wedge B\, =\, dF_1 \wedge B
\label{dF3B}
\ee
where we have used that
\be
dF_1\, =\, \sum_k \delta_2(S^k_{D7})
\ee
So when we move a D7-brane as $S_4 \raw S_4'$, the RR fluxes $(F_1, F_3)$ change to $(F_1', F_3')$ and we can define $(\Delta F_1, \Delta F_3)$ as their difference. In particular we have that
\be
d \Delta F_3 \, =\, \d_2(S_4') \wedge B - \d_2(S_4) \wedge B  \, =\,  d \Delta F_1 \wedge B
\ee
Now it is $\Delta F_1$ the flux that is quantised, and applying the reasoning of \cite{Marchesano:2014bia} we get
\be
\int_{X_6} F_1 \wedge \omega_5 \, =\, - \int_{\Sigma_5} \omega_5
\ee
for any closed 5-form $\omega_5$ and $\Sigma_5$ defined as above. In particular we can take $\omega_5 = B \wedge H_3$. Putting all these things together we arrive at the following variation for the background flux D3-brane charge
\be
\int_{X_6} F_3' \wedge H_3 - \int_{X_6} F_3 \wedge H_3 \, =\,  \int_{X_6} \Delta F_3 \wedge H_3\, =\, \int_{X_6} \Delta F_1 \wedge B \wedge H_3\, =\, - \int_{\Sigma_5} B \wedge H_3
\ee
which again cancels the variation (\ref{intB2}) and guarantees D3-brane tadpole cancellation.

\subsection{Decoupling of moduli fixing from inflation sector}

The  DBI+CS derived inflaton scalar potential that we used assumes implicitly that all the
other moduli of the theory, in particular the complex dilaton $S$ and K\"ahler  ($T^i$) and complex structure ($U^a$) moduli are fixed at
a scale well above the inflation scale. That is, we are assuming a full scalar potential of the form
\beq
V(\sigma,\theta;S,T^i,U^a)\ =\   V_{\rm inflation}(\sigma,\theta;S,T^i,U^a)\ +\  V_{\rm moduli}(S,T^i,U^a)
\eeq
In particular we are assuming that  the potential barriers fixing
$S,T^i,U^a$  are such that the inflaton scalar potential does not  modify in a substantial manner the moduli dynamics.
This may proof hard for an inflaton scale $\simeq 10^{16}$ GeV as suggested by BICEP2,  since that would require the 
compactification $M_c$ and string scale $M_s$  not much  below the reduced Planck scale $M_p\simeq 10^{18}$ GeV.
This is a general problem for all string inflation models with large field inflation, see \cite{Blumenhagen:2014nba,Hayashi:2014aua,Hebecker:2014kva}.

Here we would only like to add that the string models with the inflaton identified with open string moduli may be more flexible than 
closed string axion models in this regard. Indeed, the inflaton dynamics is localised in a D-brane sector of the theory  rather than in the bulk. 
Then, as shown in eq.(\ref{warpping}), the local $G_3$ flux felt by the  D7's  (fixing the inflaton mass) may be suppressed compared to
the flux felt by the moduli in the bulk by a warp factor $Z^{-1/2}$. In this way the barriers of  the potential $V_{\rm moduli}$ 
could be substantially higher than those in $V_{\rm inflation}$. This would help in understanding the decoupling of the
moduli fixing dynamics from the inflaton dynamics in a natural way.

\section{Some further cosmological issues}

Our study of the cosmological perturbations induced in the Higgs-otic scenario has been incomplete in several
respects. In particular, while 
single inflaton models predict a Gaussian and adiabatic spectrum, it is well known that 
multi-inflaton models may in general give rise to non-Gaussianities  as well as 
isocurvature (entropy) perturbations \cite{Wands:2007bd,GarciaBellido:1995qq}.
The Higgs inflaton potential here studied has two fields involved in inflation, $\sigma$ and $\theta$, so that 
in principle one can think that non-Gaussianities and/or isocurvature perturbations could arise.
Concerning non-Gaussianities,  one does not expect any effect in our scheme since it is known that 
2-field models yield non-linear parameters $f_{NL}$ proportional to the slow roll parameters $\epsilon,\eta$, 
see e.g. \cite{Vernizzi:2006ve,Yokoyama:2007uu}.
On the other hand two field models can yield in general isocurvature perturbations \cite{Gordon:2000hv}.
 In some simple cases, 
like the  so called double chaotic inflation and others,   such effects are suppressed
\cite{Huston:2011fr,Huston:2013kgl,Price:2014xpa}. In our case, for small fields the structure is that of double chaotic inflation
but this is corrected in a sizeable way for the relevant case with large fields, with strong rescaling effects.
 It would be interesting to study 
the possible generation of non-adiabatic perturbations in our scheme. 
 We leave a more complete analysis of these issues for future work.

Another interesting issue is that of reheating, which is expected to be quite efficient in this
Higgs-otic scenario.
At the end of inflation the universe is extremely cold and a reheating 
process occurs in which the inflaton oscillates around its minimum. The inflaton transfers all its energy
through its decay into relativistic particles. The inflaton must couple to the SM particles 
which will end up in thermal
equilibrium and give rise to the big-bang initial conditions.
  A generic problem in string cosmologies   in which the inflaton is identified with a closed string mode
  (like e.g. an axion) is that the inflaton reheats predominantly into hidden sector fields or moduli rather than
  into SM fields.
  In our case, obviously, the inflaton is a Higgs field which will decay predominantly into top quarks and  gauge bosons
  and this problem is automatically avoided. 
  The decay rate will typically be of order
  \beq
  \Gamma_H\ \simeq \ \frac {h^2m_I}{8\pi} \ ,
  \eeq
  with $h$ the top Yukawa coupling or a gauge coupling and $m_I\simeq M_{SS}\simeq 10^{13}$ GeV is the inflaton mass,
  which is of the order of the SUSY breaking scale $M_{SS}$. Perturbative reheating ends when the expansion rate
  of the universe given by the Hubble constant $H=\sqrt{8\pi\rho/3M_p^2}$ is of order of the total inflaton decay rate. 
  The SM interactions are strong enough so that thermal equilibrium is reached with a reheating  temperature
  (see e.g. \cite{Kofman:1996mv,Kofman:1997yn,Bassett:2005xm})
  \beq
  T_R\ \simeq \ 0.2 \sqrt{\Gamma_H  M_p} \ \simeq \ 10^{13}\ GeV \ ,
  \eeq
where we have set $h\simeq 1/2, m_I\simeq 10^{13}$ GeV. This is high enough so that leptogenesis may take place in the
usual way at an intermediate scale.

\section{Final comments and conclusions}\label{final}

In this paper we have completed in several directions the proposal in \cite{Ibanez:2014kia} of identifying the inflaton with 
a heavy MSSM Higgs field in a chaotic-like inflation fashion, dubbing the resulting scenario as {\em Higgs-otic inflation}.  
In this scheme, the inflaton mass scale is identified with the size of the SUSY breaking soft terms,
$m_I\simeq M_{SS}\simeq 10^{12}-10^{13}$ GeV. Such large value of $M_{SS}$ requires the SM Higgs doublet to have a fine-tuned 
mass. As a result, the role of supersymmetry is not to solve the hierarchy problem but instead to stabilise the SM Higgs potential in the ultraviolet
as in \cite{hebecker1,imrv,Ibanez:2013gf,hebecker2}, this being nicely consistent with a SM Higgs mass $m_H \simeq 126$ GeV. 
The implementation of inflation requires trans-Planckian excursions of the inflaton/Higgs field, which implies that we need to have
a certain control over Planck scale corrections, i.e. a theory of quantum gravity.  Our leading theory of quantum gravity is
string theory, which we take as the underlying fundamental theory in which our explicit realisations are based. 
Notice that the fact that the SM Higgs is fine-tuned also points in the 
direction of string theory, where a large landscape of solutions may justify the fine-tuning in terms of anthropic considerations.

The vevs of MSSM Higgs doublets in string compactifications may be embedded into string theory as either Wilson lines or 
D$p$-brane position moduli.  We have discussed in detail a toy example realised in terms of a IIB compactifications where
the MSSM is realised via D7-branes at singularities, and where  the Higgs vevs is realised in terms of D7-brane position moduli. 
Such Higgs vev parametrise the D7-brane position in its transverse space, which in this case is ${\bf T}^2$. 
In this setup soft terms creating a potential for the Higgs/inflaton fields are induced by ISD three-form closed string fluxes. 

A particularly important advantage of this realisation is that we can then compute the scalar potential in terms of the DBI+CS action, 
which give us the inflation potential to all orders in $\alpha'$.  The leading term of this potential for small field may be also obtained 
in terms of a $N=1$ supergravity Lagrangian assuming SUSY-breaking is induced by the auxiliary fields of the K\"ahler moduli.  
However, for large field, which is relevant for chaotic-like inflation, the $\alpha '$ 
corrections in the DBI+CS action rescale the Higgs fields, giving rise to a flattening of the scalar potential, which becomes 
almost linear for large fields. This effect is not captured by the (2-derivative) $N=1$ supergravity formulation.
We have also discussed how this protection of the potential against arbitrary Planck suppressed corrections may be
understood from a Kaloper-Sorbo effective action point of view.

The resulting inflaton/Higgs potential is a 2-field model involving the neutral components of the fields $h$ and $H$. The parameters of the model are the flux-dependent parameters $\hat G$ and $A$ defined in the main text, 
as well as the initial field values $\sigma(0)$ and ${\tilde \theta}(0)$. A distinctive feature the Higgs-otic scenario is that the flux parameters are constrained in order for a massless SM Higgs to survive.
Further imposing 50-60 e-folds constraints substantially the slow roll parameters  and the resulting adiabatic
perturbations  that  one obtains correspond to a (broad) line in the $r-n_s$ 
plane. In particular the model predicts $r>0.07$, with most inflaton initial conditions leading to values of $r$ close to this lower limit. These values of $r$ will be tested experimentally in the near future.

The Higgs-otic idea is conceptually quite attractive, since two apparent very different phenomena like
Higgs physics and cosmological inflation are intimately connected.  Getting light 
scalars is probably  a rare event in the string landscape so that the merging of two fine-tunings, one
for the SM Higgs and another for the inflaton would be economical in this sense.
The form of the inflaton potential is 
restricted by low-energy particle physics data (i.e. the SM Higgs mass) and the couplings of the inflaton 
are related to known low-energy couplings.  Efficient perturbative reheating is natural, given the large Higgs
couplings to SM particles.

There are a good number of directions in which to complete the present study.  From the string theory  model
building point of view, our detailed analysis is based on a local two-family model in which the Higgs-inflaton fields
parametrise the position of a D7-brane on a ${\bf T}^2$. It would be interesting to construct specific globally consistent 
three-family models embedding and/or extending this kind of structure to other more general geometries with D$p$-branes
travelling along more general surfaces. Another direction to explore is the construction of analogous models 
with wandering D3-branes instead of D7's. Local models in which the MSSM Higgs vevs are parametrised in terms
of D3 positions are easy to construct. However the implementation of monodromy in terms of
fluxes is more subtle in this case, and it needs of closed string fluxes of the IASD kind. However, general compactifications 
may also include IASD fluxes and it would be interesting to implementing Higgs-otic models based on D3 or D7-branes in such
backgrounds. Finally, it would also be interesting to consider globally consistent
heterotic models in which the Higgs vevs parametrise continuous Wilson lines and the potential energy
could come from geometric fluxes. 

Another important topic is the issue of the fixing of the closed string moduli of the theory, which 
we have taken as frozen degrees of freedom. It would be important to find a regime in which moduli fixing
occurs at scales well above the inflation scale, so that the inflaton potential does not substantially
modify the moduli fixing potential. In this context we have emphasised that a strong non-constant warp factor $Z$ 
may help in separating  the inflaton and moduli dynamics. 

While the Higgs-otic scheme developed here is quite concrete, some of our findings may be readily applied to 
slightly different scenarios. For instance, if SUSY particles were found at LHC, the present Higgs-otic scenario 
would be ruled out, since it relies on a heavy SUSY spectrum with large masses of order $M_{SS}\simeq 10^{12}-10^{13}$ GeV. 
Nevertheless, a similar idea could be applied to GUT Higgs multiplets or SM singlets. In that case SUSY preserving $(2,1)$ fluxes 
could give a large SUSY mass term for the GUT Higgs and a potential could be derived from the DBI+CS action
yielding a result similar to the $A=0$ limit of the Higgs-otic potential.  The results for inflation would then be similar to
the one-field limit with $A=0$ discussed in the text.

From the cosmological point of view,  it would be interesting to perform a more complete study of perturbations including 
isocurvature perturbations as well a more detailed analysis of the reheating process. We hope to address these
issues in future work.

\vspace{1cm}

\noindent 
\centerline
{\bf Acknowledgements}
\\

We would like to thank F.~Pedro and A.~Uranga for useful discussions. This work has been supported by the ERC Advanced Grant SPLE under contract ERC-2012-ADG-20120216-320421, by the grant FPA2012-32828 from the MINECO, the REA grant agreement PCIG10-GA-2011-304023 from the People Programme of FP7 (Marie Curie Action), and the grant SEV-2012-0249 of the ``Centro de Excelencia Severo Ochoa" Programme. F.M. is supported by the Ram\'on y Cajal programme through the grant RYC-2009-05096.  I.V. is supported through the FPU grant AP-2012-2690. 

\newpage


\appendix

\section{The DBI+CS computation}
\label{DBI}

From the viewpoint of the local SU(3) structure the antisymmetric flux $G_3$ transforms as $20=10+\bar{10}$ with the $\bar{10}$ and $10$ representations corresponding respectively to the Imaginary Self Dual (ISD) $G_3^+$ and  Anti Imaginary Self Dual (AISD) $G_3^-$ components of the 3-form flux, defined as
\begin{equation}
G_3^{\pm} = \frac12 (G_3 \mp i*_6 G_3)\ , \qquad *_6 G_3^\pm = \pm i G_3^\pm
\end{equation}
These components can be further decomposed into irreducible 
representations of SU(3). Thus, ISD fluxes in the $\mathbf{\bar{10}}$ are decomposed according to $\mathbf{\bar{10}} = \mathbf{\bar{6}} + \mathbf{\bar{3}} + \mathbf{\bar{1}}$, corresponding to $(2,1)$-form, $(1,2)$-form and $(0,3)$-form fluxes respectively. Throughout this paper we have only considered $G_{(2,1)}$ and $G_{(0,3)}$ fluxes since the $\mathbf{\bar{3}}$ representation corresponds to (1,2) non-primitive component of the flux, incompatible with the ${\bf Z}_4$ action of the orbifold. In tensorial notation, they are denoted by 
\beqa
G_{(0,3)}=G_{\bar 1\bar 2\bar 3}d\bar z_1\wedge d\bar z_2\wedge d\bar z_3\\
G_{(2,1)}=G_{ij\bar k}d z_i\wedge d z_j\wedge d \bar z_k
\eeqa
In addition, we only consider the component $G_{ij\bar 3}$ of $G_{(2,1)}$ since the other flux components generically lead to Freed-Witten anomalies in the worldvolume of the D7-branes and are not invariant under the ${\bf Z}_4$ action either.

The effective action for the microscopic fields of a system of D7-branes in the 10d Einstein frame is given by the Dirac-Born-Infeld (DBI) + Chern-Simons (CS) actions
\beq
S_{DBI}=-\mu_7 g_s^{-1}\text{STr}\left(\int d^8\xi\sqrt{-\text{det}(P[E_{MN}]+\sigma F_{MN})}\right)
\label{ap:DBID7}
\eeq
\beq
S_{CS}=\mu_7g_s\text{STr}\left(\int d^8\xi P\left[-C_6\wedge B_2+C_8\right]\right)
\label{ap:CSD7}
\eeq
where
\beq
E_{MN}=g_s^{1/2}G_{MN}-B_{MN}\quad ;\quad \sigma=2\pi\alpha'\quad ;\quad \mu_7=(2\pi)^{-3}\sigma^{-4}g_s^{-1}
\eeq
$P[\cdot]$ denotes
the pullback of the 10d background onto the D7-brane worldvolume and `STr' is the symmetrised trace over gauge indices. Finally, we have ignored the factor ${\rm det} (Q_{mn})$ which, as discussed in the main text, gives rise to the D-term potential.

The determinant in the DBI action  can be factorised between Minkowski and the internal space as follows
\begin{multline}
\textrm{det}(P[E_{MN}] + \sig F_{MN})=g_s^{4}\,\textrm{det}\left(\eta_{\mu\nu}+2Z\sigma^2\partial_\mu\Phi\partial_\nu\bar\Phi+Z^{1/2}g_s^{-1/2}\sigma F_{\mu\nu}\right)\\
\cdot\textrm{det}\left(g_{ab}+Z^{-1/2}g_s^{-1/2}\sigma F_{ab}-Z^{-1/2}g_s^{-1/2}B_{ab}-\sigma^2([A_a,\Phi][A_b,\bar\Phi]+[A_a,\bar\Phi][A_b,\Phi])\right)
\end{multline}
where $\mu,\nu$ label the 4d non-compact directions and $a,b$ the internal D7-brane dimensions. Then, using the matrix identity
\bea
\label{det}
\textrm{det}(1+\ep M)& = &1 + \ep\,\textrm{tr }M - \ep^2\left[\frac{1}{2}\textrm{tr }M^2 - 
\frac12(\textrm{tr }M)^2\right] \\ \nonumber
&+& \ep^3\left[\frac13\textrm{tr }M^3-\frac12(\textrm{tr }M)(\textrm{tr }M^2)+\frac16(\textrm{tr }M)^3\right]\\ \nonumber
& - &\ep^4\left[\frac14\textrm{tr }M^4-\frac18(\textrm{tr }M^2)^2-\frac13(\textrm{tr }M)(\textrm{tr }M^3)\right.\\ \nonumber
& + & \left.\frac14(\textrm{tr }M)^2(\textrm{tr }M^2)+\frac{1}{24}(\textrm{tr }M)^4\right] + \CO(\ep^5)
\eea
we obtain on the one hand that 
\be
-\textrm{det}\left(\eta_{\mu\nu}+2Z\sigma^2\partial_\mu\Phi\partial_\nu\bar\Phi+Z^{1/2}g_s^{-1/2}\sigma F_{\mu\nu}\right)
=1+2Z\sigma^2\left(\partial_\mu\Phi\partial^\mu\bar\Phi-\frac{g_s^{-1}}{4} F_{\mu\nu}F^{\mu\nu}\right) 
\label{DBIext}
\ee
where we have neglected terms with more than two derivatives in 4d, in agreement with the slow-roll condition that will be imposed on this system. On the other hand we have that 
\be
\textrm{det}\left(g_{ab}+Z^{-1/2}g_s^{-1/2}\CF_{ab}\right)\, =\, \textrm{det}(g_{ab})\, f(\CF)^2
\label{DBIint}
\ee
where $\CF_{ab} = \sig F_{ab} - B_{ab}$ and 
\beq
f(\CF)^2=1+\frac12 Z^{-1}g_s^{-1}\CF_{ab}\CF^{ab}-\frac{g_s^{-2}}{4}Z^{-2}\CF_{ab}\CF^{bc}\CF_{cd}\CF^{da}
+\frac{g_s^{-2}}{8}Z^{-2}\left[\CF_{ab}\CF^{ab}\right]^2
\label{fF}
\eeq
Notice that for simplicity in the l.h.s. of  (\ref{DBIint}) we have not included couplings of the form $[A,\Phi]$ which will not be relevant for the scalar potential of the moving D7-brane analysed in the main text. Moreover, unlike in (\ref{DBIext}), when deriving (\ref{DBIint}) we have not made any approximation. Indeed by taking 
\be
M \, =\,  g^{-1} \CF \quad \quad {\rm and}\quad \quad \ep\, =\, (g_sZ)^{-1/2}
\ee
and using the fact that $M$ is a $4 \times 4$ matrix it is easy to see that the expansion of eq.(\ref{det}) ends at order $\ep^4$. Finally, using that 
\be
\tr g^{-1} \CF\, =\, \tr \CF^t g^{-1 \, t}\, =\, - \tr g^{-1} \CF 
\ee
so that $\tr M = \tr M^3\, =\, 0$, we are led to the above result, and then eqs.(\ref{detP}) and (\ref{fB0}) in the main text follow by simply replacing $\CF \raw - B$. 

In fact for a $4 \times 4$ matrix $M$ with these properties we also have the identity 
\be
\text{det}(1+\ep M)\, =\,1 - \ep^2\frac{1}{2}\textrm{tr }M^2+ \ep^4 {\rm det} M
\label{finaldet}
\ee
which is easy to prove by looking at the characteristic polynomial of $M$. This allows us to write
\be
\textrm{det}(1+\ep M)\, =\,  1 + \ep^2 \CF^2  + \ep^4 \frac{1}{4} \left(\CF \wedge \CF\right)^2
\ee
where the square of a $p$-form $\omega$ is defined as $\omega \cdot \omega$ with
\be
\omega_p \cdot \chi_p\, =\, \frac{1}{p!}  \omega_{a_1\dots a_p} \chi^{a_1\dots a_p}
\ee
Now, whenever $\CF$ is a self or antiselfdual two form
\be
\CF \, =\, \pm *_4 \CF
\ee
we will have that
\be
\left(\CF \wedge \CF\right)^2 \, =\,  \left(\CF \wedge *_4 \CF\right)^2\, =\,   \left(\CF^2 d{\rm vol}_{S_4} \right)^2\, =\, (\CF^2)^4 
\ee
and so
\be
\textrm{det}(1+\ep M)\, =\, \left(1 + \oh \ep^2 \CF^2\right)^2
\ee
obtaining a perfect square. This is will be the case for our wandering D7-brane system, since there $\CF = - B$ will be a $(2,0)+(0,2)$ form due to (\ref{Bcomp}).\footnote{To connect with the derivation of the perfect square in eq.(\ref{fB}) notice that in our case we have the identity 
\be
\nonumber
{\rm det} M\, =\, - \frac14\textrm{tr }M^4+\frac18(\textrm{tr }M^2)^2
\ee
and that $\CF$ (anti)selfdual translates into $4 \tr M^4 = (\tr M^2)^2$ so that finally $16\, {\rm det} M = (\tr M^2)^2 = 4B^2$.} 

Putting everything together we find that the relevant part of the DBI action is given by
\beq
S_{DBI}=-\mu_7 g_s{\rm STr}\int d^8\xi \sqrt{ \textrm{det}(g_{ab}) f(\CF)^2\left(1+2Z\sigma^2 D_\mu\Phi D_\mu\bar\Phi+\frac12 Zg_s^{-1}\sigma^2F_{\mu\nu}F^{\mu\nu}\right)}
\eeq
Expanding this expression to second order in 4d derivatives and setting $\CF = -B$ we obtain
\be
S_{DBI}=-\mu_7 g_s\text{STr}\int d^8\xi \sqrt{ \textrm{det} g}  f(B)\left[1+Z\sigma^2 D_\mu\Phi D^\mu\bar\Phi+\frac14 Zg_s^{-1}\sigma^2F_{\mu\nu}F^{\mu\nu} )\right]
\ee
which is the expression used in the main text (c.f.(\ref{DBIfin})) where for simplicity $\sqrt{ \textrm{det} g} = 1$ has been taken.

\end{document}